\documentclass[iop]{emulateapj}
\usepackage{amssymb,amsmath}
\usepackage{graphicx,natbib}
\usepackage{epstopdf}

\shorttitle{High contrast with arbitrary apertures}
\shortauthors{Pueyo $\&$ Norman}

\begin{document}

\title{High Contrast Imaging with an Arbitrary Aperture: Active Correction of Aperture Discontinuities}

\slugcomment {ApJ Accepted 11-26-2012}

\author{Laurent Pueyo\altaffilmark{1}, Colin Norman \altaffilmark{1}}

\altaffiltext{1}{Department of Physics and Astronomy, Johns Hopkins University, Baltimore, MD, USA}

\email{email: lap@pha.jhu.edu}

\begin{abstract}
We present a new method to achieve high-contrast images using segmented and/or on-axis telescopes. Our approach relies on using two sequential Deformable Mirrors to compensate for the large amplitude excursions in the telescope aperture due to secondary support structures and/or segment gaps. In this configuration the parameter landscape of Deformable Mirror Surfaces that yield high contrast Point Spread Functions is not linear, and non-linear methods are needed to find the true minimum in the optimization topology. We solve the highly non-linear Monge-Ampere equation that is the fundamental equation describing the physics of phase induced amplitude modulation. We determine the optimum configuration for our two sequential Deformable Mirror system and show that high-throughput and high contrast solutions can be achieved using realistic surface deformations that are accessible using existing technologies. We name this process Active Compensation of Aperture Discontinuities (ACAD). We show that for geometries similar to JWST, ACAD can attain at least $10^{-7}$ in contrast and an order of magnitude higher for both the future Extremely Large Telescopes and on-axis architectures reminiscent of HST. We show that the converging non-linear mappings resulting from our Deformable Mirror shapes actually damp near-field diffraction artifacts in the vicinity of the discontinuities. Thus ACAD actually lowers the chromatic ringing due to diffraction by segment gaps and strut's while not amplifying the diffraction at the aperture edges beyond the Fresnel regime. This outer Fresnel ringing can be mitigated by properly designing the optical system. Consequently, ACAD is a true broadband solution to the problem of high-contrast imaging with segmented and/or on-axis apertures. We finally show that once the non-linear solution is found, fine tuning with linear methods used in wavefront control can be applied to further contrast by another order of magnitude. Generally speaking, the ACAD technique can be used to significantly improve a broad class of telescope designs for a variety of problems. 
\end{abstract}


\maketitle 
\section{Introduction}\label{sect:intro}
Exo-planetary systems that are directly imaged using existing facilities \citep{2008Sci...322.1348M,2008Sci...322.1345K,2010Sci...329...57L} give a unique laboratory to constrain planetary formation at wide separations \citep{2005ApJ...621L..69R,2009ApJ...707...79D,2010ApJ...710.1375K,2010PASP..122..905J}, to study the planetary luminosity distribution at critical young ages \citep{2012ApJ...745..174S,2008ApJ...683.1104F} and the atmospheric properties of low surface gravity objects \citep{2011ApJ...735L..39B,2011ApJ...733...65B,2010ApJ...723L.117M,2011ApJ...737...34M}. Upcoming surveys, conducted with instruments specifically designed for high-contrast \citep{2006SPIE.6269E..24D,2007arXiv0704.1454G,2011PASP..123...74H}, will unravel the bulk of this population of self-luminous jovian planets and provide an unprecedented understanding of their formation history. Such instruments will reach the contrast required to achieve their scientific goals by combining Extreme Adaptive Optics systems (Ex-AO, \cite{2005JOSAA..22.1515P}), optimized coronagraphs  \citep{2011ApJ...729..144S,2003A&A...404..379G,2000PASP..112.1479R} and  nanometer class wavefront calibration \citep{2007JOSAA..24.2334S,wallace:74400S,pueyo:77362A}. In the future, high-contrast instruments on Extremely Large Telescopes will focus on probing planetary formation in distant star forming regions \citep{2006SPIE.6272E..20M},  characterizing both the spectra of cooler gas giants \citep{2010SPIE.7736E..55V} and the reflected light of planets in the habitable zone of low mass stars. The formidable contrast necessary to investigate the presence of biomarkers at the surface of earth analogs ($ > 10^{10}$) cannot be achieved from the ground beneath atmospheric turbulence and will require dedicated space-based instruments \citep{2005ApJ...629..592G}.

The coronagraphs that will equip upcoming Ex-AO instruments on 8 meter class telescopes have been designed for contrasts of at most $\sim10^{-7}$. Secondary support structures (or spiders: 4 struts each $1$ cm wide, $\sim 0.3 \%$ of the total pupil diameter in the case of Gemini South) have a small impact on starlight extinction at such levels of contrasts. In this case, coronagraphs have thus been optimized on circularly symmetric apertures, which only take into account the central obscuration \citep{2011ApJ...729..144S}. However,  high-contrast instrumentation on  future observatories will not benefit from such gentle circumstances. ELTs will have to support a substantially heavier secondary than 8 meter class observatories do, and over larger lengths: as a consequence the relative area covered by the secondary support will increase by a factor of 10 ($30$ cm wide spiders, occupying $\sim 3 \%$ of the pupil diameter in the case of TMT). This will degrade the contrast of coronagraphs only designed for circularly obscured geometries by a factor $ \sim 100$, when the actual envisioned contrast for an ELT exo-planet imager can be as low as $\sim10^{-8}$ \citep{2006SPIE.6272E..20M}. While the trade-offs associated with minimization of spider width in the space-based case have yet to be explored, secondary support structures will certainly hamper the contrast depth of coronagraphic instruments of such observatories at levels that are well above the $10^{10}$ contrast requirement. As a consequence, telescope architectures currently envisioned for direct characterization of exo-earths consist of monolithic, off-axis, and thus un-obscured, telescopes \citep{2008SPIE.7010E..59G,2010SPIE.7731E..67T}. Coronagraphs for such architectures take advantage of the pupil symmetry to reach a theoretical contrast of ten orders of magnitude \citep{2005ApJ...622..744G,2003ApJ...599..686V,2003ApJ...590..593V,2005ApOpt..44.1117K,mawet:773914,2002ApJ...570..900K,2003A&A...397.1161S}. However, using obscured on-axis and/or segmented apertures take full advantage of the limited real estate associated with a given launch vehicle and can allow larger apertures that increase the scientific return of space-based direct imaging survey. Recent solutions can mitigate the presence of secondary support structures in on-axis apertures. However these concepts present practical limitations: APLCs on arbitrary apertures \citep{2009ApJ...695..695S} and Shaped Pupils \citep{2011OExpr..1926796C} suffer from throughput loss for very high contrast designs, and PIAAMCM \citep{2010ApJS..190..220G} rely on a phase mask technology whose chromatic properties have not yet been fully characterized. Moreover segmentation will further complicate the structure of the telescope's pupil: both the amplitude discontinuities created by the segments gaps and the phase discontinuities resulting from imperfect phasing will thus further degrade coronagraphic contrast. Devising a practical solution for broadband coronagraphy on asymmetric, unfriendly apertures is an outstanding problem in high contrast instrumentation. The purpose of the present paper is to introduce  a family of practical solutions to this problem. As their ultimate performances depend strongly on the pupil structure we limit the scope of this paper to a few characteristic examples. Full optimization for specific telescope geometries can be conducted as needed.

\begin{figure}[h!]
\begin{center}
\includegraphics[width=3in]{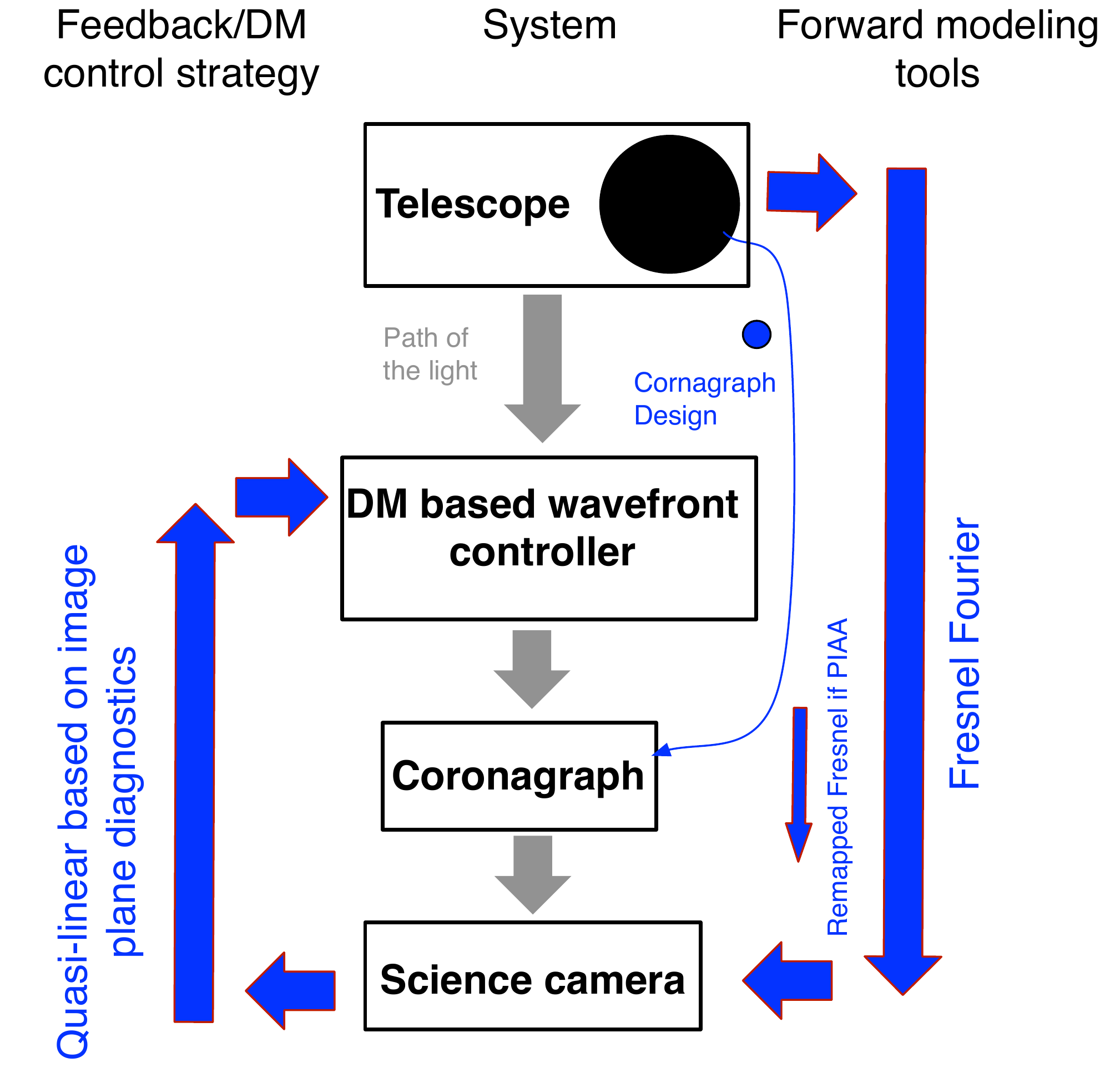}
\includegraphics[width=3in]{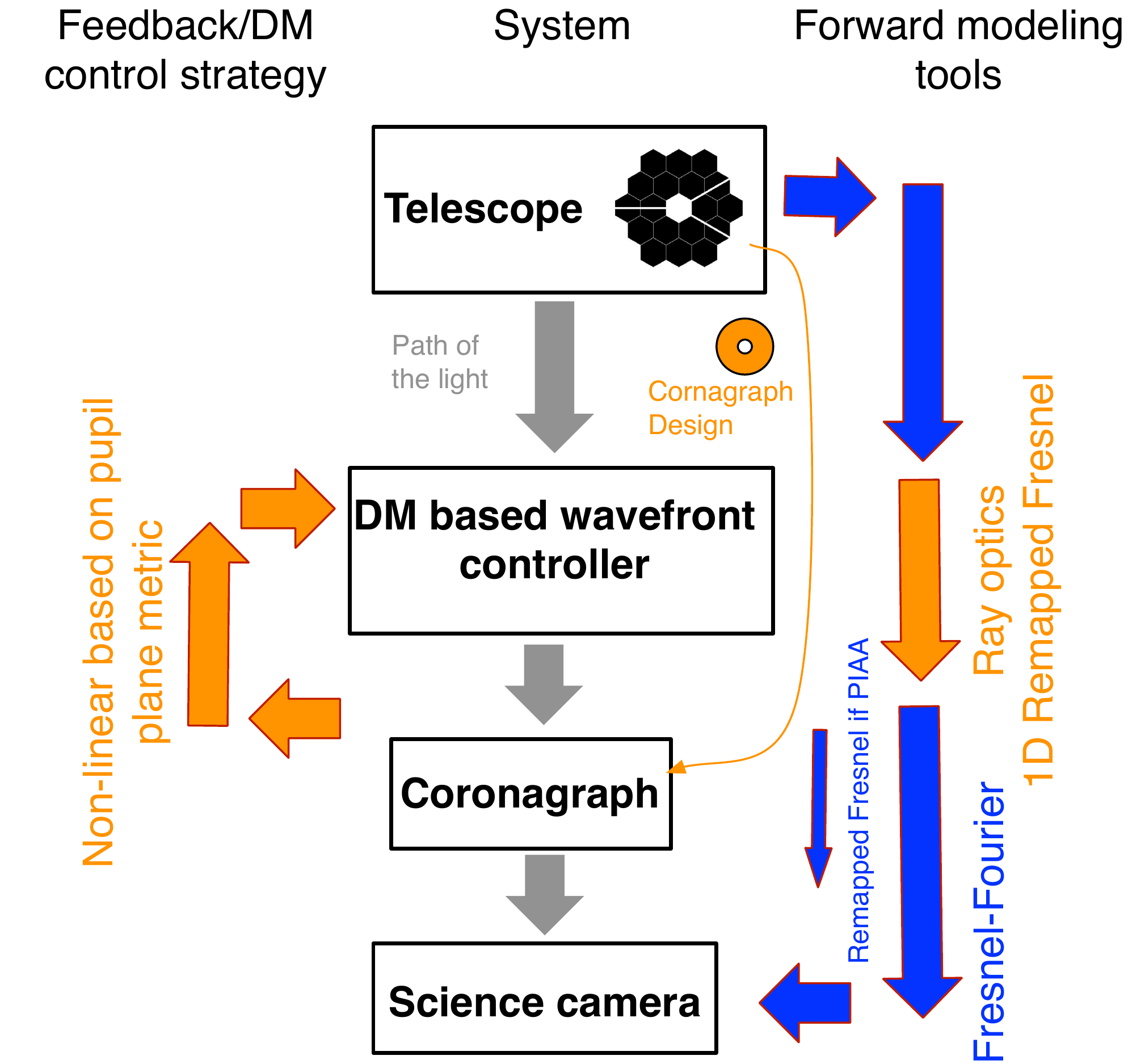}
\end{center}
\caption[]{{\bf Top, blue:} Envisioned architecture of future exo-earth imaging missions: a monolithic un-obscured telescope feeds a coronagraph designed on a circular aperture. The wavefront errors are corrected using two sequential Deformable Mirrors (DMs) that are controlled using a quasi-linear feedback loop based on image plane diagnostics. The propagation between optical surfaces,  between the DMs in particular is assumed to occur in the Fresnel regime. {\bf Bottom, orange:} ACAD solution: the coronagraph is designed for a circular geometry around the central obscuration. The two sequential DMs are controlled in the non-linear regime based on a pupil plane cost function. The propagation between the DMs is modeled using ray optics, and a we conduct a quantitative one-dimensional analysis of  the diffraction artifacts.}
\label{fig:State}
\end{figure}

The method proposed here takes advantage of state-of-the art Deformable Mirrors in modern high-contrast instruments to address the problem of pupil amplitude discontinuities for on-axis and/or segmented telescopes. Indeed, coronagraphs are not sufficient to reach the high contrast  required to image faint exo-planets: wavefront control is needed to remove the light scattered by small imperfections on the optical surfaces \citep{BrownBurrows}. Over the past eight years, significant progress has been made in this area, both in the development of new algorithms \citep{2006ApJ...638..488B,Giveon:07} and in the experimental demonstration of high-contrast imaging with a variety of coronagraphs \citep{Giveon:07,2007Natur.446..771T,2010PASP..122...71G,2011SPIE.8151E...1B}. These experiments rely on a system with a single Deformable Mirror which is controlled based on diagnostics downstream of the coronagraph, either at the science camera or as close as possible to the end detector \citep{wallace:74400S,pueyo:77362A}. Such configurations are well suited to correct phase wavefront errors arising from surface roughness but have limitations in the presence of pure amplitude errors (reflectivity), or phase-induced amplitude errors, which result from the propagation of surface errors in optics that are not conjugate to the telescope pupil \citep{2006ApOpt..45.5143S,2007ApJ...666..609P}. Indeed a single DM can only mimic half of the spatial frequency content of amplitude errors  and compensate for them only on one half of the image plane (thus limiting the scientific field of view)  over a moderate bandwidth. In theory, architectures with two sequential Deformable Mirrors, can circumvent this problem and create a symmetric broadband high contrast PSF  \citep{2006ApOpt..45.5143S,2007ApJ...666..609P}. The first demonstration of symmetric dark hole was reported in \cite{Pueyo:09} and has since been generalized to broadband by \cite{2011SPIE.8151E..31G}. In such experiments the coronagraph has been designed over a full circular aperture, the DM control strategy is based on a linearization of the relationship between surface deformations and electrical field at the science camera, and the modeling tools underlying the control loop consist of classic Fourier and fresnel propagators. This is  illustrated on the left panel of  Fig.~\ref{fig:State}. As a consequence, a wavefront control system composed of two sequential Deformable Mirrors is currently the baseline architecture of currently envisioned coronagraphic space-based instruments \citep{Stuart2DMsSPIE,2011SPIE.8151E..12K} and ELT planet imagers \citep{2006SPIE.6272E..20M}. One can thus naturally be motivated to investigate if such wavefront control systems can be used to cancel the light diffracted by secondary supports and segments in large telescopes, since such structures are amplitude errors, albeit large amplitude errors. 

The purpose of our study is to demonstrate that indeed a two Deformable Mirror  (DM for the remainder of this paper) wavefront control system can mitigate the impact of the pupil asymmetries, such as spiders and segments, on contrast and thus enable high contrast on unfriendly apertures. In \S.~\ref{sec:CentralObsc} we first present a new approach to coronagraph design in the presence of a central obscuration, but in the absence of spiders or segments. We show that for coronagraphs with a pupil apodization and an opaque focal plane stop, contrasts of $10^{-10}$ can be reached for any central obscuration diameter, provided that the Inner Working Angle is large enough. Naturally the secondary support structures, and in the segmented cases, segment gaps, will degrade this contrast. As our goal is to use two DMs as an amplitude modulation device, we first briefly review in \S.~\ref{sec:physicsofAM} the physics of such a modulation. In \S.~\ref{sec:MADM} we introduce a solution to this problem: we show how to compute DM surfaces that mitigate spiders and segment gaps. Current algorithms used for amplitude control operate under the assumption that amplitude errors are small, and thus they cannot be readily applied to the problem of compensating aperture discontinuities, which have inherently large reflectivity non-uniformities. Fig.~\ref{fig:State} illustrates how the present manuscript introduces a control strategy for the DMs that is radically different from previously published amplitude modulators in the high-contrast imaging literature. Our technique, which we name Active Compensation of Aperture Discontinuities (hereafter ACAD), finds the adequate DM shapes in the true non-linear large amplitude error regime. In this case the DMs' surfaces are calculated as the solution of a non-linear partial differential equation, called the Monge-Ampere Equation. We describe our methodology to solve this equation in \S.~\ref{sec:MADM} and illustrate each step using an obscured and segmented geometry similar to JWST. As ACAD DM surfaces are prescribed in the ray-optics approximation this is a fundamentally broadband technique provided that chromatic diffractive artifacts, edge ringing in particular, do not significantly impact the contrast. This is what we  discuss in \S.~\ref{sec:Chrom}. We find that when remapping small discontinuities with Deformable Mirrors,  the spectral bandwidth is only limited by wavelength-dependent edge-diffraction ringing in the Fresnel approximation (as discussed in \citet{2007ApJ...666..609P} for instance). High-contrast instruments where this ringing is mitigated have already been designed; while future work in high precision optical modeling is necessary to fully quantify the true chromatic performances of ACAD, we do not expect these effects to be a major limitation to broadband operations. In \S.~\ref{sec:Resuls} we present the application of our method to various observatory architectures. Note that the contrast levels stated in \S.~\ref{sec:Resuls}  represent a non-optimal estimate of ACAD performances with on-axis and/or segmented apertures. Our calculations are carried out in the absence of atmospheric turbulence, quasi-static wavefront errors or coronagraphic manufacturing defects. We discuss these limitations in \S.~\ref{sec:Discuss}, with a specific emphasis on quasi-static phase errors in a segmented telescope. We show that when the aperture discontinuities are thin enough field distortion is negligible for spatial frequencies within the field of view defined by the DMs controllable spatial frequencies. We then discuss issues associated with phase discontinuities when applying ACAD to a segmented telescope. We show that they can be corrected by superposing single DM classical wavefront control solutions to the ACAD shape of the second DM. We finally argue, that should  high precision diffractive models be developed, then the solutions presented herein can be used as the starting point of dual DMs iterative algorithms relying on an image-plane based metric and thus lead to higher contrast than reported herein. Most of the future exo-planet imagers, either on ELTs or on future space missions, are envisioned to control their wavefront in real time with two sequential DMs. The method presented in this manuscript thus renders high contrast coronagraphy possible on {\em any} observatory geometry without adding any new hardware.

\section{Coronagraphy with a Central Obscuration}
\label{sec:CentralObsc}

\subsection{Optimizing pupil apodization in the presence of a central obscuration}
Because the pupil obscuration in an on-axis telescope is large it will be very difficult to mitigate its impact with DMs with a limited stroke. Indeed, the main hindrance to high-contrast  coronagraphy in on-axis telescope is the presence of the central obscuration: it often shadows much more than $10 \%$ of the aperture width while secondary supports and segments gaps cover $\sim 1 \%$. We thus first focus of azimuthally symmetric coronagraphic designs in the presence of a central obscuration.  This problem (without the support structures) has been addressed in previous publications either using circularly symmetric pupil apodization \citep{2011ApJ...729..144S} or a series of phase masks \citep{2011OptL...36.1506M}. Both solutions however are subject to limitations. The singularity at the center of the Optical Vector Vortex Coronagraph might be difficult to manufacture and a circular opaque spot thus lies in the  central portion of the phase  mask ($< \lambda/D$) which results in a degradation of the ideal contrast of such a coronagraph \citep{2011SPIE.8151E..12K}. The solutions in \citet{2011ApJ...729..144S} result from an optimization seeking to maximize the off-axis throughput for a given focal plane stop diameter: the final contrast is absent from the optimization metric and is only a by-product of the chosen geometry. Higher contrasts are then obtained by increasing the size of the focal plane mask, and thus result in a loss in IWA. 

\subsection{The optimization problem}

Here we revisit the solution proposed by \citet{2011ApJ...729..144S} in a slightly different framework. We recognize that, in the presence of wavefront errors, high contrast can only  be achieved in an area of the field of view that is bounded by the spatial frequency corresponding to the DM's actuator spacing. We thus consider the design of an Apodized Pupil Lyot Coronagraph which only aims at generating high contrast between the Inner Working Angle (IWA) and Outer Working Angle (OWA).  In order to do so, we rewrite  coronagraphs described by \cite{2003A&A...397.1161S} as an operator $\mathcal{C}$ which relates the entrance pupil $P(r)$  to the electrical field in the final image plane. We first call $\hat{P}(\xi)$ the Hankel transform of the entrance pupil:
\begin{equation}
\hat{P}(\xi) =  \int_{D_{S}/2}^{D/2}  P(r) J_0(r \xi) r dr 
\end{equation}
where $D$ is the pupil diameter, $D_S$ the diameter of the secondary and $\xi$ the coordinate at the science detector expressed in units of angular resolution ($\lambda_0/D$).  $\lambda_0$ is the design wavelength of the coronagraph chosen to translate the actual physical size of the focal plane mask in units of angular resolution (often at the center of the bandwidth of interest). $\lambda$ is then the wavelength at which the coronagraph is operating (e.g. the physical size of the focal plane mask remains constant as the width of the diffraction pattern changes with wavelength). For the purpose of the monochromatic designs presented herein $\lambda = \lambda_0$. Then the operator is given by:
\begin{equation}
\mathcal{C}\left[P(r)\right](\xi) = \hat{P}(\xi) - \frac{\lambda^2}{\lambda_0^2} \int_{D_{S}/2}^{D/2} \hat{P}(\eta) K(\xi,\eta) \eta d \eta  
\label{eq:TransferCorono}
\end{equation}
where $ K(\xi,\eta)$ is the convolution kernel that captures the effect of the focal plane stop of diameter $M_{stop}$:
\begin{equation}
K(\xi,\eta) = \int_{0}^{M_{stop}/2} J_0(u \eta) J_0(u \xi) u du 
\end{equation}
An analytical closed form for this kernel can be calculated using Lommel functions. Note that this Eq.~\ref{eq:TransferCorono} assumes that the Lyot stop is not undersized. Since we are  interested in high contrast regions that only span radially all the way up to a finite OWA, we seek pupil apodization of the form:
\begin{equation}
P(r) = \sum_{k = 0}^{N_{modes}} p_k J_{Q}(\frac{r}{\alpha_{k}^{Q}})
\end{equation}
where $J_{Q}(r)$ denotes the Bessel function of the first kind of order $Q$ and $\alpha_{k}^{Q}$ the $k$th zero of this Bessel function. In order to devise optimal apodizations over-obscured pupils, $Q$, can be chosen to be large enough so that  $J_{Q}(r) \ll 1$ for $r < D_{S}/2$ (in practice we choose $Q = 10$). The $\alpha_{k}^{Q}$ corresponds  to the spatial scale of oscillations in the coronagraph entrance pupil, and such a basis set yields high contrast regions all the way to $OWA \simeq N_{modes} \lambda_0/D$. Since the operator in Eq.~\ref{eq:TransferCorono} is linear, finding the optimal $p_k$ can be written as the following linear programming problem:

\begin{subequations}
\begin{align}
&\max_{\{p_k\}} \left[ \min_{r}(P(r)) \right]  \mbox{Under the constraints:}   \\
 &|\mathcal{C} \left[ (P(r)) \right](\xi)| <10^{-\sqrt{C}} \; \mbox{for}\; IWA < \xi <OWA   \\
 &\max_{r}(P(r))= 1  \\
 &|\frac{d}{dr} [P(r)] | < b \label{eq:linprog}.
\end{align}
\end{subequations}


Our choice of cost function and constraints has been directed by the following rationale: 
\begin{itemize}
\item[5.a] We maximize the smallest value on the apodization function in an attempt to maximize throughput. The actual throughput is a quadratic function of the $p_k$. Maximizing it requires the solution of a non-linear optimization problem  (as described in \citet{2003ApJ...590..593V})
\item[5.b] The contrast constraint is enforced between the IWA and the OWA ($<N_{modes}$).
\item[5.c] The maximum of the apodization function is set to one (otherwise the $p_k$ will be chosen to be sufficiently small so that the contrast constraint is met). 
\item[5.d] The absolute value of the derivative across the pupil cannot be larger than a limit, denoted as $b$ here. As the natural solutions of such problem are very oscillatory (or ``bang bang'',  \cite{2003ApJ...590..593V,2003ApJ...599..686V}), a smoothness constraint has to be enforced (see \citet{2007ApJ...665..794V} for a similar case). 
\end{itemize} 

Note that the linear transfer function in Eq.~\ref{eq:TransferCorono} can also be derived for other coronagraphs, with grayscale and phase image- plane masks, or for the case of under-sized Lyot stops. As general coronagraphic design in obscured circular geometries is not our main purpose, we limit the scope of the paper to coronagraphs represented by Eq.~\ref{eq:TransferCorono}.

\begin{figure*}[h!]
\begin{center}
\includegraphics[width=7in]{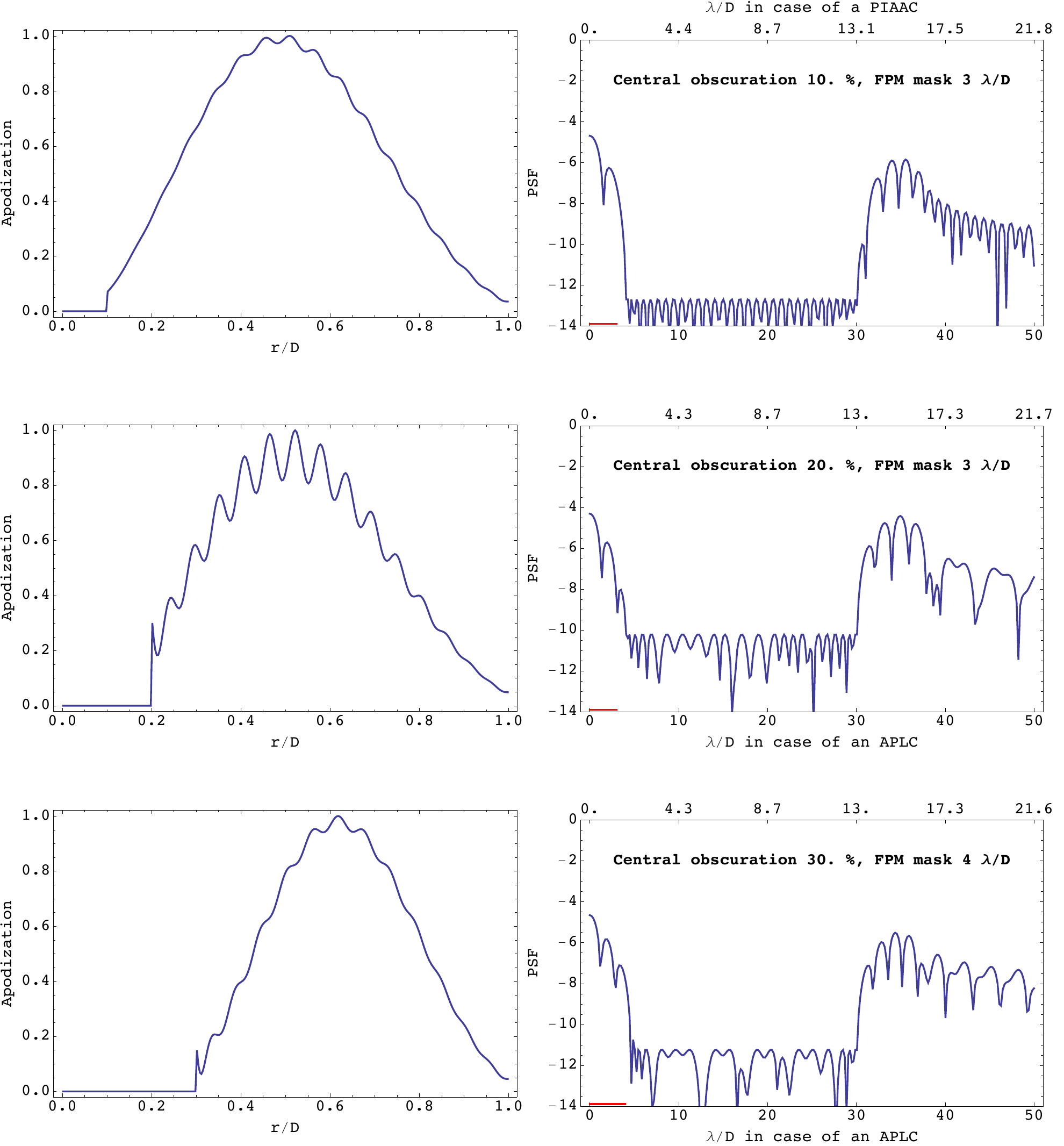}
\end{center}
\caption[]{Optimal design Apodized Pupil Lyot Coronagraphs on circularly obscured apertures. With fixed obscuration ratio, size of opaque focal plane mask, IWA and OWA, our linear programming approach yields solutions with theoretical contrast below $10^{10}$. All PSFs shown on this figure are monochromatic for $\lambda = \lambda_0$. When all other quantities remain equal and the central obscuration ratio increases (from $10 \%$ in the top panel to $20 \%$ in the middle panel), then the solution becomes more oscillatory (e.g less feasible) and the contrast constraint has to be relaxed. Eventually the optimizer does not find a solution and the size of the opaque focal plane mask (and thus the IWA) has to be increased (central obscuration of $30 \%$ in the bottom panel). On the right hand side, we present our results in two configurations: when the apodization is achieved using at grayscale screen (APLC), ``on-sky'' $\lambda / D$ bottom x-axis, and when the apodization is achieved via two pupil remapping mirrors (PIAAC), ``on-sky'' $\lambda / D$ top x-axis. We adopt this  presentation to show that ACAD is ``coronagraph independent'' and that it can be applied to coronagraphs with high throughput and small IWA.}
\label{fig:corono}
\end{figure*}

\subsection{Results of the optimization}

Typical results of the monochromatic optimization in Eqs.~\ref{eq:linprog}, with $\lambda = \lambda_0$, are shown in Fig.~\ref{fig:corono} for central obscurations of $10$, $20$ and $30 \%$. In the first two cases the size of the focal plane stop is  equal to $3 \lambda/D$, the IWA is $4 \lambda/D$ and the OWA is $30 \lambda /D$. As the size of the central obscuration increases the resulting optimal apodization becomes more oscillatory and the contrast constraint has to be loosened in order for the linear programming optimizer to converge to a smoother solution. Alternatively increasing the size of the focal plane stop yield smooth apodizers with high contrast, at cost in angular resolutions  (bottom panel with a central obscuration of $30 \%$, a focal plane mask of radius $4 \lambda/D$, an IWA of $5 \lambda/D$ and an OWA of $30 \lambda /D$).  These trade-offs were described in \cite{2011ApJ...729..144S}, however our linear programming approach to the design of pupil apodizations now imposes the final contrast instead of having it be a by product of  fixed central obscuration and focal plane stop. These apodizations can either be generated using grayscale screen (at a cost in throughput and angular resolution) or a series of two aspherical PIAA mirrors (for better throughput and angular resolution).  In order not to lose generality, we present our results on Fig.~\ref{fig:corono} considering the two types of practical implementations (classical apodization and PIAA apodization). In the case of a grayscale amplitude screen the angular resolution units are as defined in Eq.~\ref{eq:TransferCorono} and the throughput is smaller than unity. In the case of PIAA apodisation the throughput is unity and the angular resolution units have been magnified by the field independent centroid based angular magnification defined in \citet{Pueyo:11}. We adopt this presentation for the remainder of the paper where one dimensional PSFs will be presented with  ``APLC angular resolution units'' in the bottom horizontal axis and``PIAAC angular resolution units'' in the top horizontal axis.

Note that this linear programming approach only optimizes the contrast for a given wavelength. However, since the solutions presented in Fig.~\ref{fig:corono} feature contrasts below $10^{10}$, we choose not to focus on coronagraph chromatic optimizations. Instead, in order to account for the chromatic behavior of the coronagraph, the monochromatic simulations in \S.~\ref{sec:Resuls} are carried out under the {\em conservative} assumption that the physical  size of the focal plane stop is somewhat smaller than optimal (or that the operating wavelength of the coronagraph is slightly off, $\lambda = 1.2 \lambda_0$). As a consequence the raw contrast of the coronagraphs presented in \S.~\ref{sec:Resuls} is $\sim 10^{-9}$. Note that this choice is not representative of all possible Apodized Pupil Coronagraph chromatic configurations. It is merely a shortcut we use to cover the variety of cases presented in \S.~\ref{sec:Resuls}.  In \S.~\ref{sec:wavefrontcontrol} we present a set of broadband simulations that include wavefront errors and the true coronagraphic chromaticity for a specific configuration and show that bandwidth is more likely to be limited by the spectral bandwidth of the wavefront control system than by the coronagraph. However, future studies aimed at defining the true contrast limits of a given telescope geometry will have to rely on solutions of the linear problem in Eqs.~\ref{eq:linprog} which has been augmented to accommodate for broadband observations. In theory, the method presented here can also be applied to asymmetric pupils. However, the optimization quickly becomes computationally intensive as the dimensionality of the linear programming increases  (in particular when the smoothness constraint and the bounds on the apodization have to be enforced at all points of a two dimensional array). This problem can be somewhat mitigated when seeking for binary apodizations, as shown in \citet{2011OExpr..1926796C}, at a cost in throughput and angular resolution.

\section{Physics of amplitude modulation}
\label{sec:physicsofAM}

\subsection{General equations}

\begin{figure*}[t!]
\begin{center}
\includegraphics[width=6in]{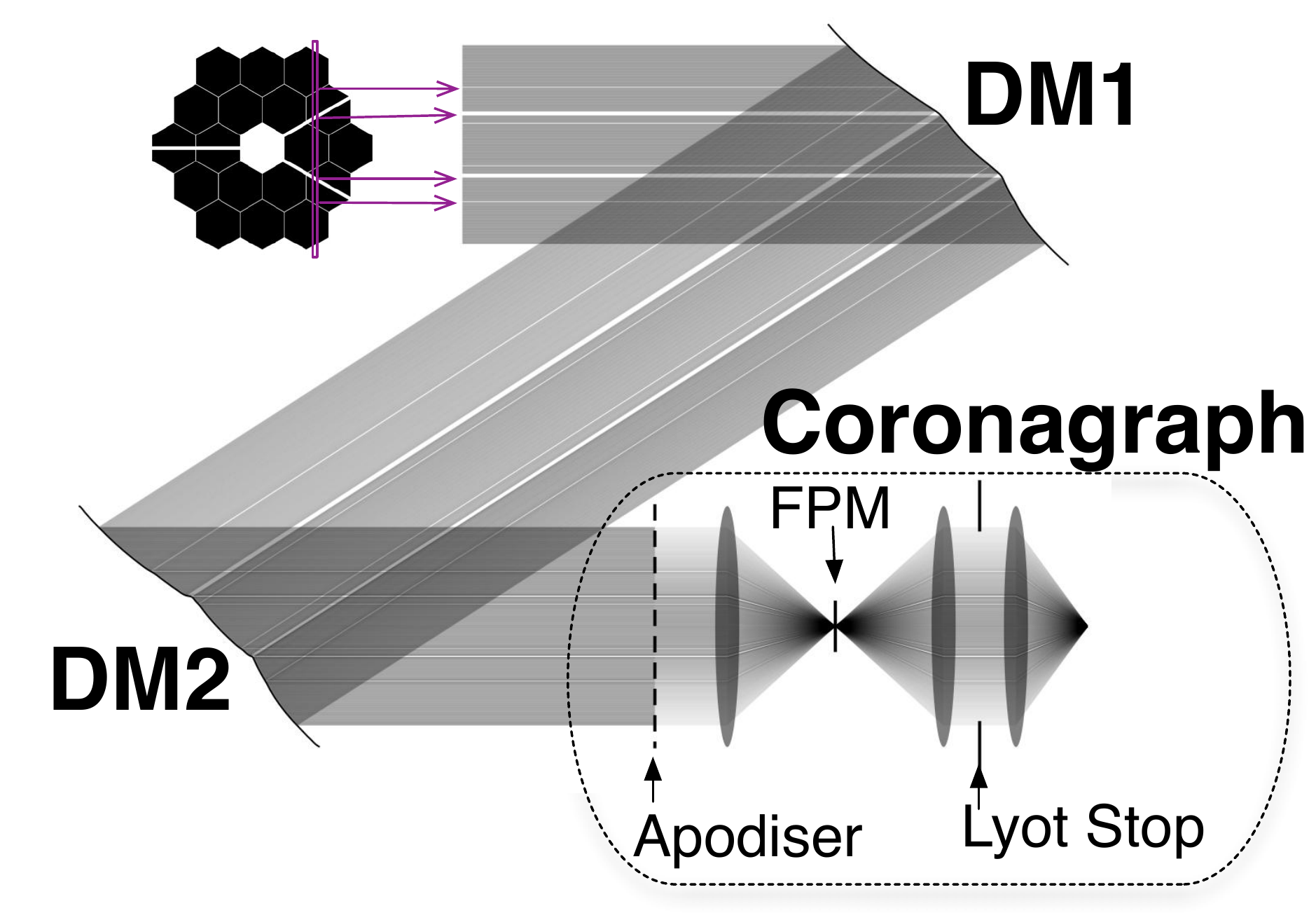}
\end{center}
\caption[]{Schematic of the optical system considered: the telescope apertures is followed by two sequential Defromable Mirrors (DMs) in non-conjugate planes whose purpose is to remap the pupil discontinuities. The beam then enters a coronagraph to suppress the bulk of the starlight: in this figure we show an Apodized Lyot Pupil Coronagraph (APLC). This is the coronagraphic architecture we consider for the remainder of the paper but we stress that the method presented herein is applicable to {\em any} coronagraph.}
\label{fig:layout}
\end{figure*}

\begin{figure}[h!]
\begin{center}
\includegraphics[width=3.5in]{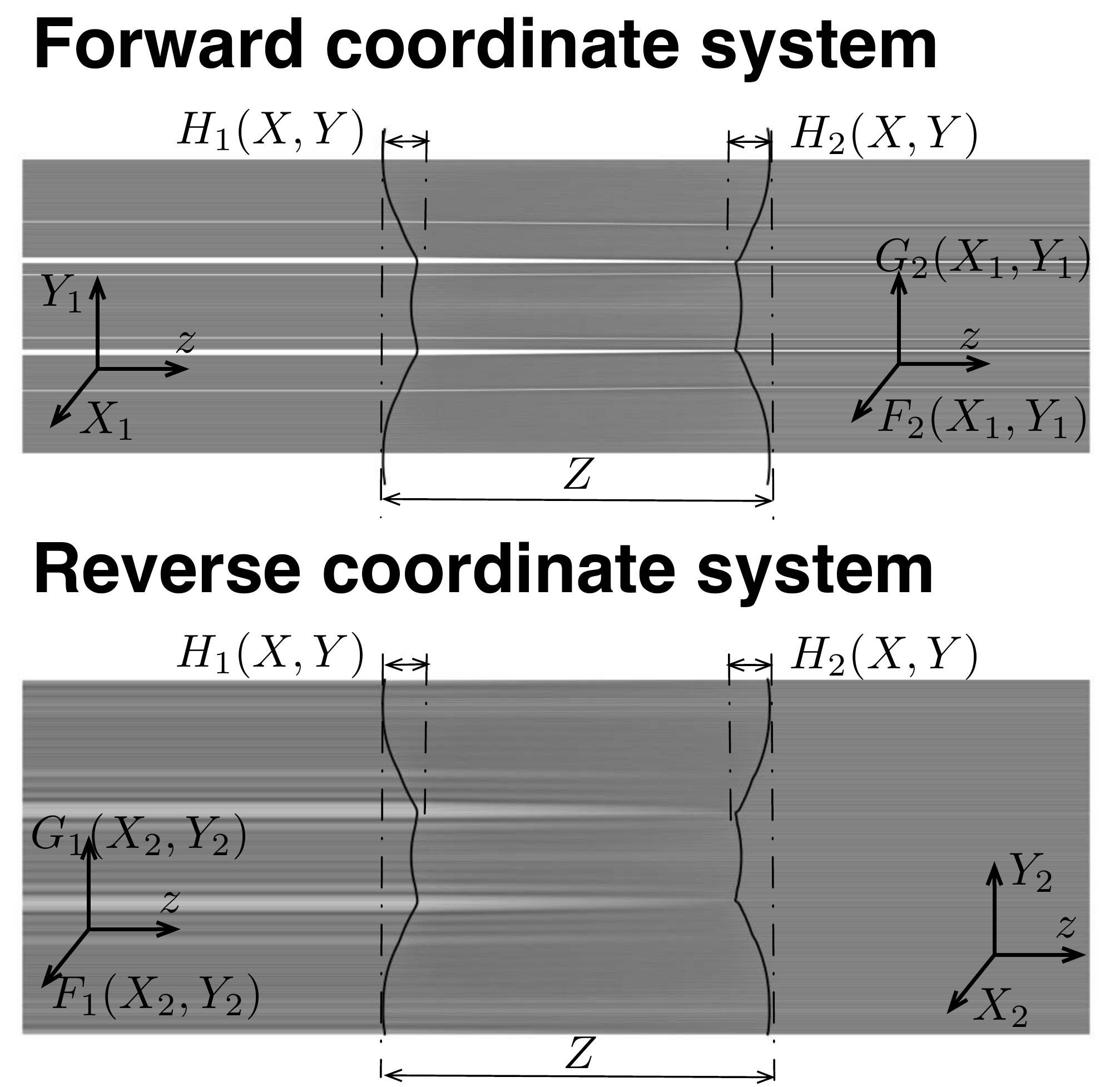}
\end{center}
\caption[]{Forward and reverse coordinate systems, with their respective forward and reverse coordinate transforms, in the case of the folded system studied in this paper. Note that this figure shows normalized units. As explained in the body of the text the correspondence between normalized and real units scales as follow: $(x_{i},y_{i}) = (DX_{i},DY_{i})$, $(f_{i},g_{i}) = (DF_{i},DG_{i})$, $H_{i} = \frac{D^2}{Z} h_{i}$, where $i =1,2$, $D$ is the aperture diameter and $Z$ the separation between DMs. We solve Monge Ampere Equation in normalized coordinates and then apply the scalings in order to find the true DM shapes.}
\label{fig:coordinates}
\end{figure}

We have shown in \S.~\ref{sec:CentralObsc} that by considering the design of pupil apodized coronagraphs in the presence of a circular central obscuration as a linear optimization problem, high contrast can be reached provided that the focal plane mask is large enough. In practice, the secondary support structures and the other asymmetric discontinuities in the telescope aperture (such as segment gaps) will prevent such levels of starlight suppression. We demonstrate that well controlled DMs can circumvent the obstacle of spiders and segment gaps. In this section, we first set-up our notations and review the physics of phase to amplitude modulation. We consider the system represented on Fig.~\ref{fig:layout} where two sequential DMs are located between the telescope aperture and the entrance pupil of the coronagraph. In this configuration, the telescope aperture and the pupil apodizer are not in conjugate planes. This will have an impact on the chromaticity of the system and is discussed in \S.~\ref{sec:Chrom}. Without loss of generality we work under the ``folded'' assumption illustrated on Fig.~\ref{fig:coordinates} where the DMs are not tilted with respect to the optical axis and can be considered as lenses of index of refraction $-1$ (as discussed in \cite{2005ApJ...626.1079V} . In the scalar approximation the relationship between the incoming field, $E_{DM1}(x,y)$, and the outgoing field, $E_{DM2}(x_2,y_2)$, is given by the diffractive Huygens Integral:
\begin{equation}
E_{DM2}(x_2,y_2)=\frac{1}{ i \lambda Z} \int_{\mathcal{A}} E_{DM1}(x,y) e^{i\frac{2 \pi}{\lambda}Q(x,y,x_2,y_2)} dx dy  \label{huygens}
\end{equation}
where $\mathcal{A}$ corresponds to the telescope aperture and $Q(x,y,x_2,y_2)$ stands for the optical path length between any two points at DM1 and DM2:
\begin{equation}
Q(x,y,x_2,y_2) = h_1(x,y)+S(x,y,x_2,y_2)-h_2(x_2,y_2)  
\label{DefOPL}
\end{equation}
$S(x,y,x_2,y_2)$ is the free space propagation between the DMs:

\begin{equation}
\sqrt{(x-x_2)^{2}+(y-y_2)^{2}+(Z+h_1(x,y)-h_2(x_2,y_2))^{2}} \label{DefS}
\end{equation}

where $Z$ is the distance between between the two DMs, $h_1$ and $h_2$ are the shapes of DM1 and DM2 respectively (as shown on Fig.~\ref{fig:coordinates}) and $\lambda$ is the wavelength. We recognize that two sequential DMs act as a pupil remapping unit similar to PIAA coronagraph \citep{2003A&A...404..379G} whose ray optics equations were first derived by \citet{2003ApJ...599..695T}. We briefly state the notation used to describe such an optical system as introduced in \citet{2011ApJS..195...25P}: 
\begin{itemize}
\item For a given location at DM2, $(x_2,y_2)$, the location at DM1 of the incident ray according to ray optics is given by:
\begin{subequations}
\begin{align}
& x_1(x_2,y_2) = f_1(x_2,y_2) \\
& y_1(x_2,y_2) = g_1(x_2,y_2). \label{eq:fone}
\end{align}
\end{subequations}
\item Conversely, for a given location at DM1, $(x_1,y_1)$, the location at DM2 of the outgoing ray according to geometric optics is given by:
\begin{subequations}
\begin{align}
& x_2(x_1,y_1) = f_2(x_1,y_1) \\
& y_2(x_1,y_1) = g_2(x_1,y_1).\label{eq:ftwo}
\end{align}
\end{subequations}

\item Fermat's principle dictates the following relationships between the remapping functions and the shape of DM1:

\begin{subequations}
\begin{align}
& \frac{\partial h_1}{\partial x}\Big|_{(x_1,y_1)} =  \frac{x_1 - f_2(x_1,y_1)}{Z}  \\
& \frac{\partial h_1}{\partial y}\Big|_{(x_1,y_1)} =  \frac{y_1 - g_2(x_1,y_1)}{Z}. \label{eq:DM1EquaDiff}
\end{align}
\end{subequations}

\item Conversely, if we choose the surface of  DM2 to ensure that the outgoing on-axis wavefront is flat, we then find:

\begin{subequations}
\begin{align}
& \frac{\partial h_2}{\partial x}\Big|_{(x_2,y_2)} =  \frac{x_2 - f_1(x_2,y_2)}{Z} \\
& \frac{\partial h_2}{\partial y}\Big|_{(x_2,y_2)} =  \frac{y_2 - g_1(x_2,y_2)}{Z}.  \label{eq:DM2EquaDiff}
\end{align}
\end{subequations}
\end{itemize}

\subsection{Fresnel approximation and Talbot Imaging}
In \citet{2011ApJS..195...25P} we showed that one could approximate the propagation integral in Eq.~\ref{huygens} by taking in a second order Taylor expansion of $Q(x,y,x_2,y_2)$ around the rays that trace $(f_1(x_2,y_2),g_1(x_2,y_2))$ to $(x_2,y_2)$. In this case the relationship between the fields at DM1 and DM2 is:
%
%
%
\begin{widetext}
\begin{equation}
E_{DM2}(x_2,y_2) = \frac{e^{\frac{2 i \pi}{\lambda} Z}}{ i \lambda Z} \left \{\int_{\mathcal{A}}   E_{DM1}(x,y)  e^{\frac{ i \pi }{\lambda Z}  \Big[ \frac{\partial f_2}{\partial x}(x- x_1)^{2} + 2 \frac{ \partial g_2}{\partial x}(x- x_1 )(y- y_1) +\frac{\partial g_2}{\partial y}(y- y_1 )^{2}  \Big]}dx dy \right \} \Bigg|_{(x_1,y_1)} \label{eq::SecondExp}
\end{equation}
When the mirror's deformations are very small compared to both the wavelength and $D^2/Z$, the net effect of the wavefront disturbance created by  DM1 can be captured in $E_{DM1}(x,y)$ and the surface of DM2 can be factored out of Eq.~\ref{huygens}. In this case $x_1 = x_2$, $y_1 = y_2$, $\frac{\partial f_2}{\partial x}\big |_{x_1,y_1} = 1$, $\frac{\partial f_2}{\partial y}\big |_{x_1,y_1}=0$, $\frac{\partial g_2}{\partial y}\big |_{x_1,y_1} = 1$. Then, Eq.~\ref{eq::SecondExp} reduces to:
\begin{equation}
E_{DM2}(x_2,y_2) = \frac{e^{\frac{2 i \pi}{\lambda}(Z-h_2(x_2,y_2))}}{ i \lambda Z} \int_{Aperture}  e^{\frac{2 i \pi}{\lambda}(h_1(x,y))} e^{\frac{ i \pi }{\lambda Z}  \big( (x- x_2)^{2} +(y- y_2)^{2}  \big)}dx dy 
\end{equation}
 \end{widetext}
 which is the Fresnel approximation. If moreover $h_1(x,y) = \lambda \epsilon \cos (\frac{2 \pi}{D}(m x+ n y))$, $h_2(x,y) =  - h_1(x,y)$, with  $\epsilon \ll 1$, then the outgoing field is to first order:
 \begin{equation}
 E_{DM2}(x_2,y_2) \propto  \frac{\pi \lambda Z (m^2+n^2) }{D^2}\lambda \epsilon \cos (\frac{2 \pi}{D}(m x_2+ n y_2)). \label{eq::talbot}
 \end{equation}
This phase-to-amplitude coupling is a well known optical phenomenon called Talbot imaging and was introduced to the context of high contrast imaging by \citet{2006ApOpt..45.5143S}. In the small deformation regime, the phase on DM1 becomes an amplitude at DM2 according to the coupling in Eq.~\ref{eq::talbot}. When two sequential DMs are controlled to cancel small  amplitude errors, as in \cite{Pueyo:09}, they operate in this regime. Note, however, that the coupling factor scales with wavelength (the resulting amplitude modulation is wavelength independent, but the coupling scales as $\lambda$): this formalism is thus not applicable to our case, for which we are seeking to correct large amplitude errors (secondary support structures and segments) with the DMs. In practice, when using Eq.~\ref{eq::talbot} in the wavefront control scheme outlined in \cite{Pueyo:09} to correct aperture discontinuities, this weak coupling results in large mirror shapes that lie beyond the range of the linear assumption made by the DM control algorithm. For this reason, methods outlined on the left panel of Fig.~\ref{fig:State} to correct for aperture discontinuities  do not converge to high contrast.  Because phase to amplitude conversion is fundamentally a very non-linear phenomena, these descending  gradient methods  \citep{2006ApJ...638..488B,Giveon:07,Pueyo:09} are not suitable to find DM shapes that mitigate apertures discontinuities. We circumvent these numerical limitations by calculating DMs shapes that are based on the full non-linear problem, right panel of Fig.~\ref{fig:State}. 
\subsection{The SR-Fresnel approximation}
In the general case, starting from Eq.~\ref{eq::SecondExp} and following the derivation described in \S.~\ref{sec:Chrom}, the field at DM2 can be written as follows:
\begin{widetext}
\begin{equation}
E_{DM2}(x_2,y_2) = \left \{ \sqrt{|\det[J]|}
  \int_{\mathcal{F}\mathcal{P}}  \widehat{ E_{DM1}}(\xi,\eta)  e^{i 2 \pi (\xi f_1+\eta g_1)}e^{- i\frac{\pi \lambda Z}{\det[J]}\left( \frac{\partial g_1}{\partial y} \xi^2+ \frac{\partial f_1}{\partial x}\eta^2 \right)} \; d\xi d \eta \right \}\bigg|_{(x_2,y_2)}\label{eq::propint}
\end{equation}
\end{widetext}

where $ \widehat{ E_{DM1}}(\xi,\eta) $ is the Fourier transform of the telescope aperture, $\mathcal{F}\mathcal{P}$ and stands for the Fourier plane. We call this integral the Stretched-Remapped Fresnel approximation (SR-Fresnel). Moreover $\det[J(x_2,y_2)]$ is the determinant of the Jacobian of the change of variables that maps $(x_2,y_2)$ to $(x_1,y_1)$:
\begin{equation}
det[J(x_2,y_2)] = \left \{\frac{\partial f_1}{\partial x}\frac{\partial g_1}{\partial y}- \left(\frac{ \partial g_1}{\partial x}\right)^2 \right \}\bigg|_{(x_2,y_2)}.
\end{equation}
In the ray optics approximation, $\lambda \sim 0 $, the non linear transfer function between the two DMs becomes:
\begin{eqnarray}
E_{DM2}(x_2,y_2)  & = & \left \{ \sqrt{|\det[J]|} E_{DM1}\left(f_1,g_1\right) \right \}\bigg|_{(x_2,y_2)} \\
\left[E_{DM2}(x_2,y_2)\right]^{2} &=& \left \{\det[J] \left[ E_{DM1}\left(f_1,g_1\right) \right]^2 \right \}\bigg|_{(x_2,y_2)}\label{eq::mongeremap}
\end{eqnarray}
The square form (e.g Eq.~\ref{eq::mongeremap}) of this transfer function can also be derived based on conservation of energy principles and is a generalization to arbitrary geometries of the equation driving the design of PIAA coronagraphs \citep{2005ApJ...626.1079V}. A full diffractive optimization of the DM surfaces requires use of the complete transfer function shown in Eq.~\ref{eq::propint}. However, there do not exist yet tractable numerical  method to evaluate Eq.~\ref{eq::propint} efficiently enough in order for this model to be included  in an optimization algorithm. Moreover even solving the ray optics problem is extremely complicated: it requires to find the mapping function $(f_1,g_1)$ which solves the non-linear partial differential equation in Eq.~\ref{eq::mongeremap}. Substituting for$(f_1,g_1)$ and using Eqs.~\ref{eq:DM2EquaDiff} yields a second order non-linear partial differential equation in $h_2$. This is the problem that we set ourselves to tackle in the next section, and is the cornerstone of our Adpative Compensation of Aperture Discontinuities. As a check, one can verify that in the small deformation regime (e.g. if $h_1(x,y) = \lambda \epsilon \cos (\frac{2 \pi}{D}(m x+ n y))$ and  $h_2(x,y) =  - h_1(x,y)$)  Eq.~\ref{eq::mongeremap} yields the same phase-to-amplitude coupling as in Talbot imaging \citep{2008PhDT.........3P}.
Eq.~\ref{eq::mongeremap} is a well know optimal transport problem \citep{MongeVieux}, which has already been identified as underlying optical illumination optimizations \citep{Glimm02}. While the existence and uniqueness of solutions in arbitrary dimensions have been extensively discussed in the mathematical literature (see \cite{dacorogna90} for a review), there was no practical numerical solution published up until recently. In particular, to our knowledge, not even a dimensional solution for which the DM surfaces can be described using a realistic basis-set has been published yet. We now introduce a method that calculates  solutions to Eq.~\ref{eq::mongeremap} which can be represented by feasible DM shapes.

\begin{figure}[h!]
\begin{center}
\includegraphics[width=3.5in]{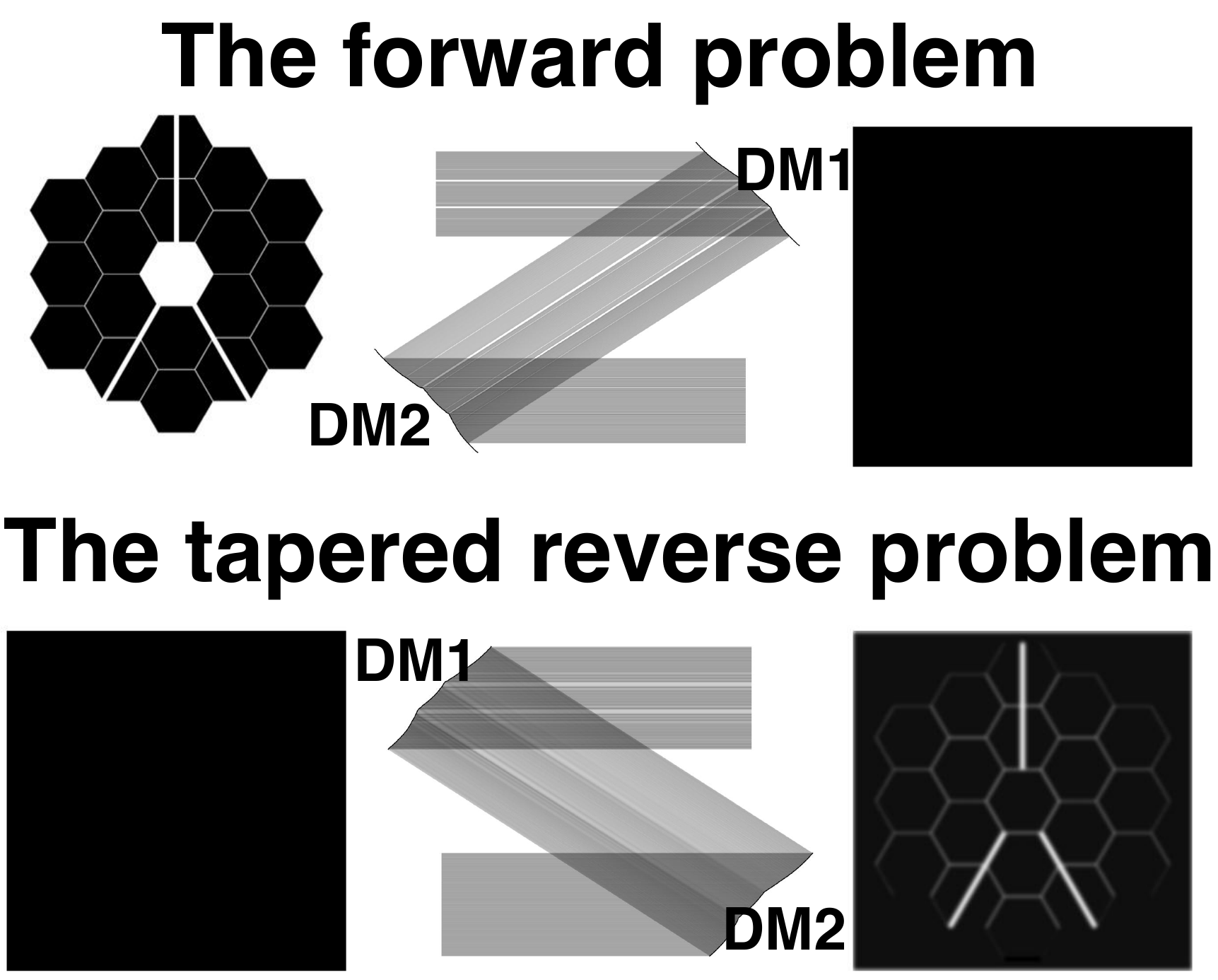}
\end{center}
\caption[]{{\bf Top}: In the ideal case the two DMs would fully remap all the discontinuities in the telescope's aperture to feed a fully uniform beam to the coronagraph. However this would require discontinuities in the mirror's curvatures which cannot be achieved in practice. Moreover solving the Monge-Ampere Equation in this direction is a difficult exercise as the right hand side of Eq.~\ref{eq::mongeremap} presents an implicit dependence on the solution $h_2$. {\bf Bottom}: We circumvent this problem by solving the reverse problem, where the the input beam is now uniform and the implicit dependence drops out. Moreover we taper the edges of the discontinuities by convolving the target field $A(x_1,y_1)$ by a gaussian of full width at half maximum $\omega$ ($\omega = 50$ cycles per aperture in this figure).}
\label{fig:reverse}
\end{figure}

\section{Calculation of the Deformable Mirror shapes}
\label{sec:MADM}

\subsection{Statement of the problem}

Ideally, we seek DM shapes that fully cancel all the discontinuities at the surface of the primary mirror and yield a uniform amplitude distribution, as shown in the top panel of Fig.~\ref{fig:reverse}. A solutions for a particular geometry with four secondary support has been derived by \citet{2009PASP..121.1232L}. It relies on reducing the dimensionality of the problem to the direction orthogonal to the spiders. It is implemented using a transmissive correcting plate that is a four-faced prism arranged such that the vertices coincides with the location of the spiders. The curvature discontinuities at the location of the spiders are responsible for the local remapping that removes the spiders in the coronagraph pupil. However such a solution cannot be readily generalized to the case of more complex apertures, where the secondary support structures might vary in width, or in the presence of segment gaps. Moreover it is transmissive and thus highly chromatic. Here we focus on a different class of solutions and seek to answer a different question. How well can we mitigate the effect of pupil discontinuities using DMs with smooth surfaces, a limited number of actuators (e.g a limited maximal curvature), and a limited stroke? Under these constraints directly solving  Eq.~\ref{eq::mongeremap} (e.g. solving the forward problem  illustrated in the top panel of Fig.~\ref{fig:reverse}) is not tractable as both factors on the left hand side of Eq.~\ref{eq::mongeremap} depend on $h_2$. More specifically, the implicit dependence of $E_{DM1}\left(f_1(x_2,y_2),g_1(x_2,y_2) \right)$ on $h_2$ can only be addressed using finite elements solvers, whose solutions might not be realistically representable using a DM. 
However this can be circumvented using the reversibility of light and solving the reverse problem, where the two mirrors have been swapped. Indeed, since
\begin{eqnarray}
x_2 = f_2(f_1(x_2,y_2),g_1(x_2,y_2)) \label{eq:x2self} \\
y_2 = g_2(f_1(x_2,y_2),g_1(x_2,y_2)) \label{eq:y2self} 
\end{eqnarray} 
then we have the following relationship between the determinants of the forward and reverse remappings:
\begin{equation}
1 = \left \{ \frac{\partial f_2}{\partial x}\frac{\partial g_2}{\partial y} - \left(\frac{ \partial g_2}{\partial x}\right)^2 \right \}\bigg|_{(x_1,y_1)} \left \{ \frac{\partial f_1}{\partial x}\frac{\partial g_1}{\partial y} \left(\frac{ \partial g_1}{\partial x}\right)^2 \right \}\bigg|_{(x_2,y_2)} \label{eq:remapunity}
\end{equation}
We thus focus on the inverse problem, bottom panel of Fig.~\ref{fig:reverse}, that consists of first finding the surface of $h_1$ as the solution of: 
\begin{eqnarray}
&&\left \{ \left[ E_{DM2}(f_2,g_2) \right]^2  \left[ \frac{\partial f_2}{\partial x}\frac{\partial g_2}{\partial y}- \left(\frac{ \partial g_2}{\partial x}\right)^2\right] \right \}\bigg|_{(x_1,y_1)} \nonumber \\
&=& \left[ E_{DM1}(x_1,y_1) \right]^2 
\end{eqnarray}
Since our goal is to obtain a pupil as uniform a possible we seek a field at DM2 as uniform as possible:
\begin{eqnarray}
  && E_{DM2}(f_2(x_1,y_1),g_2(x_1,y_1)) = \nonumber \\
  && \sqrt{\int_{\mathcal{A}}E_{DM1}(x,y)^2 dx dy} = \mbox{Constant}.
\end{eqnarray}
Moreover we are only interested in compensating asymmetric structures located between the secondary and the edge of the primary. We thus only seek to find $(f_2,g_2)$ such that:
\begin{eqnarray}
&&E_{DM1}(x_1,y_1) = A(x_1,y_1)= \nonumber \\
&&\left( \left[ P(x,y) - (1 - P_{O}(x,y)) \right] \ast e^{- \frac{x^2+y^2}{\omega^2}} \right) \Big|_{(x_1,y_1)}
\end{eqnarray} 
 where $P_{O}(x_1,y_1)$ is the obscured pupil, without segments or secondary supports. Finally, we focus on solutions with a high contrast only up to a finite OWA. We artificially taper the discontinuities by convolving the control term in the Monge Ampere Equation, $\left[ P(x,y) - (1 - P_{O}(x,y)) \right]$, with a gaussian of width $\omega$. Note that this tapering is only applied when calculating the DM shapes via solving the reverse problem. When the resulting solutions are propagated through Eq.~\ref{eq::mongeremap} we use the true telescope pupil  for $E_{DM1}(x_1,y_1)$. The parameter $\omega$ has a significant impact on the final post-coronagraphic contrast. Indeed we are here working with a merit function that is based on a pupil-plane residual, while ideally our cost function should be based on image-plane intensity. By convolving the control term in the reverse Monge Ampere Equation, we low pass filter the discontinuities. This is equivalent to giving a stronger weight to low-to-mid spatial frequencies of interest in the context of exo-planets imaging . For each case presented in \S.~\ref{sec:Resuls} we calculate our DM shapes over a grid of values of $\omega$ which correspond to low pass filters with cutoff frequencies ranging  from $\sim$ OWA to $\sim \; 2$ OWA and we keep the shapes which yield the best contrast. This somewhat ad-hoc approach can certainly be optimized for higher contrasts. However such an optimization is beyond the scope of the present manuscript.
 
The problem we are seeking to solve is illustrated in the second panel of Fig.~\ref{fig:reverse}. In this configuration the full second order Monge Ampere Equation can be written as: 
\begin{equation}
(1+Z\frac{\partial^2 h_1}{\partial x^2}) (1+Z\frac{\partial^2 h_1}{\partial y^2}) - \left(Z\frac{ \partial^2 h_1}{\partial x \partial y}\right)^2  = A(x,y)^2
\end{equation}
where we have dropped the $(x_1,y_1)$ dependence for clarity.  Since we are interested in surface deformations which can realistically be created using a DM, we seek for a Fourier representation of the DMs surface:
\begin{eqnarray}
&& h_1(x,y) = \frac{D^2}{Z}H_1(X,Y) \nonumber \\
&=& \frac{D^2}{Z} \sum_{n=-N/2}^{N/2} \sum_{m=-N/2}^{N/2} a_{m,n} e^{i \frac{2 \pi}{D} (m X +n X)}
\label{eq:H1Four}
\end{eqnarray}
with $a_{-m,-n}= a_{-m,-n}^{\star}$, where $N$ is the limited number of actuators across the DM. Note that we have normalized the dimensions in the pupil plane $X = x/D$, $Y = y/D$. The normalized second order Monge Ampere Equation is then:
\begin{equation}
(1+\frac{\partial^2 H_1}{\partial X^2}) (1+\frac{\partial^2 H_1}{\partial Y^2}) - \left(\frac{ \partial^2 H_1}{\partial X \partial Y}\right)^2  = A(X,Y)^2 \label{eq::normMA}
\end{equation}
For each configuration in this paper we first solve Eq.~\ref{eq::normMA} and then transform the normalized solution in physical units, which depends on the DMs diameter $D$ and their separation $Z$.

\subsection{Solving the Monge-Ampere equation to find $H_1$}
Over the past few years a number of  numerical algorithms aimed at solving Eq.~\ref{eq::normMA} have emerged in the literature \citep{Loeper2005319,benamou2010two}. Here we summarize our implementation of two of them: an explicit Newton method \citep{Loeper2005319}, and a semi-implicit method \citep{froese2012accurate}. We do not delve into the proof of convergence of each method, they can be found in \cite{Loeper2005319,benamou2010two,froese2012accurate}.  Note that \citet{Zheligovsky20105043} discussed both approaches in a cosmological context and devised Fourier based solutions. Here we are interested in a two dimensional problem and we outline below the essence of each algorithm.
\subsubsection{Explicit Newton algorithm}
This method was first introduced by \citet{Loeper2005319} and relies on the fact that Eq.~\ref{eq::normMA} can be re-written as
\begin{equation}
\det \left[ 
\left(
\begin{array}{cc}
1+\frac{\partial^2 H_1}{\partial X^2} & \frac{ \partial^2 H_1}{\partial X \partial Y}  \\
\frac{ \partial^2 H_1}{\partial X \partial Y} & 1+\frac{\partial^2 H_1}{\partial Y^2} 
\end{array}
\right)
\right] = \det[Id+\mathcal{H}(H_1(X,Y))]
\end{equation} 
where $\mathcal{H}(\cdot)$ is the two dimensional Hessian of a scalar field and $Id$ the identity matrix. If one writes $H_1 = u +  v$ with $||v|| \ll ||u||$ then:
\begin{widetext}
\begin{equation}
\det[Id+\mathcal{H}( u + \delta v)] = \det[Id+\mathcal{H}(u)] +\delta \;  Tr\left[ \left(  Id+\mathcal{H}(u)\right)^{\dagger^T} \mathcal{H}(v)  \right] +o(\delta^2)
\end{equation} 
where $(\cdot)^{\dagger^T} $ denotes the transpose of the comatrix.  Eq.~\ref{eq::normMA} can thus be linearized as: 
\begin{equation}
(1+\frac{\partial^2 u}{\partial Y^2}) \frac{\partial^2 v}{\partial X^2} +
(1+\frac{\partial^2 u}{\partial X^2}) \frac{\partial^2 v}{\partial Y^2}
- 2\frac{ \partial^2 u}{\partial X \partial Y} \frac{ \partial^2 v}{\partial X \partial Y}   = \left( A(X,Y)^2  -  \det[\mathcal{H}(\frac{X^2+Y^2}{2}+u)] \right) \label{eq:linearizedMA}
\end{equation}
The explicit Newton algorithm relies on Eq.~\ref{eq:linearizedMA} and can then be summarized as carrying out the following iterations:
\begin{itemize}
\item Choose a first guess $H_1^{0}$.
\item At each iteration $k$ we seek for a solution of the form $H_1^{k+1} = H_1^{k} + V^{k}$, where $V^{k}$ is the DM shape update.
\item In order to find $V^{k}$ we write:
\begin{eqnarray}
L_{E}( H_1^{k},V^{k}) &=& (1+\frac{\partial^2 H_1^{k}}{\partial Y^2}) \frac{\partial^2 V^{k}}{\partial X^2} +
(1+\frac{\partial^2 H_1^{k}}{\partial X^2}) \frac{\partial^2 V^{k}}{\partial Y^2}
- 2\frac{ \partial^2 H_1^{k}}{\partial X \partial Y} \frac{ \partial^2 V^{k}}{\partial X \partial Y} \\
R_{E}( H_1^{k}) &=& \frac{1}{\tau}\left( A^2  -  \det[Id+\mathcal{H}(H_1^{k})] \right)
\end{eqnarray}
and solve 
\begin{equation}
L_{E}( H_1^{k},V^{k})  = R_{E}( H_1^{k}). 
\label{eq:linloeper}
\end{equation}
Eq.~\ref{eq:linloeper} is a linear partial differential equation in $V^{k}$. Since we are interested in a solution which can be expanded in a Fourier series we write $V^{k}$ as: 
%
\begin{equation}
 V_1^{k}(X,Y) =  \sum_{n=-N/2}^{N/2} \sum_{m=-N/2}^{N/2} v_{m,n}^{k} e^{i \frac{2 \pi}{D} (m X +n X)}.
 \end{equation}
Both the right hand side and the left hand side of Eq.~\ref{eq:linloeper} can be written as a Fourier series, with a spatial frequency content between $-N$ and $N$ cycles per aperture. Equating each Fourier coefficient in these two series yields the following linear system of $(2 N+1)^2$ equations with  $(N+1)^2$ unknowns. 

\begin{equation}
\mbox{For all} \; m_0,n_0 \; \in \; [-N, N] \mbox{: }\int_{DM} e^{i 2 \pi (m_0 X+n_0 Y)} \left[ L_{E}( H_1^{k},V^{k})  - R_{E}( H_1^{k}) \right]  dX \;dY  = 0
 \label{eq:inverseLoeper}
\end{equation}
When searching for $V^{k}$as a Fourier series over the square geometry chosen here, this inverse problem is always well posed.
\item Update the solution $H_1^{k+1} =  H_1^{k} +  V_1^{k+1}$
\end{itemize}
\end{widetext}
The convergence of this algorithm relies on the introduction of a damping constant $\tau>1$. \citet{Loeper2005319} showed that as long as  $\frac{X^2+Y^2}{2}+H_1^{k}$ remains convex, which is always true for ACAD with reasonably small aperture discontinuities, there exists a $\tau$ large enough so that this algorithm converge towards a solution of Eq.~\ref{eq::normMA}. However since this algorithm is gradient based, it is not guaranteed that it converges to the global minimum of the underlying non-linear problem. In order to avoid having this solver stall in a local minimum we follow the methodology outlined by \citet{froese2012accurate} and first carry out a series of implicit iterations to get within a reasonable neighborhood of the global minimum.

\begin{figure*}[t!]
\begin{center}
\includegraphics[width=8in]{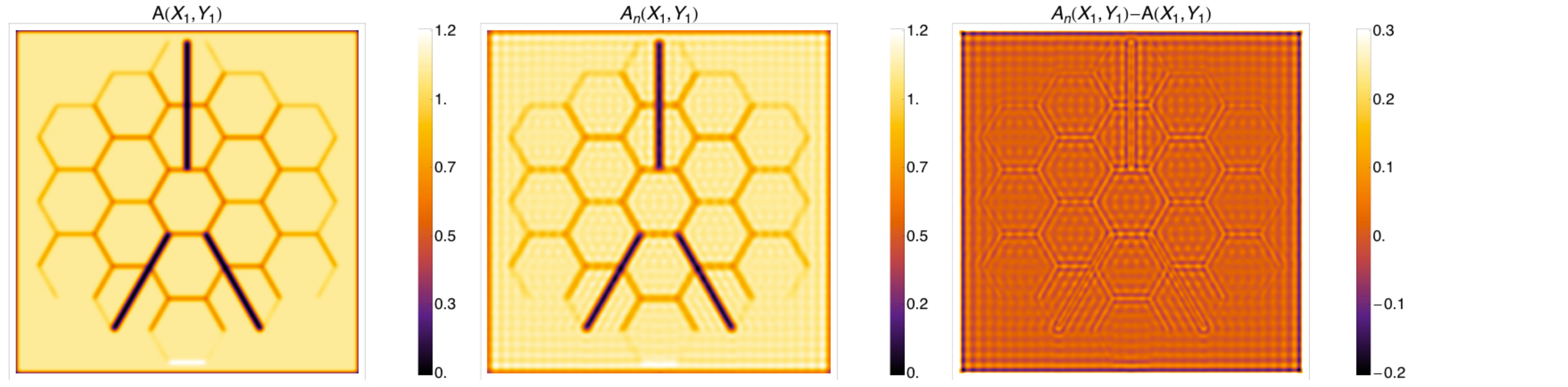}
\end{center}
\caption[]{Virtual field at M1 when solving the reverse problem. $A(X_1,Y_1)$ is the desired apodization (top),  $A_n(X_1,Y_1)$ is the apodization obtained after solving the Monge Ampere Equation (center). The bottom panel shows the difference between the two quantities: the bulk of the energy in the residual is located in {\em high spatial frequencies} that cannot be controlled by the DMs.}
\label{fig:results}
\end{figure*}
\subsubsection{Implicit algorithm}
This algorithm, along with its convergence proof, is thoroughly explained in \citet{froese2012accurate} . It relies on rewriting Eq.~\ref{eq::normMA} as:
\begin{equation}
\frac{\partial^2 H_1}{\partial X^2}+\frac{\partial^2 H_1}{\partial Y^2} = \sqrt{\det[Id+\mathcal{H}(H_1(X,Y))]^2+2 A(X,Y)^2}
\end{equation}
\begin{widetext}
The implicit method consists of carrying out the following iterations:
\begin{itemize}
\item Choose a first guess $H_1^{0}$
\item In order to find $H^{k+1}$ we write:
\begin{eqnarray}
&& L_{I}( H_1^{k},V^{k})  =\frac{\partial^2 H_1^{k+1}}{\partial X^2}+\frac{\partial^2 H_1^{k+1}}{\partial Y^2} \nonumber \\
&& R_{E}( H_1^{k}) = \sqrt{\det[Id+\mathcal{H}(H_1(X,Y))]^2+2 A(X,Y)^2} \nonumber
\end{eqnarray}
and solve 
\begin{equation}
L_{I}( H_1^{k+1})  = R_{I}( H_1^{k}). 
\label{eq:linloeper}
\end{equation}
This problem is a linear system of $(N+1)^2$ equations with $(N+1)^2$ unknowns and can be solved using projections on a Fourier Basis:
 \begin{equation}
\mbox{For all} \; m_0,n_0 \; \in \; [-N, N] \mbox{: }\int_{DM} e^{i 2 \pi (m_0 X+n_0 Y)} \left[ L_{I}( H_1^{k},V^{k})  - R_{I}( H_1^{k}) \right]  dX \;dY  = 0
 \label{eq:inversePoisson}
\end{equation}
Note that the term under the square root in $R_{I}( H_1^{k})$  is guaranteed to be positive at each iteration.
\item Iterate over $k$
\end{itemize}
\end{widetext}
The inverse problem in Eq.~\ref{eq:inversePoisson} is {\em always} well posed, for any basis set or pupil geometry, while the explicit Newton method runs into convergence issues when not using a Fourier basis over a square. When seeking to use a basis set that is more adapted to the geometry of the spiders and segments or when using a trial influence function basis for the DM,  the implicit method is the most promising method.  In this paper we have limited our scope to solving the reverse problem in the bottom panel of Fig.~\ref{fig:reverse}, and used a Fourier representation for the DM, we are able to use both methods. In order to make sure that the algorithm converges towards the true solution of Eq.~\ref{eq::normMA} we first run a few tens of iterations of the implicit method and, once it has converged, we seek for a more accurate solutions using the Newton algorithm. Typical results are shown in Fig.~\ref{fig:results} where most of the residual error resides in the high spatial frequency content (e.g. above $N$ cycles per aperture). Our solutions are limited by the non- optimality of the Fourier basis to describe the mostly radial and azimuthal structures present in telescopes's apertures. Moreover the DM shape is the result of the minimization of a least squares residual in the virtual end-plane of the reverse problem, with little regard to the spatial frequency content of the solution in the final image plane of the coronagraph. While this method yields significant contrast improvements, as reported in \S.~\ref{sec:Resuls}, we discuss in  \S.~\ref{sec:Discuss} how it can be refined for higher contrast.

\begin{figure}[h!]
\begin{center}
\includegraphics[width=3.5in]{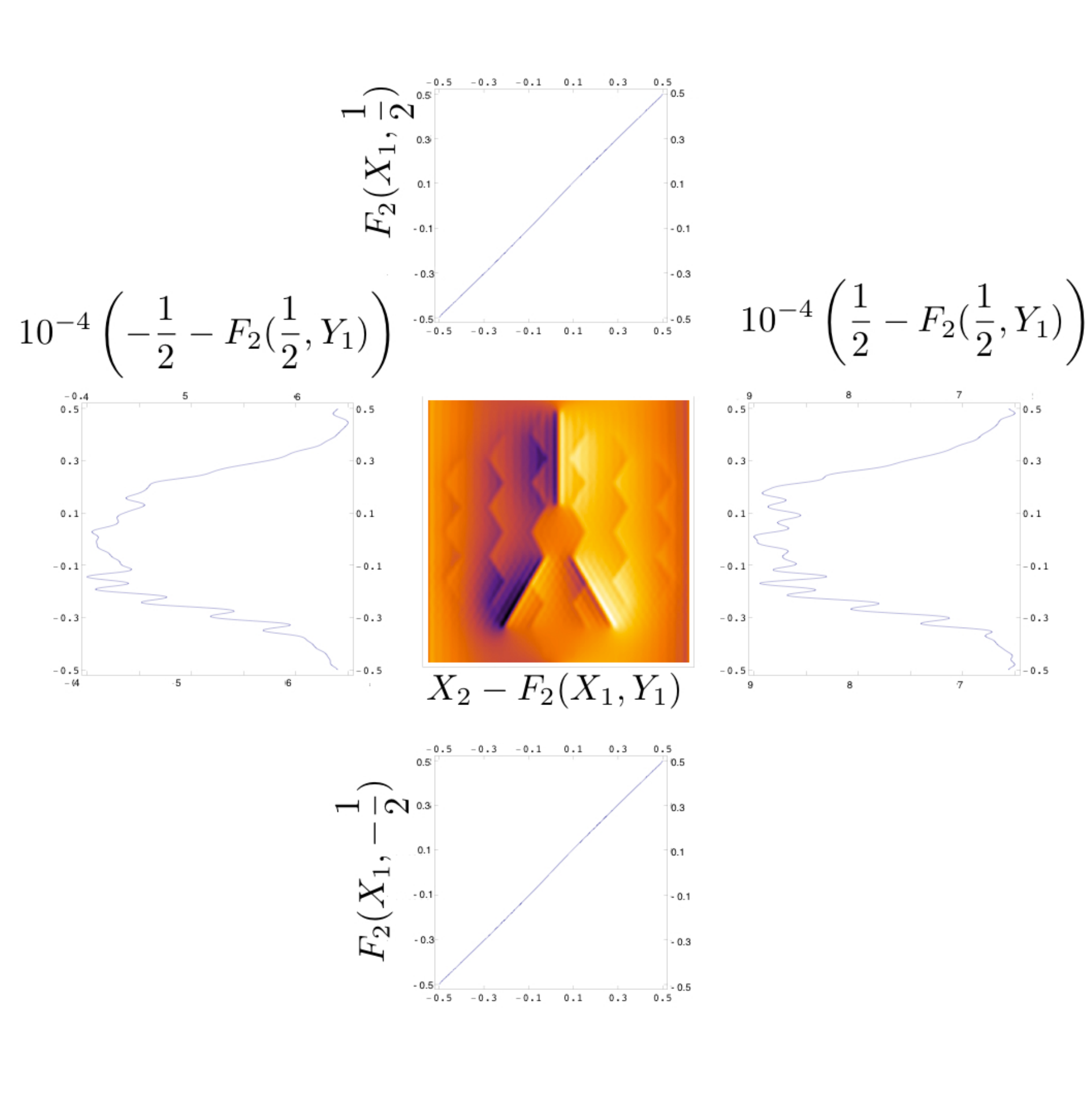}
\end{center}
\caption[]{Boundary conditions seen in the horizontal remapped space. Should the boundary conditions have strictly been enforced by our solver then $F_2(-\frac{1}{2},Y) = -\frac{1}{2}$, $F_2(\frac{1}{2},Y) = \frac{1}{2}$,  $F_2(X,-\frac{1}{2}) = X$, $F_2(X,\frac{1}{2}) = X$. The remapping function obtained with our solutions are very close to these theoretical boundary conditions and the residuals can easily be mitigated by sacrificing the edge rows and columns of actuators on each DM.}
\label{fig:BCX}
\end{figure}

\subsection{Deformation of the Second Mirror}
Once the surface of DM1 has been calculated as a solution of Eq.~\ref{eq::mongeremap}, we compute the surface of DM2 based on Eqs.~\ref{eq:DM2EquaDiff}, which stem from enforcing flatness of the outgoing on-axis wavefront. We seek a Fourier representation for the surface of DM2:
\begin{eqnarray}
&& h_2(x,y) = \frac{D^2}{Z}H_2(X,Y) = \nonumber \\
&& \frac{D^2}{Z} \sum_{n=-N/2}^{N/2} \sum_{m=-N/2}^{N/2} b_{m,n} e^{i \frac{2 \pi}{D} (m X +n X)}
\end{eqnarray}
Plugging the solution found in the previous step for $h_1$ into Eqs.~\ref{eq:DM1EquaDiff} yields a closed form for the normalized remapping functions, $(F_2,G_2)$:
\begin{eqnarray}
F_2(X_1,Y_1) &=& X_1 -  \sum_{n=-N/2}^{N/2} \sum_{m=-N/2}^{N/2} i 2 \pi m \; a_{m,n} \; e^{i 2 \pi (m X_1 +n Y_1)}\nonumber \\
G_2(X_1,Y_1) &=& Y_1 -  \sum_{n=-N/2}^{N/2} \sum_{m=-N/2}^{N/2} i 2 \pi n \; a_{m,n} \; e^{i 2 \pi (m X_1 +n Y_1)} \nonumber
\end{eqnarray}
Then the normalized version of Eqs.~\ref{eq:DM2EquaDiff} can be rewritten as: 
\begin{eqnarray}
L_{x}(X_1,Y_1) &=& \sum_{m=-N/2}^{N/2} i 2 \pi m\; b_{m,n} \; e^{i 2 \pi (m F_2(X_1,Y_1)  +n F_2(X_1,Y_1) )} \nonumber \\
R_{x}(X_1,Y_1) &=& X_1 - F_2(X_1,Y_1) \nonumber \\
L_{x}& =& R_{x} \label{eq:invremap1} \\
L_{y}(X_1,Y_1)&=& \sum_{m=-N/2}^{N/2} i 2 \pi n\; b_{m,n} \; e^{i 2 \pi (m F_2(X_1,Y_1)  +n F_2(X_1,Y_1) )} \nonumber \\
R_{y}(X_1,Y_1)&=& Y_1 - G_2(X_1,Y_1) \nonumber \\
L_{y} &=& R_{y} \label{eq:invremap2}
\end{eqnarray}
We then multiply each side of Eq.~\ref{eq:invremap1} and Eq.~\ref{eq:invremap2} by:
\begin{equation}
 e^{i 2 \pi (m_0 F_2(X_1,Y_1)  +n_0 F_2(X_1,Y_1) )} \det \left[ Id+ \mathcal{H}\left(H_1^{k}(X_1,Y_1)\right) \right] \nonumber
 \end{equation}
 where $(m_0,n_0)$ corresponds to a given DM spatial frequency. 
 \begin{widetext}
 Integrating over the square area of the DM and using the orthogonality of the Fourier basis yields the following system of $2*(N+1)^2$ equations with $(N+1)^2$ real unknowns:
\begin{eqnarray}
&&\mbox{ For all }  (m_0,n_0) \mbox{:}   \nonumber\\
 && 2 \pi i \; m_0 b_{m_0,n_0} = \int_{DM} R_{x}(X,Y)\det[ Id +\mathcal{H}(H_1^{k}(X,Y))]  e^{i 2 \pi (m_0 F_2(X,Y)  +n_0 F_2(X,Y) )}\; dX dY \nonumber \\
&&\mbox{ For all } \; (m_0,n_0) \mbox{:} \nonumber \\
 && i 2 \pi i \;  n_0 b_{m_0,n_0} = \int_{DM} R_{x}(X,Y) \det[Id + \mathcal{H}( H_1^{k}(X,Y))]  e^{i 2 \pi (m_0 F_2(X,Y)  +n_0 F_2(X,Y) )}\; dX dY \nonumber
\end{eqnarray}
We then find $H_{2}$ , the normalized surface of DM2, by solving this system in the least squares sense.
 \end{widetext}
 Once the Monge Ampere Equation has been solved, the calculation of the surface of the second mirror is a much easier problem. Indeed, by virtue of the conservation of the on-axis optical path length, finding the surface of DM2 only consists of solving a linear system (see \cite{2003ApJ...599..695T}). 

\begin{figure}[t!]
\begin{center}
\includegraphics[width=3.5in]{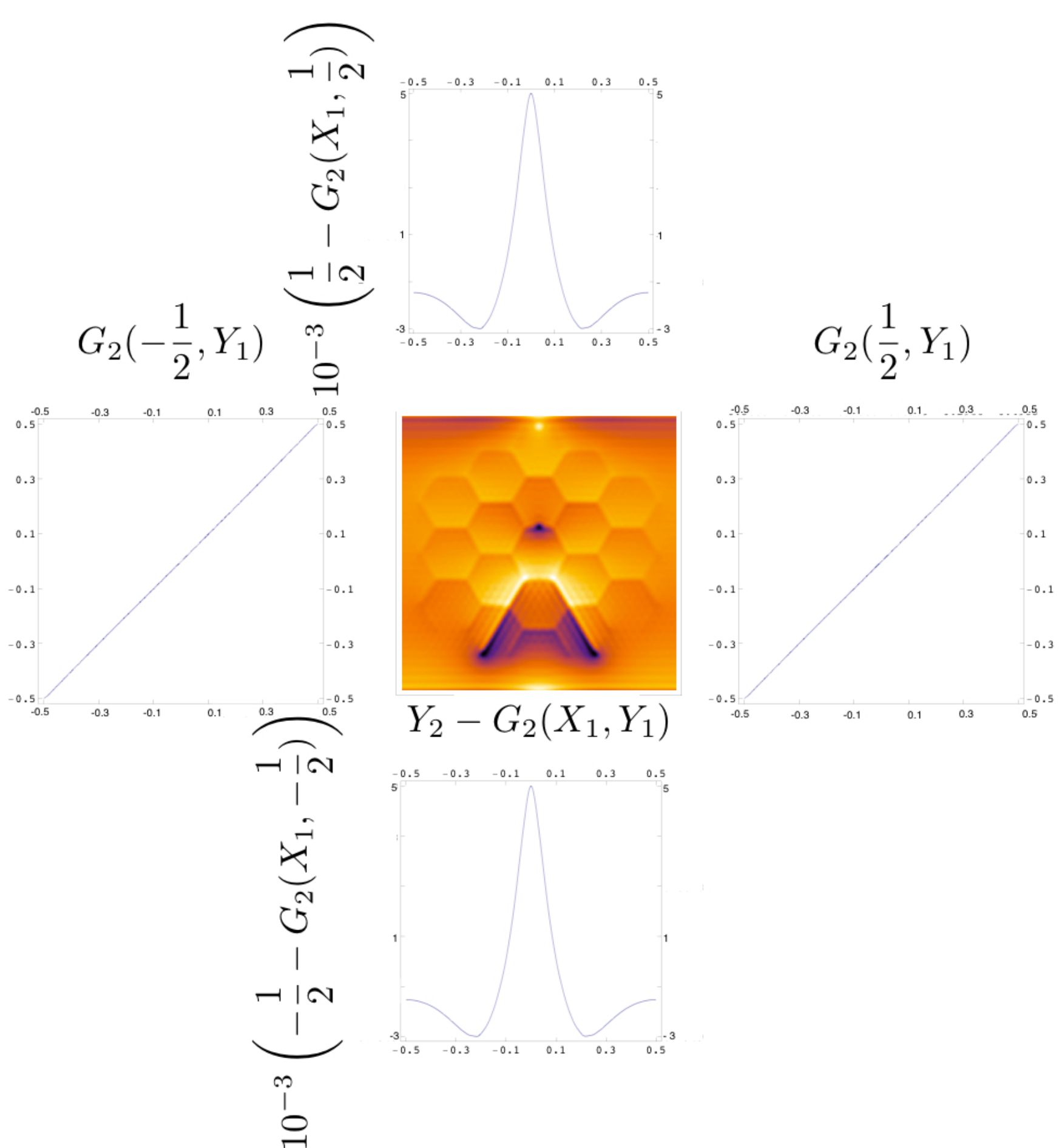}
\end{center}
\caption[]{Boundary conditions seen in the vertical remapped space. Should the boundary conditions have strictly been enforced by our solver then $G_2(-\frac{1}{2},Y) =Y$, $G_2(\frac{1}{2},Y) = Y$,  $G_2(X,-\frac{1}{2}) =  -\frac{1}{2}$, $G_2(X,\frac{1}{2})  = \frac{1}{2}$. The remapping function obtained with our solutions are very close to these theoretical boundary conditions and the residuals can easily be mitigated by sacrificing the edge rows and columns of actuators on each DM.}
\label{fig:BCY}
\end{figure}

\subsection{Boundary Conditions}
The method described above does not enforce any boundary conditions associated with Eq.~\ref{eq::normMA}. One practical set of boundary conditions consists of forcing the edges of each DM to map to each other:
\begin{eqnarray}
F_i(\pm \frac{1}{2},Y) &=& \pm\frac{1}{2}\\
F_i(X,\pm\frac{1}{2}) &=& X\\
G_i(\pm\frac{1}{2},Y) &=&Y\\
G_i(X,\pm\frac{1}{2}) &=&  \pm\frac{1}{2}
\end{eqnarray}
with $i=1,2$. These correspond to a set of Neumann boundary conditions in $H_1(X,Y)$ and $H_2(X,Y)$. These boundary conditions can be enforced by augmenting the dimensionality of the linear systems on Eq.~\ref{eq:inverseLoeper} and Eq.~\ref{eq:inversePoisson}, however  doing so increases the residual least squares errors and thus hampers the contrast of the final solution. Moreover Fig.~\ref{fig:BCX} and Fig.~\ref{fig:BCY} show that, because of the one to one remapping near the DM edges in the control term of the reverse problem, the boundary conditions are  almost met in practice. For the remainder of this paper we thus do not include boundary conditions when calculating the DM shapes, when solving for $H_1(X,Y)$ in Eq.~\ref{eq::normMA} since, in the worse case, only the edge rows and columns of the DMs actuator will have to be sacrificed in order for the edges to truly map to each other.

\subsection{Remapped aperture}

For a given pupil geometry we have calculated $(H_1,H_2)$. We then convert the DM surfaces to real units, $(h_1,h_2)$, by multiplication with $D^2/Z$. We evaluate the remapping functions using Eqs.~\ref{eq:DM1EquaDiff} and \ref{eq:DM2EquaDiff} and obtain the field at the entrance of the coronagraph in the ray optic approximation
\begin{widetext}
\begin{equation}
E_{DM2}(x_2,y_2)  =  \left \{ \sqrt{det[J]} E_{DM1}\left(f_1,g_1\right) e^{i \frac{2 \pi }{\lambda} \left( S(f_1,g_1) + h_{1}(f_1,g_1) - h_{2} \right)} \right \}\bigg|_{(x_2,y_2)}
\label{eq:remappedfield}
\end{equation}
 \end{widetext}

where the exponential factor corresponds to the Optical Path Length through the two DMs. Even if the surface of the DMs has been calculated using only a finite set of Fourier modes, we check that the optical path length is conserved. \\

\begin{figure}[t!]
\begin{center}
\includegraphics[width=3in]{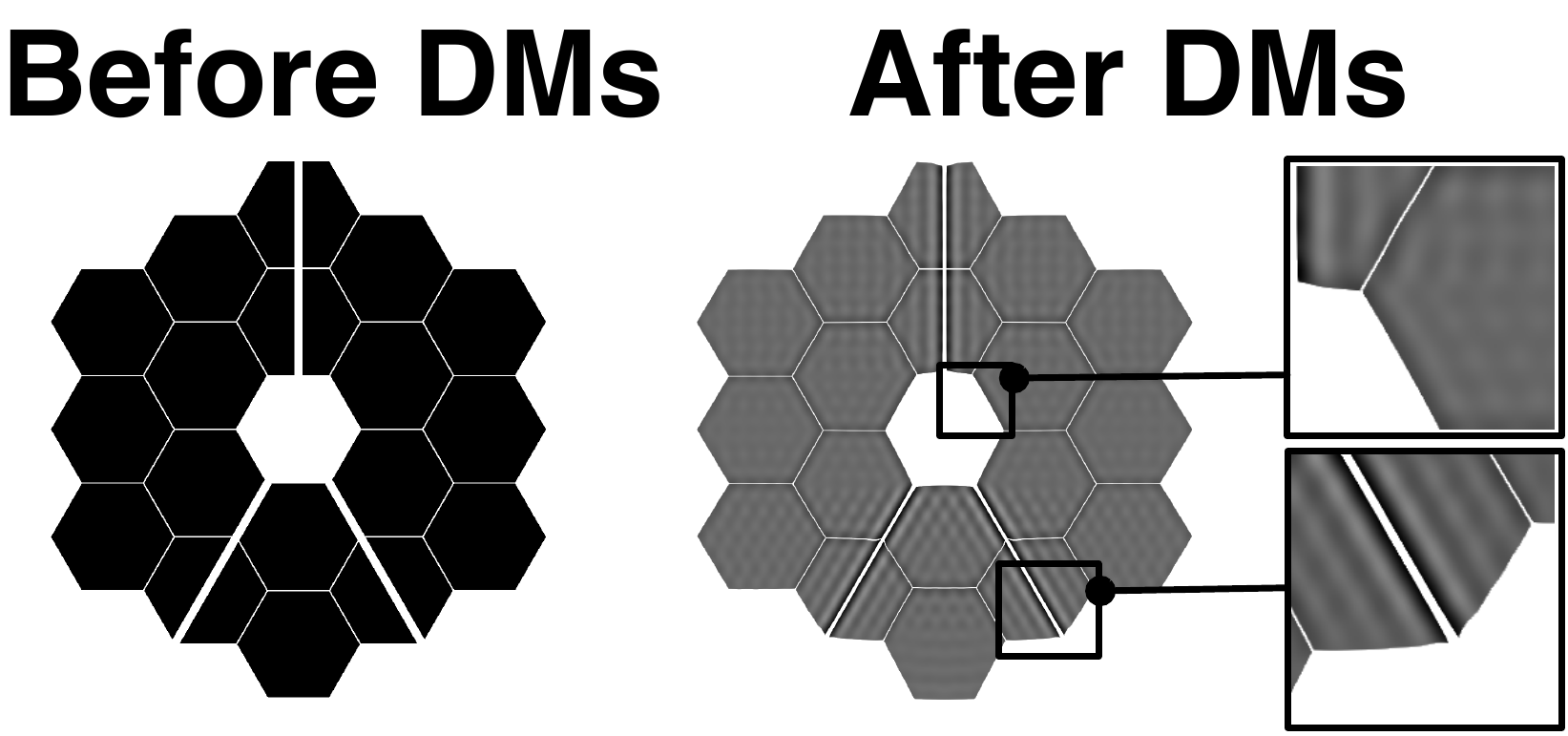}
\end{center}
\caption[]{Beam amplitude before and after the DMs in the case of a geometry similar to JWST.  We chose to solve the reverse problem over a square using a Fourier basis-set and not stricly enforcing boundary conditions. This results in small distortions of the edges of the actual aperture in the vicinity of  segments gaps and secondary supports. We address this problem by slightly oversizing the secondary obscuration and undersizing the aperture edge in the coronagraph.}
\label{fig:RepammedJWSTPupil}
\end{figure}

Fig.~\ref{fig:RepammedJWSTPupil} shows that since the curvature of the DMs is limited by the number of modes $N$, our solution does not fully map out the discontinuities induced by the secondary supports and the segments. However, they are significantly thinner and one can expect that their impact on contrast will be attenuated by orders of magnitude. In order to quantify the final coronagraphic contrasts of our solution we then propagate it through an APLC coronagraph designed using the method in \S.~\ref{sec:CentralObsc}. In the case of  a hexagon based primary (such as JWST), we use a coronographic apodizer with a slightly oversized secondary obscuration and undersize outer edge in order to circularize the pupil. Note that this choice is mainly driven by the type of coronagraph we chose in \S.~\ref{sec:CentralObsc} to illustrate our technique. Since the DM control strategy presented in this section is independent of the coronagraph, it can be generalized to any of the starlight suppression systems which have been discussed in the literature. For succinctness we present our results using coronagraph solely based on using pupil apodization (either in an APLC or in a PIAAC configuration). Results for a JWST geometry are shown on Fig.~\ref{fig:ResultsJWST} and Fig.~\ref{fig:SlicesJWST} and discussed in \S.~\ref{sec:Resuls}. Eq.~\ref{eq:remappedfield} assumes that the propagation between the two DMs occurs according to the laws of ray optics. In the next section we derive the actual diffractive field at DM2, e.g Eq.~\ref{eq::propint}, and show that in the pupil remapping regime of ACAD, edge ringing due to the free space propagation is actually smaller than  in the Fresnel regime. 

\section{Chromatic properties}
\label{sec:Chrom}

\subsection{Analytical expression of the diffracted field}
ACAD is  based on ray optics. It is an inherently broadband technique, and provided that the coronagraph is optimized for broadband performance ACAD will provide high contrast over large spectral windows. However, when taking into account the edge diffraction effects that are captured by the quadratic integral in Eq.~\ref{eq::propint}, the true propagated field at DM2 becomes wavelength dependent. More specifically, when $\lambda$ is not zero then the oscillatory integral superposes on the ray optics field a series of high spatial frequency oscillations. In theory, it would be best to use this as the full transfer function to include chromatic effects in the computation of the DMs shapes. However, as discussed in S.~\ref{sec:MADM}, solving the non-linear Monge-Ampere Equation is already a delicate exercise, and we thus have limited the scope of this paper to ray optics solutions. Nonetheless, once the DMs' shapes have been determined using ray optics, one should check whether or not the oscillations due to edge diffraction will hamper the contrast. This approach is reminiscent of the design of PIAA systems where the mirror shapes are calculated first using geometric optics  and are then propagated through the diffractive integral in order to check a posteriori whether or not the chromatic diffractive artifacts are below the design contrast \citep{2005astro.ph.12421P}. In this section we detail the derivation of Eq.~\ref{eq::propint} that is the diffractive integral for the two DMs remapping system and use this formulation to discuss the diffractive properties of ACAD. 

\begin{widetext}
We start with  the expression of the second order diffractive field at DM2 as derived in \citet{2011ApJS..195...25P}, Eq.~\ref{eq::SecondExp} . 
\begin{equation}
E_{DM2}(x_2,y_2) = \frac{1}{ i \lambda Z} \left \{ \int    E_{DM1}(x,y)  e^{\frac{ i \pi }{\lambda Z}  \Big[ \frac{\partial f_2}{\partial x}(x- x_1)^{2} + 2 \frac{ \partial g_2}{\partial x}(x- x_1 )(y- y_1) +\frac{\partial g_2}{\partial y}(y- y_1 )^{2}  \Big]}dx dy  \right \}\Big|_{(x_1,y_1)}\nonumber
\end{equation}
 We write $E_{DM1}(x,y) $ as its inverse Fourier transform 
\begin{equation}
 E_{DM1}(x,y)  =  \int \widehat{E}_{DM1}(\xi,\eta) e^{i 2 \pi (x \xi + y \eta)} \; d \xi d \eta
 \end{equation}
 
 and insert this expression in Eq.~\ref{eq::SecondExp}. Completing the squares in the quadratic exponential factor then yields:
 \begin{equation}
 E_{DM2}(x_2,y_2) = \frac{1}{ i \lambda Z} \left \{ \int   \widehat{E}_{DM1}(\xi,\eta) I_1(f_1,g_1)e^{i 2 \pi [ f_1 \xi + g_1\eta]}e^{- i \pi \lambda Z \left( \frac{\partial f_2}{\partial y}|_{(f_1,g_1)} \xi^2+ \frac{\partial g_2}{\partial x}|_{(f_1,g_1)} \eta^2 \right)}
 \; d \xi d \eta \right \}\bigg|_{(x_2,y_2)}
\end{equation}
with
\begin{equation}
 I_1(y_1,x_1) = \left \{ \int e^{\frac{ i \pi }{\lambda Z}  \Big[ \frac{\partial f_2}{\partial x}(x- x_1 - \frac{\xi \lambda Z}{\frac{\partial f_2}{\partial x}})^{2} + 2 \frac{ \partial g_2}{\partial x}(x- x_1 )(y- y_1) +\frac{\partial g_2}{\partial y}(y- y_1 - \frac{\eta \lambda Z}{\frac{\partial g_2}{\partial x}})^{2}  \Big]} dx dy \right \}\bigg|_{(x_1,y_1)}.
\end{equation}

The integral over space, $ I_1(y_1,x_1)$, is the integral of a complex gaussian and can be evaluated analytically. This yields:
 \begin{equation}
 E_{DM2}(x_2,y_2) =   \left \{ \frac{1}{\sqrt{|\frac{\partial f_2}{\partial x} \frac{\partial g_2}{\partial y} - \left(\frac{ \partial g_2}{\partial x} \right)^2|} } \int   \widehat{E}_{DM1}(\xi,\eta) e^{i 2 \pi [ x_1 \xi + y_1\eta]}e^{- i \pi \lambda Z \left( \frac{\partial f_2}{\partial y}\xi^2+ \frac{\partial g_2}{\partial x} \eta^2 \right)} d \xi d \eta \right \}\bigg|_{(x_1,y_1)}.  
\end{equation}
We thus have expressed $E_{DM2}(x_2,y_2)$ as a function of $(x_1,y_1) = (f_1(x_2,y_2),g_1(x_2,y_2))$. This expression can be further simplified: using Eqs.~\ref{eq:x2self} to \ref{eq:remapunity} one can derive
$\frac{\partial f_2}{\partial x}|_{(x_1,y_1)} =\frac{1}{\det[J(x_2,y_2)]} \frac{\partial g_1}{\partial y}|_{(x_2,y_2)} $ and  $\frac{\partial g_2}{\partial y}|_{(x_1,y_1)}=\frac{1 }{\det[J(x_2,y_2)]}\frac{\partial f_1}{\partial x}|_{(x_2,y_2)}$. Which finishes to prove Eq.~\ref{eq::propint}:
\begin{equation}
E_{DM2}(x_2,y_2) = \left \{ \sqrt{|\det[J]|}
  \int_{\mathcal{F}\mathcal{P}}  \widehat{ E_{DM1}}(\xi,\eta)  e^{i 2 \pi (\xi f_1+\eta g_1)}e^{- i\frac{\pi \lambda Z}{\det[J]}\left( \frac{\partial g_1}{\partial y} \xi^2+ \frac{\partial f_1}{\partial x}\eta^2 \right)} \; d\xi d \eta \right \}\bigg|_{(x_2,y_2)}
\end{equation}
This expression is very similar to a modified Fresnel propagation and can be rewritten as such:
 \begin{equation}
E_{DM2}(x_2,y_2) = \left \{
  \int_{\mathcal{A}} E_{DM1}(x,y)  e^{- i\frac{\pi det[J]} {\lambda Z} \left( (\frac{\partial g_1}{\partial y})^{-1}  (x-f_1)^2+ (\frac{\partial f_1}{\partial x})^{-2}(y - g_1)^2 \right)} \; dx d y \right \}\bigg|_{(x_2,y_2)}
\end{equation}
Because of this similarity we call this integral the Stretched- Remapped Fresnel approximation (SR-Fresnel). Indeed in this approximation the propagation distance is stretched by $(\frac{\frac{\partial g_1}{\partial y}}{ det[J]},\frac{\frac{\partial f_1}{\partial x}}{ det[J]})$ and the integral is centered around the remapped pupil $(f_1,g_1)$.
\end{widetext}
\subsection{Discussion}
\begin{figure*}[t!]
\begin{center}
\includegraphics[width=7in]{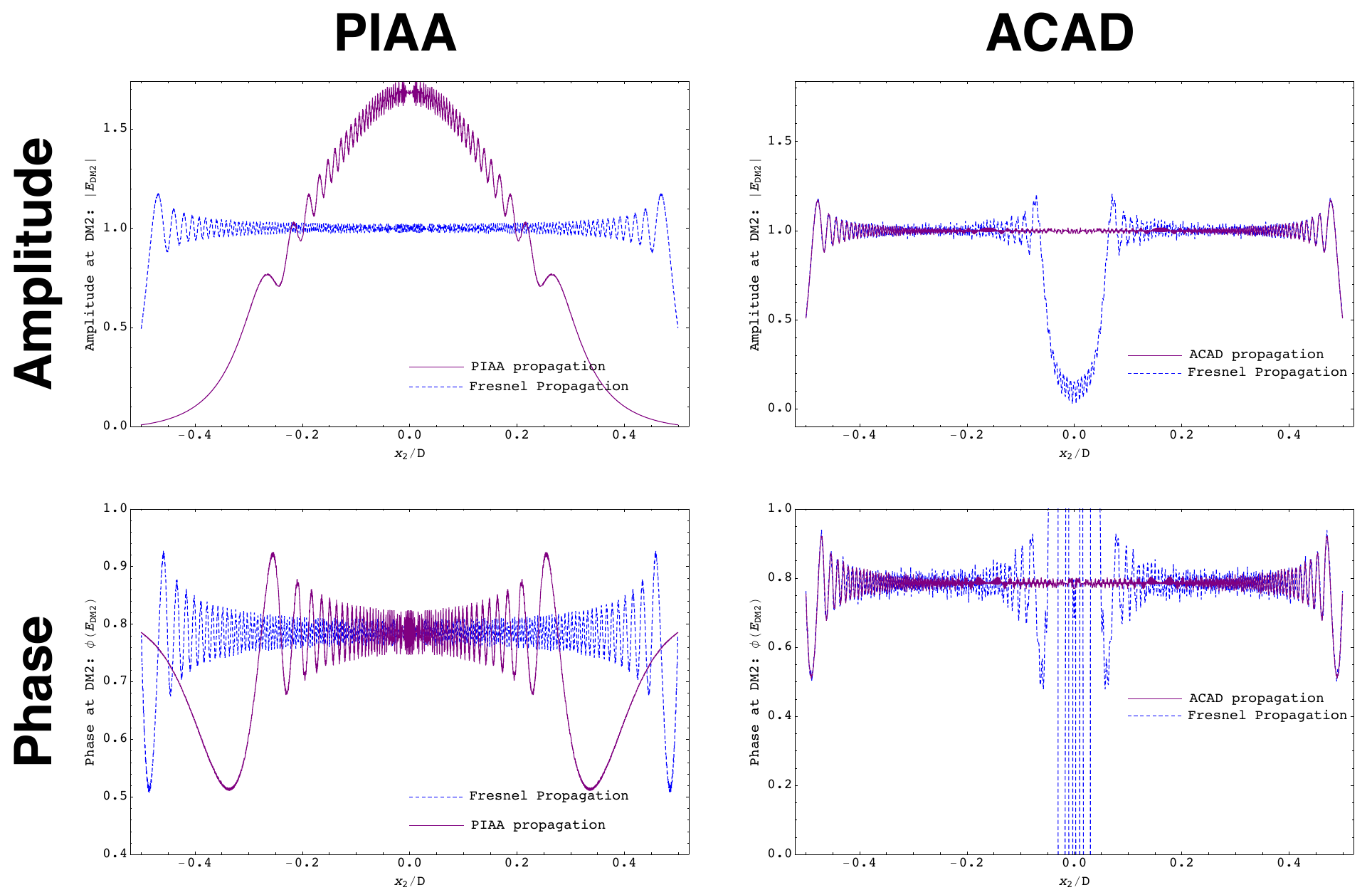}
\end{center}
\caption[]{{\bf Left:} Comparison of edge diffraction between PIAA and Fresnel propagation. Because the discontinuities in the pupil occur at the location where $\Gamma_x <1$, a PIAA amplifies the chromatic ringing when compared to a more classical propagation. {\bf Right:} Comparison of edge diffraction between ACAD and Fresnel propagation. Because the discontinuities in the pupil occur at the location where $\Gamma_x >1$, a ACAD damps the chromatic ringing when compared to a more classical propagation. These simulations were carried out for a one dimensional aperture. In this case the Monge Ampere Equation can be solved using finite elements and the calculation of the diffractive effects reduces to the evaluation of various Fresnel special functions at varying wavelength. The full two-dimensional problem is not separable and requires the development of novel numerical tools.}
\label{fig:PropCurvatures}
\end{figure*}
The integral form provides physical insight about the behavior of the chromatic edge oscillations. If we write
\begin{eqnarray}
\Gamma_{x} = \frac{ det[J]} {\frac{\partial g_1}{\partial y}}\bigg|_{(x_2,y_2)} \\
\Gamma_{y} = \frac{ det[J]} {\frac{\partial f_1}{\partial x}}\bigg|_{(x_2,y_2)}
\end{eqnarray}
we can identify a several diffractive regimes:
\begin{enumerate}
\item when $\Gamma_{x}  = \Gamma_{y} = 1$ the $E_{DM2}(x_2,y_2)$ reduces to a simple Fresnel propagation. Edge ringing can them be mitigated using classical techniques such as pre-apodization, or re-imaging into a conjugate plane using oversized relay optics.
\item when  $\Gamma_{x}, \Gamma_{y}< 1$ then it is as if the  effective propagation length through the remapping unit was {\em increased}. This magnifies the edge chromatic ringing. 
\item when  $\Gamma_{x}, \Gamma_{y} >1$ then it is as if the effective propagation length through the remapping unit was {\em decreased}. This damps the edge chromatic ringing. 
\item when  $\Gamma_{x}  >1$ and  $ \Gamma_{y} <1$, e.g. at a saddle point in the DM surface,  then it is as if the effective propagation length through the remapping unit was {\em decreased} in one direction and {\em increased} in the other. The edge chromatic ringing can either be damped or magnified depending on the relative magnitude of $\Gamma_x$ and $\Gamma_y$.
\end{enumerate}
In the case of a PIAA coronagraphs, the mirror shapes are such that $\Gamma_{x}, \Gamma_{y} >1$ at the center of the pupil and $\Gamma_{x}, \Gamma_{y}\ll 1$ at the edges of the pupil, where the discontinuities occur. As a consequence the edge oscillations are largely magnified when compared to Fresnel oscillations (see right panel of Fig.~\ref{fig:PropCurvatures}), and apodizing screens are necessary in order to reduce the local curvature of the mirror's shape (as also discussed in \citet{2005astro.ph.12421P,2011ApJS..195...25P}). In the case of ACAD, where the x-axis is chosen to be perpendicular to the discontinuity, the surface curvature is such that $\Gamma_x > 1$,  $\Gamma_y \sim 1$ at the discontinuities inside the pupil and $\Gamma_x > 1$,  $\Gamma_y  \sim 1$ elsewhere. This yields damped chromatic oscillations at the remapped discontinuities and Fresnel oscillations at the edges of the pupil (see right panel of Fig.~\ref{fig:PropCurvatures}). Note that Fig.~\ref{fig:PropCurvatures} was generated using a one dimensional toy model that assumes Eq.~\ref{eq::propint} is separable, e.g $\Gamma_y  =  1$, as described in  \citet{Pueyo:11}. In practice at the saddle points of the optical surfaces,  near the junction of two spiders for instance,  $\gamma_x > 1$,  $\Gamma_y  <\sim 1$  and thus our separable model does not guarantee than in the true 2D case chromatic edge oscillations might not be locally amplified. However even near the saddle points ACAD provides a strong converging remapping in the direction perpendicular to the discontinuity and very little diverging re-mapping in the other direction. As a consequence $\Gamma_x \gg 1$ and  $\Gamma_y$ is smaller than but close to one. We thus predict that even at these locations chromatic ringing will not be amplified. Even if ACAD based on pupil remapping, its diffraction properties are qualitatively very different from PIAA coronagraphs since edge ringing is not amplified beyond the Fresnel regime at the pupil edges, and is  attenuated near the discontinuities. We conclude that in most cases ACAD operates in a regime where edge chromatic oscillations are not larger than classical Fresnel oscillations, and sometimes actually smaller. As a consequence the chromaticity of this ringing can be mitigated using standard techniques developed in the Fresnel regime and we do not expect this phenomenon to be a major obstacle to ACAD broadband operations. 
\subsection{Diffraction artifacts in ACAD are no worse than Fresnel ringing}
\begin{figure*}[t!]
\begin{center}
\includegraphics[width=8in]{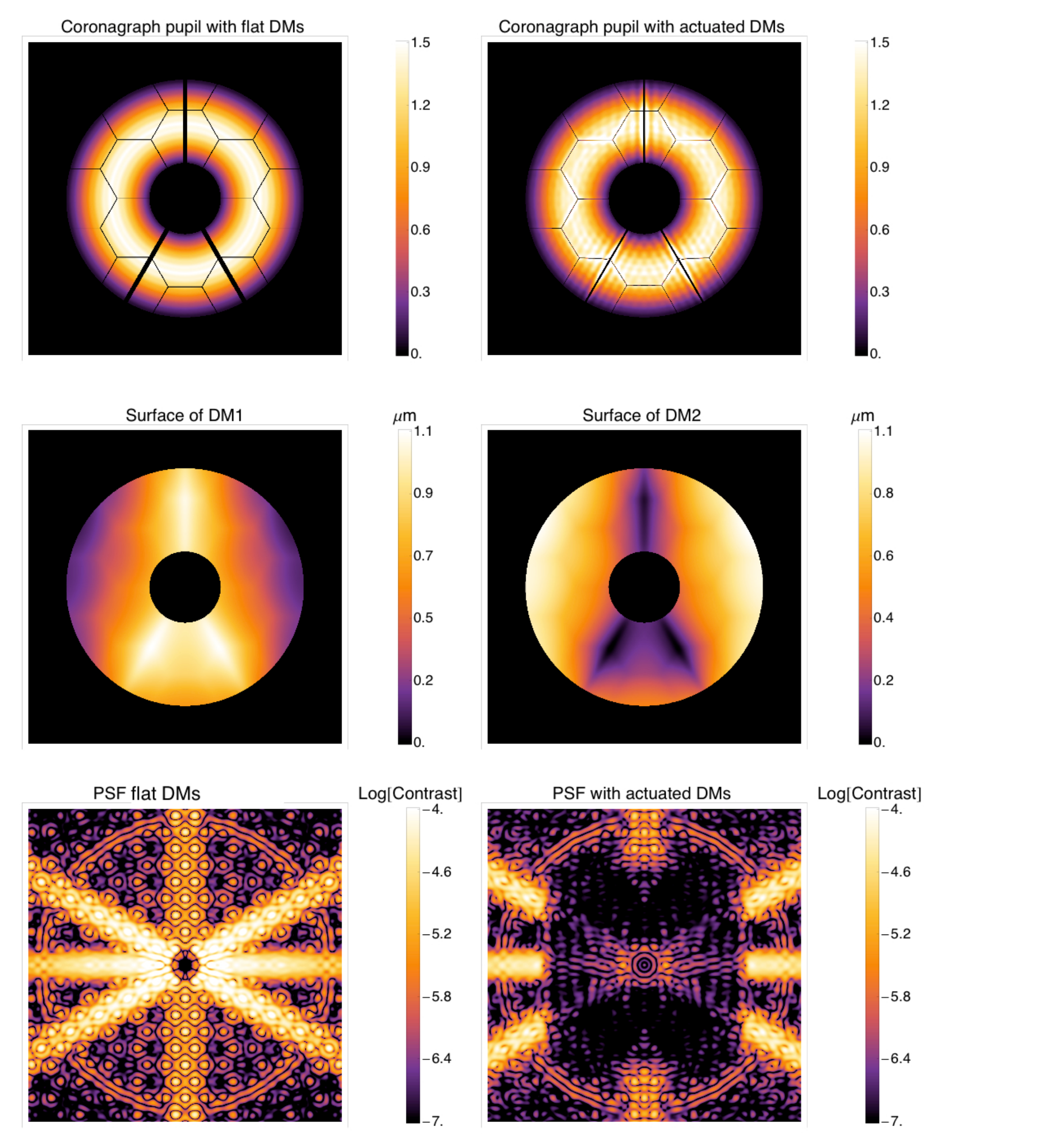}
\end{center}
\caption[]{Results obtained when applying our approach to a geometry similar to JWST . We used two 3 cm DMs of 64 actuators separated by 1 m. Their maximal surface deformation is  $1.1 \; \mu$m, well within the stroke limit of current DM technologies. The residual light in the corrected PSF follows the secondary support structures and can potentially be further cancelled by controlling the DMs using an image plane based cost function, see Fig.~\ref{fig:wavecontrolACAD}}.
\label{fig:ResultsJWST}
\end{figure*}
We have established that the diffractive chromatic oscillations introduced by the fact that DM1 and DM2 are not located in conjugate planes is no worse than classical Fresnel ringing from the aperture edges and can be mitigated using well-know techniques which have been developed for this regime. While a quantitive tradeoff study of how to design a high contrast instrument which minimizes such oscillations regime is beyond the scope of this paper, we briefly remind their qualitative essence to the reader: 
\begin{itemize}
\item The edges of the discontinuities in the telescope aperture can be smoothed via pupil apodization before DM1. This solution is not particularly appealing as it requires the introduction of a transmissive, and thus dispersive, component in the optical train. 

\item The distance between the two DMs can be reduced. Indeed the DMs deformations presented herein, for $3$ cm DMs separated by $1$ m, are all $\leq 1 \; \mu$m while current technologies allow for deformations of several microns. As the edge ringing scales as $Z/D^2$ chromatic oscillations will be reduced by decreasing $Z$. Since the DM surfaces scale as $D^2/Z$ reducing $Z$ will increase the DM deformations but have little impact on the feasibility of our solution as current DM technologies can reach $4 \; \mu$m strokes. 
\item The coronagrahic apodizer can be placed in a plane that is conjugate to the DM1. This can be achieved by re-imaging DM2 through a system of oversized optics (the over-sizing factor increases steeply when the pupil diameter decreases). By definition there are no Fresnel edge oscillations in such a plane. Alternatively a coronagraph without a pupil apodization (amplitude or phase mask in the image plane) can be used, and in this configuration it is only sufficient for the optics to be oversized.
\end{itemize} 
Note that these three solutions are not mutually exclusive and that only a full diffractive analysis, which uses robust numerical propagators that have been developed based on Eq.~\ref{eq::propint}, can quantitatively address the trade-offs discussed above. The development of such propagators is our next priority. In \citet{Pueyo:11} we laid out the theoretical foundations of such a numerical tool in the case of circularly symmetric pupil remapping and this solution has been since then practically implemented, as reported by \citet{krist:77314N}. Generalizing this method to a tractable propagator in the case of arbitrary remapping is a yet unsolved computational problem. In the meantime we emphasize that while the spectral bandwidth of coronagraphs whose incoming amplitude has been corrected using ACAD will certainly be limited by edge diffraction effects, but these effects are {\em no worse than Fresnel ringing} and can thus be mitigated using optical designs which are now routinely used in high contrast instruments (see \citet{2010SPIE.7736E..55V} for such discussions).  For the remainder of this paper we thus assume the diffractive artifacts have been adequately mitigated and we compute our results assuming a geometric propagation between DM1 and DM2. 
\begin{figure}[t!]
\begin{center}
\includegraphics[width=3.5in]{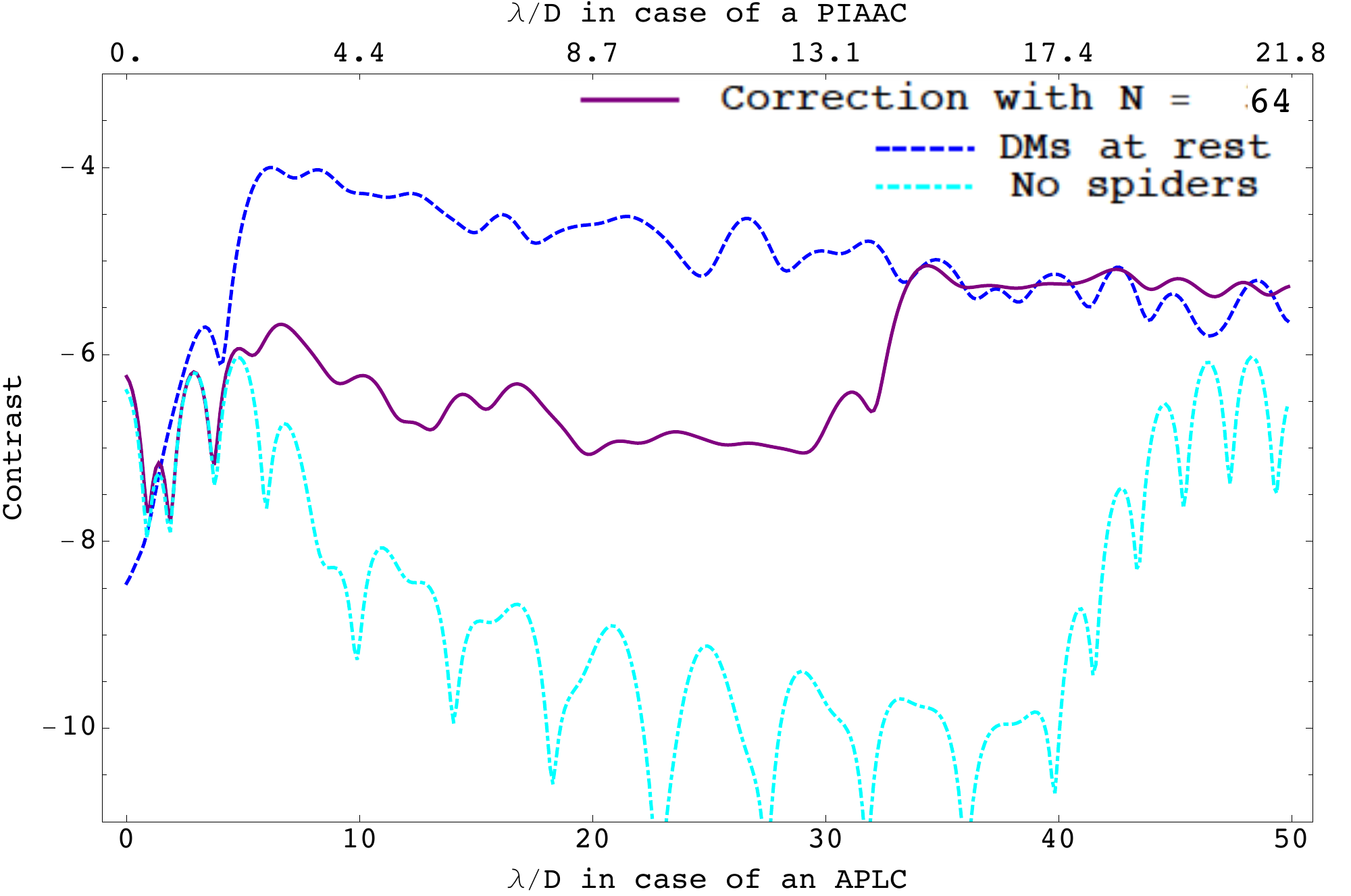}
\caption[]{{\bf Case of JWST:} Radial average obtained when applying ACAD. We used two 3 cm DMs of 64 actuators separated by 1 m. Their maximal surface deformation is  $1.1 \; \mu$m, well within the stroke limit of current DM technologies. ACAD yields a gain in contrast of two orders of magnitude, and provides contrasts levels similar to upcoming Ex-AO instruments, which are designed on much friendlier apertures geometries. Since ACAD removes the bulk of the light diffracted by the asymmetric aperture discontinuities, the final contrast can be improved by controlling the DMs using and image plane based cost function, see Fig.~\ref{fig:wavecontrolACAD}}.
\label{fig:SlicesJWST}
\end{center}
\end{figure}
\begin{figure*}[h!]
\begin{center}
\includegraphics[width=8in]{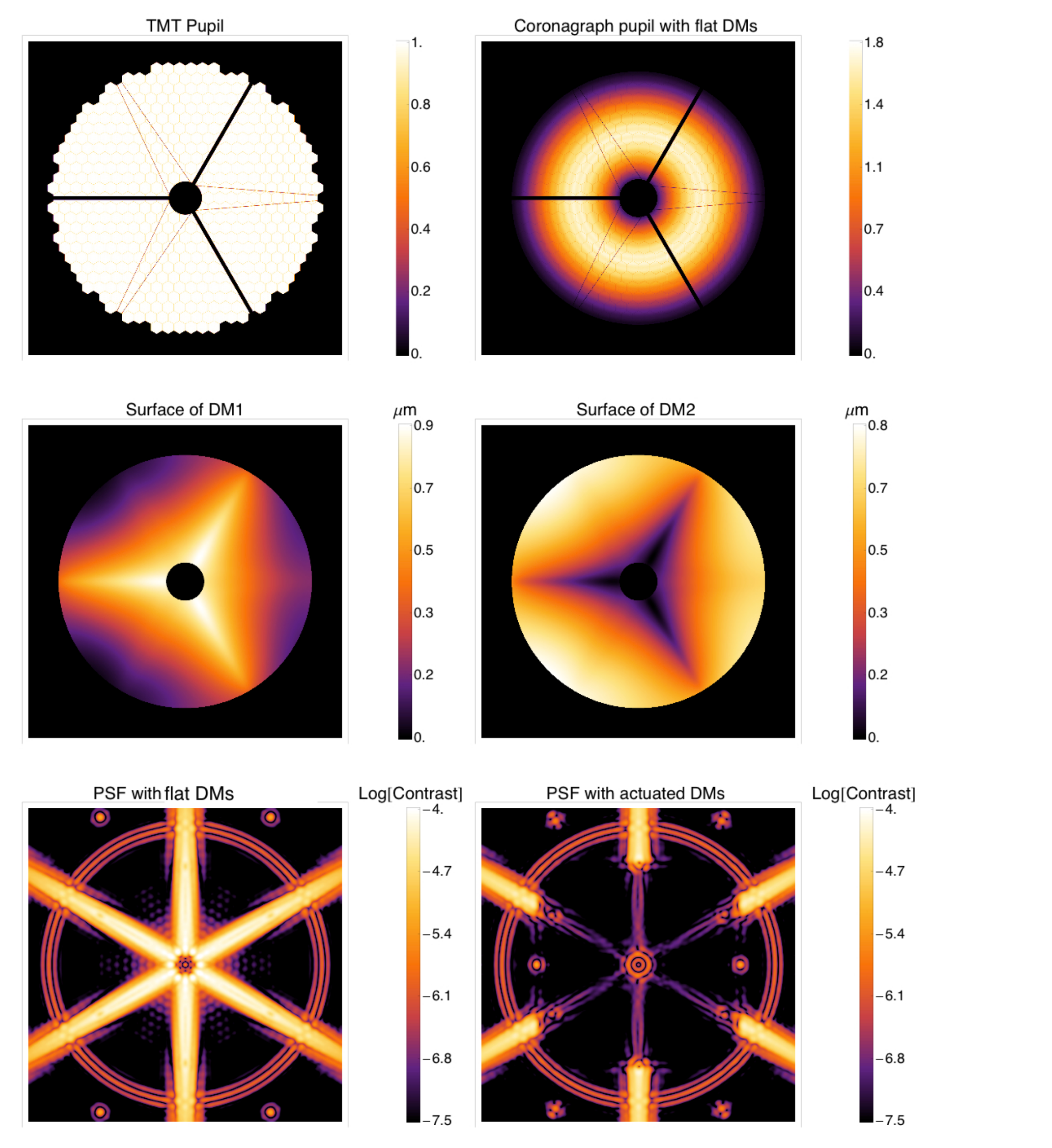}
\end{center}
\caption[]{Results obtained when applying our approach to a TMT geometry. We used two 3 cm DMs of 64 actuators separated by 1 m. Their maximal surface deformation is  $0.9 \; \mu$m, well within the stroke limit of current DM technologies. The final contrast is below $10^{7}$, in a regime favorable for direct imaging of exo-planets with ELTs.}
\label{fig:ResultsTMT}
\end{figure*}
\section{Results}
\label{sec:Resuls}
\subsection{Application to future observatories}
\subsubsection{JWST}

We have illustrated each step of the calculation of the DM shapes in \S.~\ref{sec:MADM} using a geometry similar of JWST. This configuration is somewhat a conservative illustration of an on-axis segmented telescope as it features thick secondary supports and a ``small'' number of segments whose gaps diffract light in regions of the image plane close to the optical axis (the first diffraction order of a six hexagons structure is located at $\sim 3 \lambda/D$). In order to assess the performances of ACAD on such an observatory architecture we chose to use a coronagraph designed around a slightly oversized secondary obscuration of diameter $0.25 \; D$, with a  focal plane mask of diameter $8 \lambda /D$, an IWA of $5 \lambda/D$ and an OWA of $30 \lambda / D$. The field at the entrance of the coronagraph after remapping by the DMs is shown on the top right panel of Fig.~\ref{fig:ResultsJWST}. The DM surfaces, calculated assuming 64 actuators across the pupil ($N=64$ in the Fourier expansion) and DMs of diameter $3$ cm separated by $Z = 1$ m, are shown on the middle panel of Fig.~\ref{fig:ResultsJWST}. They are well within the stroke limit of current DM technologies. The surfaces were calculated by solving the reverse problem over an even grid of 10 cutoff low-pass spatial frequencies ranging between $30$ and $70$ cycles per apertures for the tapering kernel $\omega$. The value yielding the best contrast was chosen. Note that the optimal cutoff frequency depends on the spatial scale of the discontinuities, and that higher contrasts could be obtained by choosing a set of two convolution kernels for the reverse problem and finding the optimal solution using a finer grid. However, the results in the bottom row of Fig.~\ref{fig:ResultsJWST} are extremely promising.  Fig.~\ref{fig:SlicesJWST} shows a contrast improvement of a factor of $100$ when compared to the raw coronagraphic PSF, which is quite remarkable for an algorithm which is not based on an image-plane metric. These results illustrate that even with a very unfriendly aperture similar to JWST one can obtain contrasts as high as envisioned for upcoming Ex-AO instruments, which have been designed for much friendlier apertures. While we certainly do not advocate to use such a technique on JWST, this demonstrates that ACAD is a powerful tool for coronagraphy with on-axis segmented apertures.
\begin{figure}[h!]
\begin{center}
\includegraphics[width=3.5in]{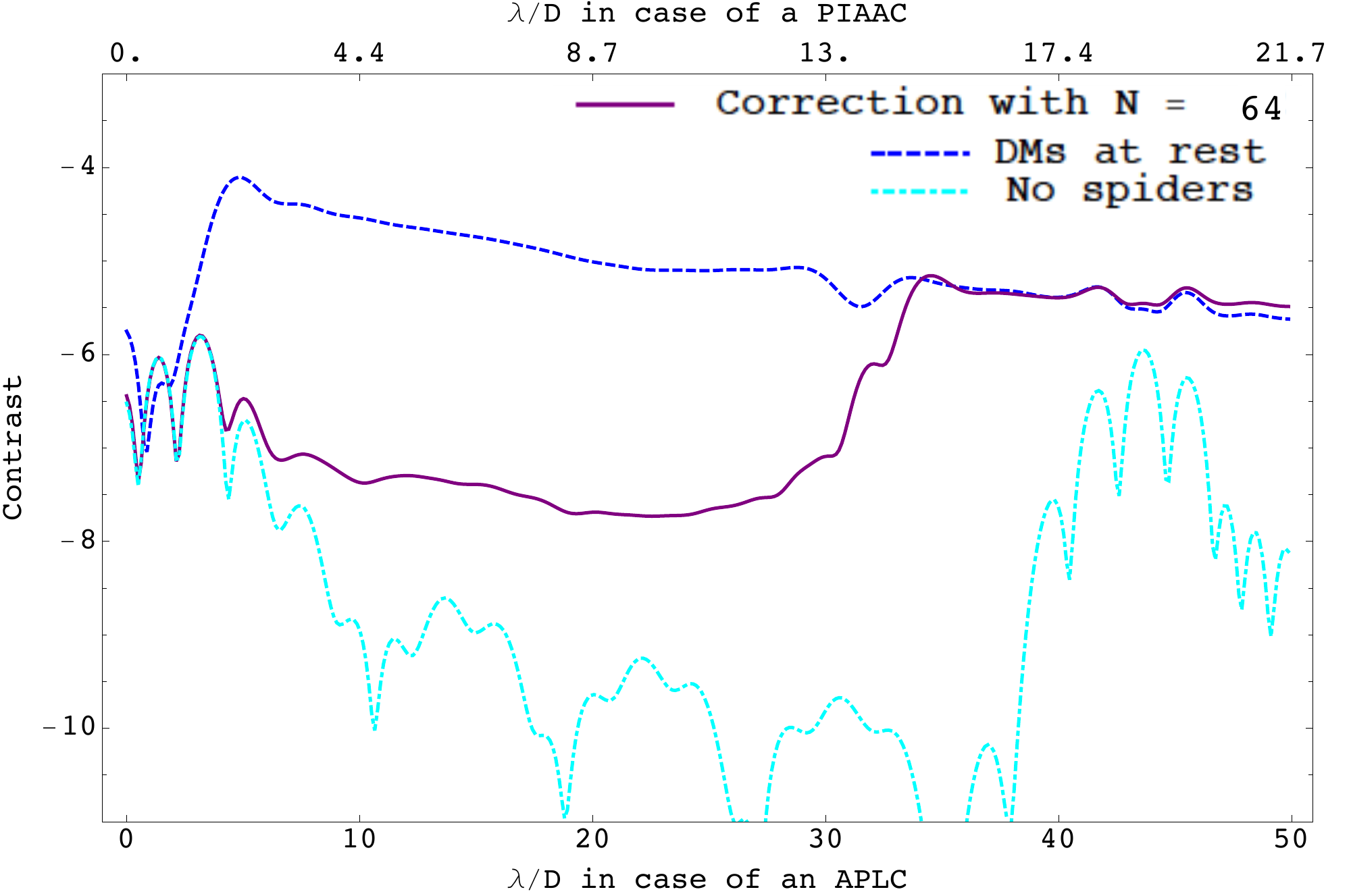}
\end{center}
\caption[]{{\bf Case of TMT:} radial average obtained when applying ACAD.  We used two 3 cm DMs of 64 actuators separated by 1 m. Their maximal surface deformation is  $0.9 \; \mu$m, well within the stroke limit of current DM technologies. The final contrast is below $10^{7}$, in a regime favorable for direct imaging of exo-planets with ELTs. Since ACAD removes the bulk of the light diffracted by the asymmetric aperture discontinuities, the final contrast can be enhanced by controlling the DMs using and image plane based metric.}
\label{fig:SlicesTMT}
\end{figure}
\subsubsection{Extremely Large Telescopes}
\begin{figure*}[h!]
\begin{center}
\includegraphics[width=8in]{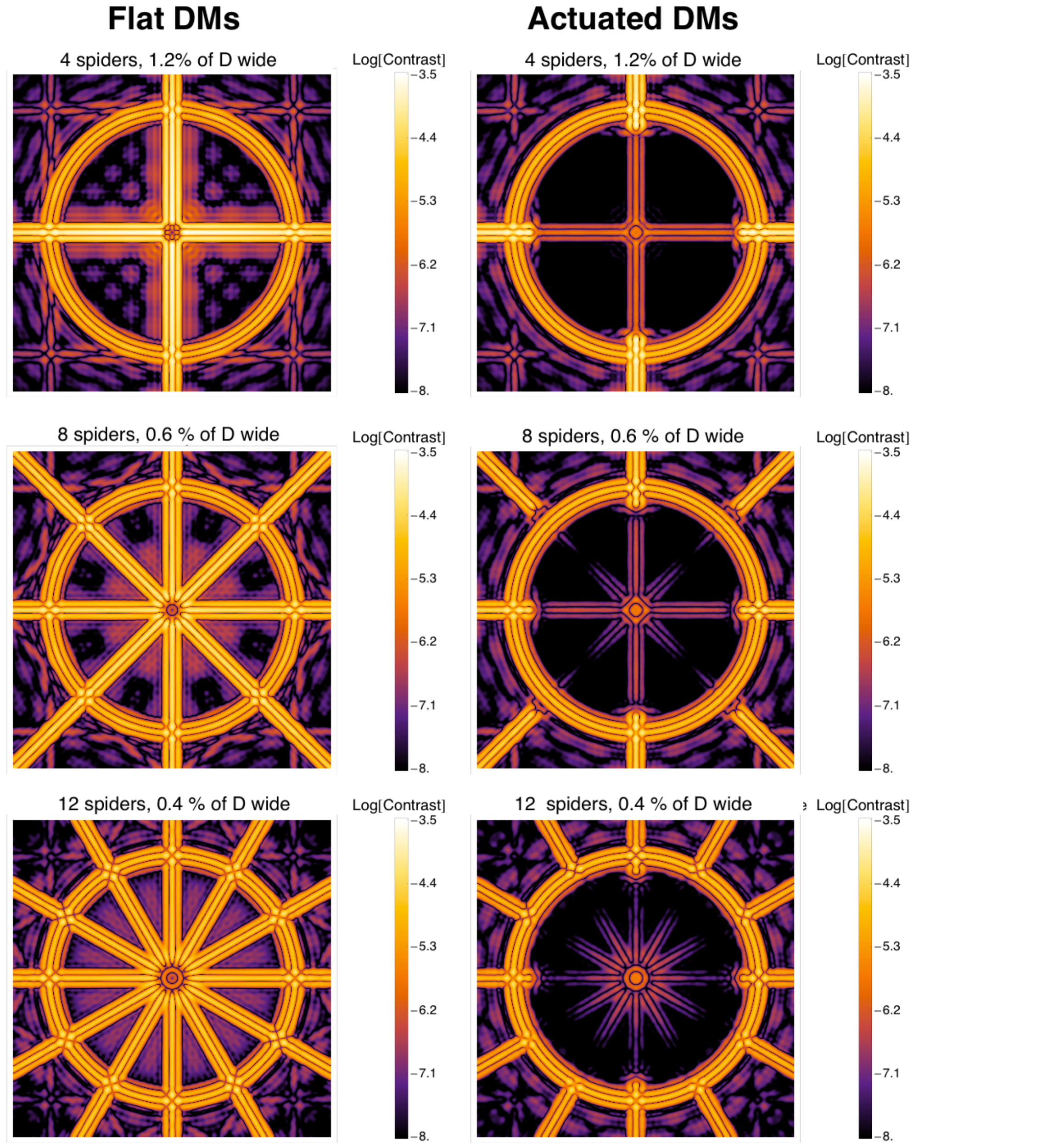}
\end{center}
\caption[]{PSFs resulting from ACAD when varying the number and thickness of secondary support structures while maintaining their covered surface constant. The surface area covered in this example is $50 \%$ greater than in the TMT example shown on Fig.~\ref{fig:ResultsTMT}. As the spiders get thinner their impact on raw contrast becomes smaller and the starlight suppression after DM correction becomes bigger. For a relatively small number of spiders ($<12$) the contrast improvement on each single structure is the dominant phenomenon, regardless of the number of spiders. ELTs designed with a moderate to large number of thin secondary support structure ($6$ to $12$) present aperture discontinuities which are easy to correct with ACAD.}
\label{fig:ResultsArea1}
\end{figure*}
We now discuss the case of Extremely Large Telescopes and provide an illustration using the example of the Thirty Meter Telescope. We considered the aperture geometry shown on the top left panel of Fig.~\ref{fig:ResultsTMT}. It consists of a pupil  37 segments across in the longest direction and a secondary of diameter $ \sim 0.12 D$  which is held by three main thick struts and six thin cables. As seen on the bottom left panel of Fig.~\ref{fig:ResultsTMT} the impact of segment gaps is minor as they diffract light beyond the OWA of the coronagraph. When using a coronagraph with a larger OWA the segment gaps will have to be taken into account, and will have to be mitigated using DMs with a larger number of actuators. In order to obtain first order estimates of the performances of ACAD on the aperture geometry shown on the top left panel of Fig.~\ref{fig:ResultsTMT}, we chose to use a coronagraph designed around a slightly oversized secondary obscuration of diameter $0.15 \; D$, with a  focal plane mask of $6 \lambda /D$ diameter, an IWA of $4 \lambda/D$ and an OWA of $30 \lambda / D$. The field at the entrance of the coronagraph after remapping by the DMs is shown on the top right panel of  Fig.~\ref{fig:ResultsTMT}.  The DM surfaces, calculated assuming 64 actuators across the pupil ($N=64$ in the Fourier expansion) and DMs of diameter $3$ cm separated by $Z = 1$ m, are shown on the middle panel of  Fig.~\ref{fig:ResultsTMT}. They are well within the stroke limit of current DM technologies. The surfaces were calculated by solving the reverse problem over an even grid of 10 cutoff low-pass spatial frequencies ranging between $30$ and $70$ cycles per apertures for the tapering kernel $\omega$. The value yielding the best contrast was chosen.  The final PSF is shown on the bottom right panel of  Fig.~\ref{fig:ResultsTMT} and features a high contrast dark hole with residual diffracted light at the location of the spiders' diffraction structures. The impact on coronagraphic contrast of secondary supports was thoroughly studied by \citet{2008A&A...492..289M}. They concluded that under a $90 \%$ Strehl ratio, the contrast in most types of coronagraphs is driven by the secondary support structures to levels ranging from $10^{-4}$ to $10^{-5}$. This, in turn, leads to a final contrast after post-processing (called Differential Imaging) of $\sim 10^{-7} - 10^{-8}$.  Fig.~\ref{fig:SlicesTMT} shows that using ACAD on an ELT pupil  yields contrasts {\em before any post-processing} which are comparable to the ones obtained by \citet{2008A&A...492..289M} after Differential Imaging. This demonstrates that should two sequential DMs be integrated into a future planet finding instrument, setting their surface deformation according to the methodology presenting above would allow this instrument to perform its scientific program at a very high contrast. Moreover the surface of the DMs could be adjusted to mitigate for the effect of missing segments at the surface of the primary (when for instance the telescope is operating while some segments are being serviced).  

\subsection{Hypothetical cases}

\subsubsection{Constant area covered by the secondary support structures}
\begin{figure}[h!]
\begin{center}
\includegraphics[width=3.5in]{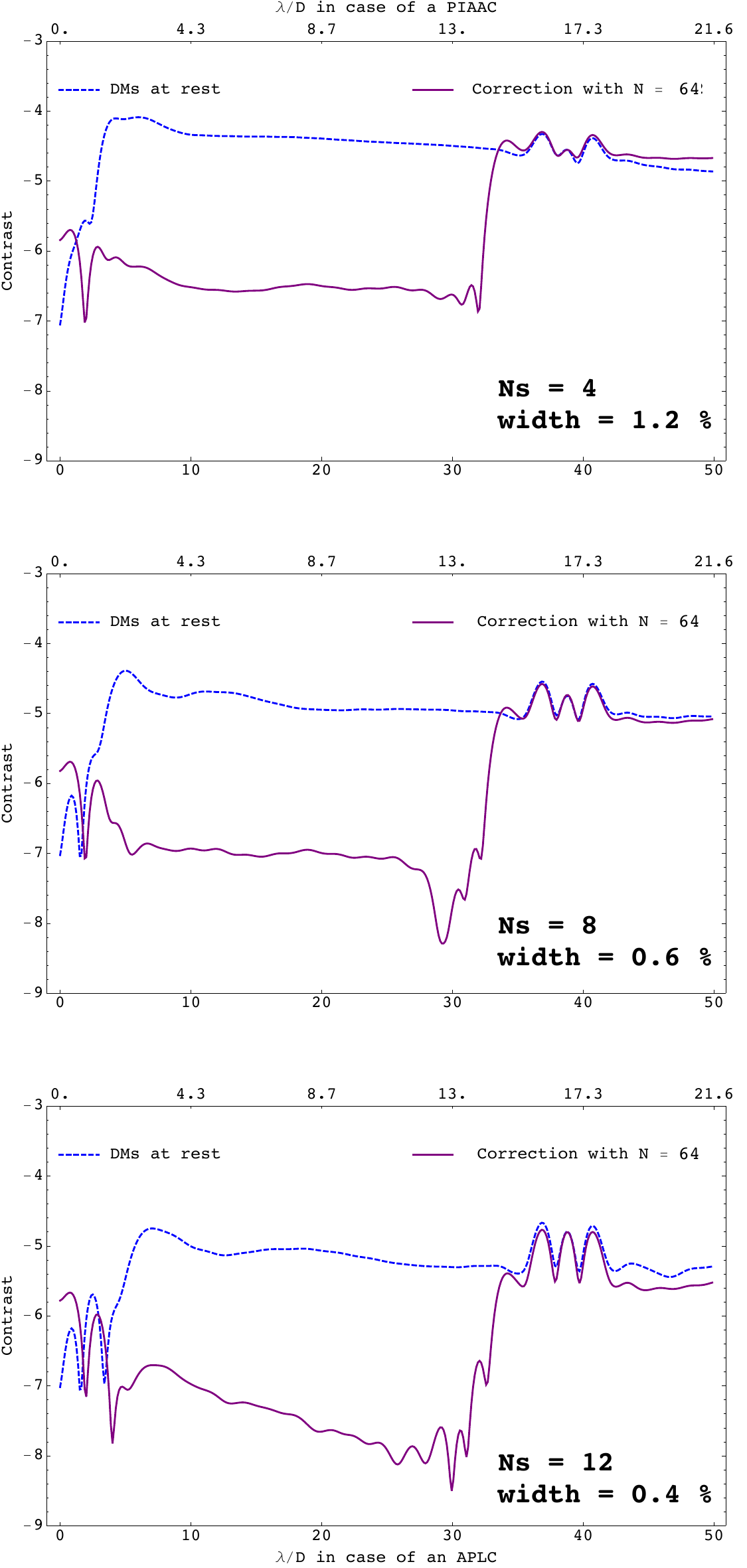}
\end{center}
\caption[]{Radial PSF profiles resulting from ACAD when varying the number and thickness of secondary support structures while maintaining their covered surface constant. The surface area covered in this example is $50 \%$ greater than in the TMT example shown on Fig.~\ref{fig:ResultsTMT}. As the spiders get thinner their impact on raw contrast becomes lesser and the starlight suppression after DM correction becomes greater. In the $12$ spiders example, at large separations, the average contrast is an order of magnitude higher than reported on Fig.~\ref{fig:SlicesTMT}.}
\label{fig:SlicesArea1}
\end{figure}

\begin{figure*}[h!]
\begin{center}
\includegraphics[width=8in]{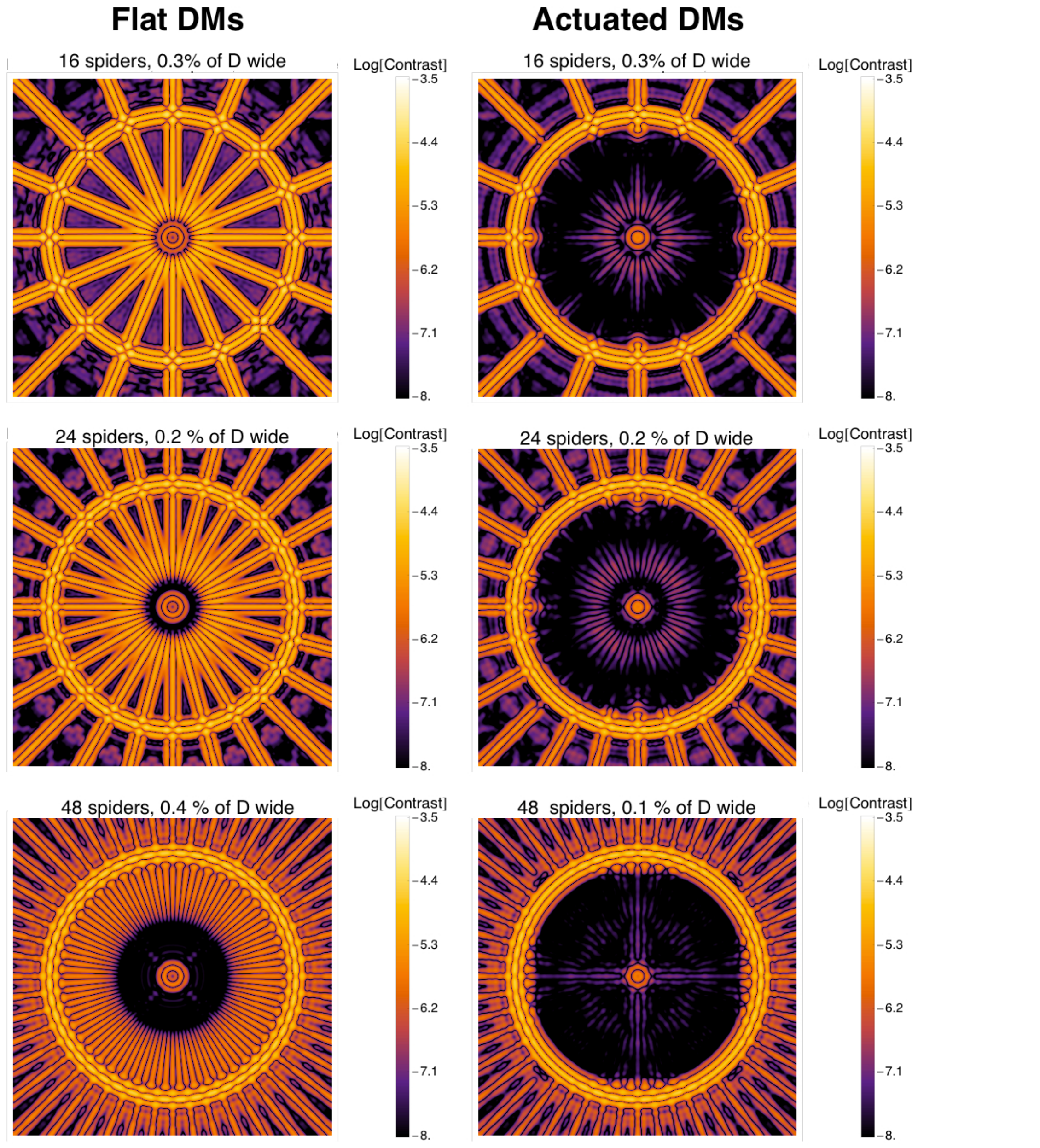}
\end{center}
\caption[]{PSFs resulting from ACAD when varying the number and thickness of secondary support structures while maintaining their covered surface constant. The surface area covered in this example is $50 \%$ greater than in the TMT example shown on Fig.~\ref{fig:ResultsTMT}. When the number of spiders increases, they produce a sharp circular diffraction feature at $N_{Spiders}/\pi \; \lambda / D$. If this number is greater than the size of the focal plane mask this structure appears in the high contrast zone and is very difficult to correct with ACAD. The brightness of this structure is mitigated by the fact that the spiders are very thin.}
\label{fig:ResultsArea2}
\end{figure*}

\begin{figure}[h!]
\begin{center}
\includegraphics[width=3.5in]{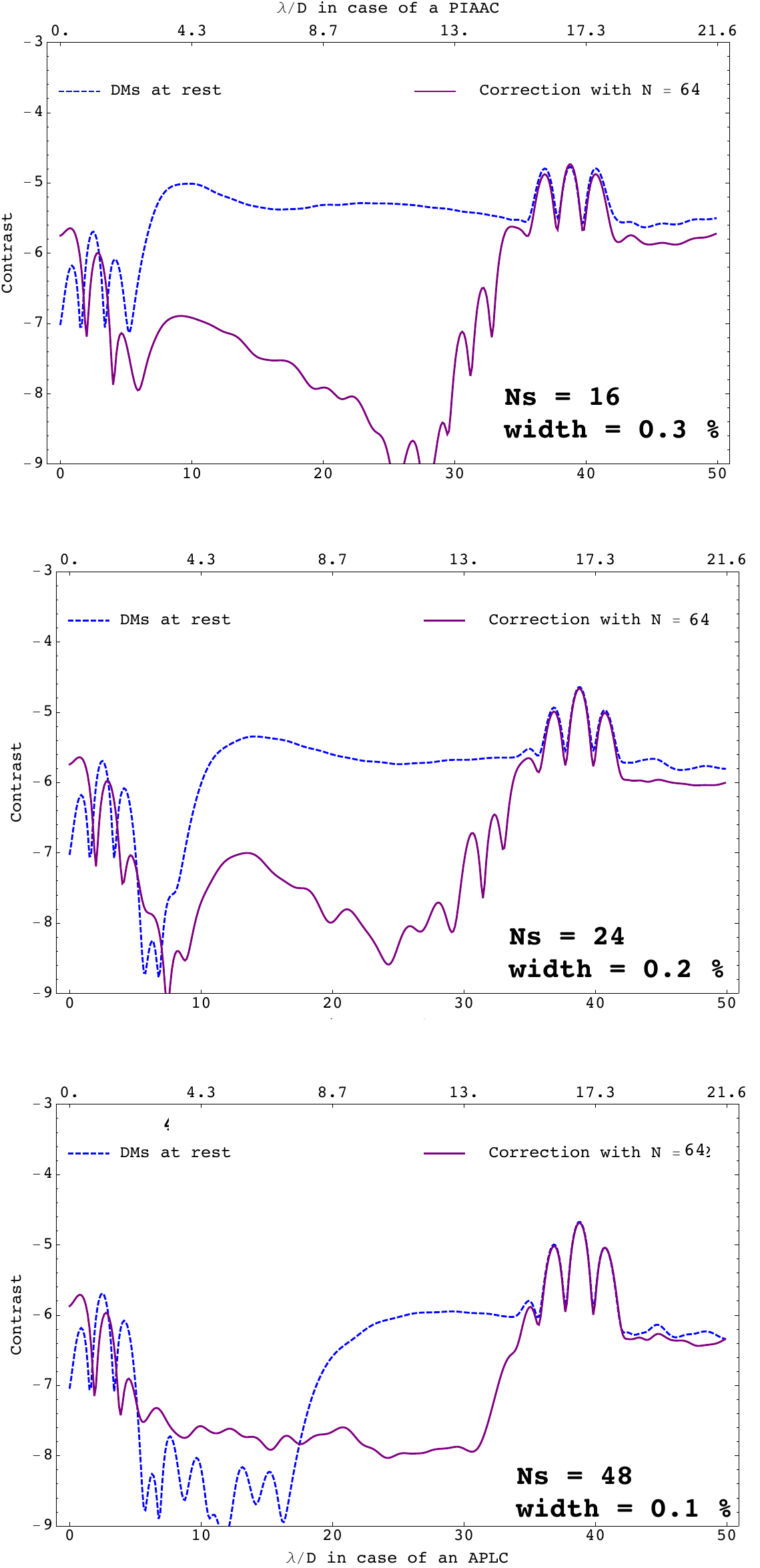}
\end{center}
\caption[]{Radial PSF profiles resulting from ACAD when varying the number and thickness of secondary support structures while maintaining their covered surface constant. The surface area covered in this example is $50 \%$ greater than in the TMT example shown on Fig.~\ref{fig:ResultsTMT}. With a large number of spiders the bright ring in the PSF structure located at $N_{Spiders}/\pi \; \lambda / D$ is difficult to correct with ACAD. However since the spiders becomes thinner its net effect on contrast after ACAD remains small.}
\label{fig:SlicesArea2}
\end{figure}

In the case of  ELTs with large number of small segments (when compared to the aperture size), gaps diffract light far from the optical axis (see  Fig.~\ref{fig:ResultsTMT} for an example). The secondary support structures are then the major source of unfriendly coronagraphic diffracted light. Under the assumption that thick structures are necessary to support the heavy secondary over the very large ELT pupils, one can use the aperture area covered by the spiders as a proxy of the secondary lift constraint. We have thus explored a series of  geometries for which the number of spiders increases as they get thinner while the overall area covered by the secondary support structures remains constant. In the examples shown from Fig.~\ref{fig:ResultsArea1} to Fig.~\ref{fig:SlicesArea2}, the area covered by the secondary support structures is $1.5$ times greater than in the TMT geometry discussed above. In all cases we used a coronagraph with a central obscuration of $0.15 \;D$, with a  focal plane mask of $6 \lambda /D$ diameter, and IWA of $4 \lambda/D$ and an OWA of $30 \lambda / D$. The surfaces were calculated by solving the reverse problem over an even grid of 10 cutoff low-pass spatial frequencies ranging between $30$ and $70$ cycles per apertures for the tapering kernel. The value yielding the best contrast was chosen. This exercise leads to several conclusions pertaining to the performances of ACAD with various potential ELT geometries. 

\vspace{0.1cm}
{\bf Clocking of the spiders with respect to the DM}
\vspace{0.1cm}

The top two panels of Fig.~\ref{fig:ResultsArea1} illustrate the importance of the clocking of the spiders with respect to the DMs actuator grid (or the Fourier grid in our case).  When the secondary support structures are clocked by $45^{\circ}$ with respect to the DM actuators they are much more attenuated by ACAD, thus yielding higher contrast. This is an artifact of the Fourier basis set chosen and would be mitigated by using DMs whose actuator placement presents circular and azimuthal symmetries \citep{2008SPIE.7015E.169W}.
\vspace{0.1cm}

{\bf Annulus in the PSF with a large number of spiders}
\vspace{0.1cm}

When the number of secondary support struts becomes very large ($>20$), an interesting phenomena occurs in the raw PSF: the spiders diffract light outside an annulus of radius  $N_{Spiders}/\pi \time \lambda / D$, just as spiderweb masks do in the case of shaped pupil coronagraphs \citep{2003ApJ...590..593V}. The ``bump'' located beyond that spatial frequency is more difficult to attenuate using the DMs (see Fig.~\ref{fig:ResultsArea1} for an illustration). ACAD creates small ripples at the edges of the remapped discontinuities and when too many discontinuities are in the vicinity of each other, then these ripples interfere constructively and hamper the starlight extinction level yielded by ACAD. 

\vspace{0.1cm}

{\bf A lot of thin spiders is more favorable than a few thick spiders}
\vspace{0.1cm}

In general decreasing the width of the spiders while increasing their number is beneficial to the contrast obtained after ACAD as illustrated on the radial averages on Fig.~\ref{fig:SlicesArea1} and Fig.~\ref{fig:SlicesArea2}. When one increases the number of spiders while decreasing their width in a classical coronagraph, the peak intensity of the diffraction pattern of one spider decreases as the squared width of the spider. The radially averaged contrast improvement without ACAD is then somewhat lesser than the square of the spider thinning factor as it is mitigated by the increasing number of spiders. When using ACAD the spiders are seen by the coronagraph as much thinner than they actually are (by a factor $\tau$) and thus the peak intensity of their diffraction pattern is lower by a factor of $\tau^2$. Our numerical experiments show that $\tau$ increases when the spider width decreases. As consequence, the  overall contrast gain after ACAD when decreasing the width of the spiders while increasing their number is greater than in the case of a classical coronagraph. When designing ELT secondary support structures and planning to correct for them using ACAD, increasing the number of spiders to 8 or even 12 has a beneficial impact on contrast as it enables each discontinuity to become thinner and thus to be corrected to higher contrast using the DMs. The PSFs of apertures with more than 12 spiders present diffraction structures which are poorly suited for correction with square DMs. While the contrast resulting from applying ACAD to such apertures is still a decreasing function of the number of spiders, Fig.~\ref{fig:SlicesArea2} shows that the net contrast gain brought by the DM based remapping is smaller than in the more gentle cases of 12 spiders. The study presented on Fig.~\ref{fig:ResultsArea1} to Fig.~\ref{fig:SlicesArea2} remains to be fully optimized for each potential design of an ELT planet finding instrument (in particular using a finer grid of cutoff spatial frequencies). It however demonstrates the flexibility of ACAD for various aperture geometries and provides a first order rule of thumb to design telescope apertures which are friendly to direct imaging of exo-planets: ``A lot of thin spiders is more favorable than a few thick spiders''. In practice the number of spiders will be limited by effects not treated in our analysis such as the mechanical rigidity, requirements on the perfection of their periodic spacing and glancing reflections from the sides of multiple spiders. We thus advocate that, should future ELTs be built with high contrast exo-planetary science as a main scientific driver, then such effect ought to be thoroughly analyzed as a large numbers of thin spiders is more favorable from a contrast standpoint when using ACAD to mitigate for pupil amplitude asymmetries. 

\subsubsection{Monolithic on-axis apertures.}
\begin{figure*}[h!]
\begin{center}
\includegraphics[width=8in]{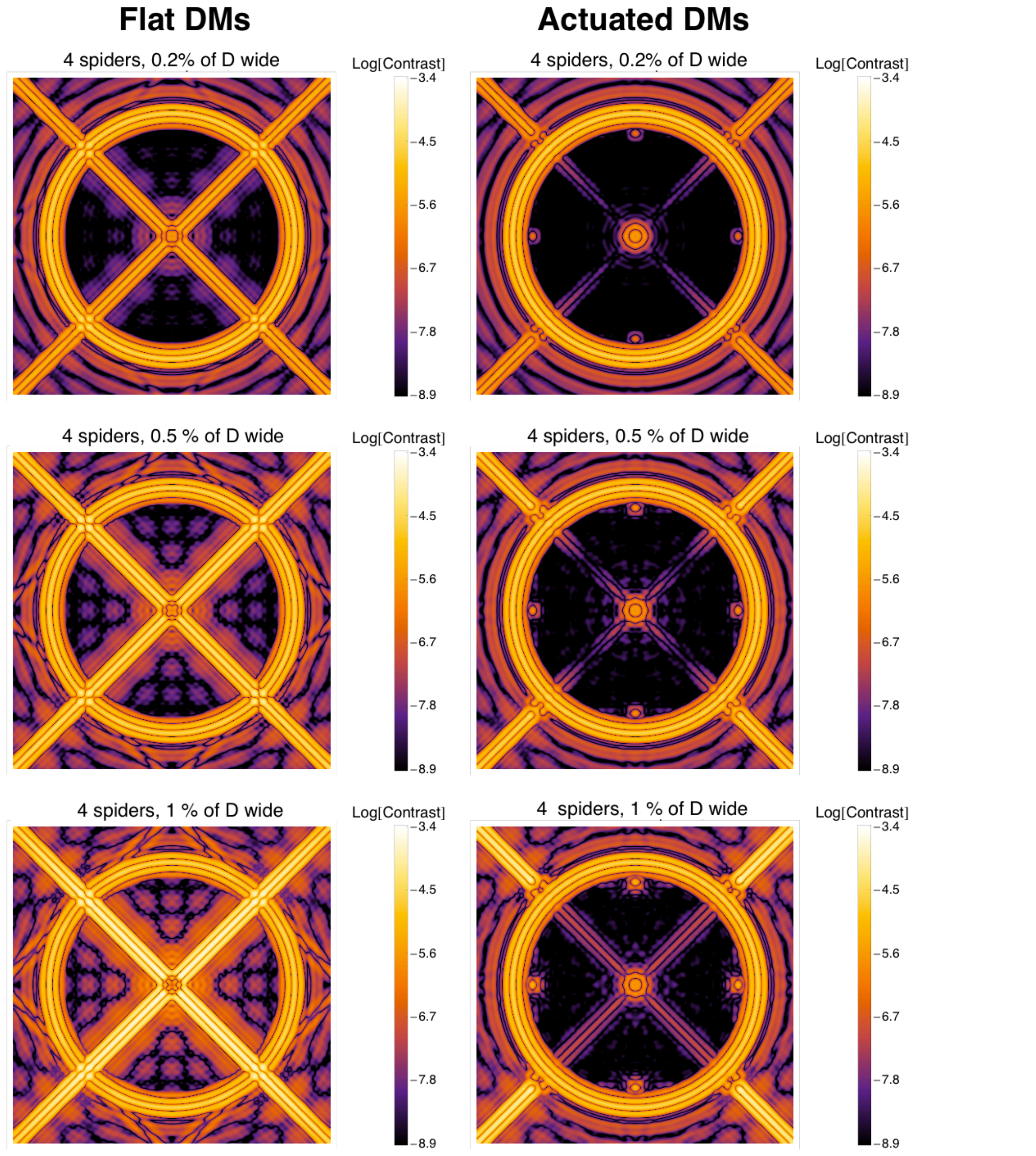}
\end{center}
\caption[]{PSFs resulting from ACAD when varying the number and thickness of secondary support structures. As the spiders get thinner their impact on raw contrast becomes lesser and the starlight suppression after DM correction becomes greater. In this case $\omega$ was optimized on a very fine grid and the aperture we clocked in a favorable direction with respect to the Fourier basis.}
\label{fig:ResultsThickness}
\end{figure*}

When discussing the case of JWST we stressed the complexity associated with the optimization of ACAD in the presence of aperture discontinuities of varying width. Carrying out such an exercise would be extremely valuable to study the feasibility of the direct imaging of exo-earth with an on-axis segmented future flagship observatory such as ATLAST \citep{2010ASPC..430..361P}. However, such an effort is computationally heavy and thus beyond the scope of the present paper, which focuses on introducing the ACAD methodology and illustrating using key basic examples.

\begin{figure}[t!]
\begin{center}
\includegraphics[width=3.5in]{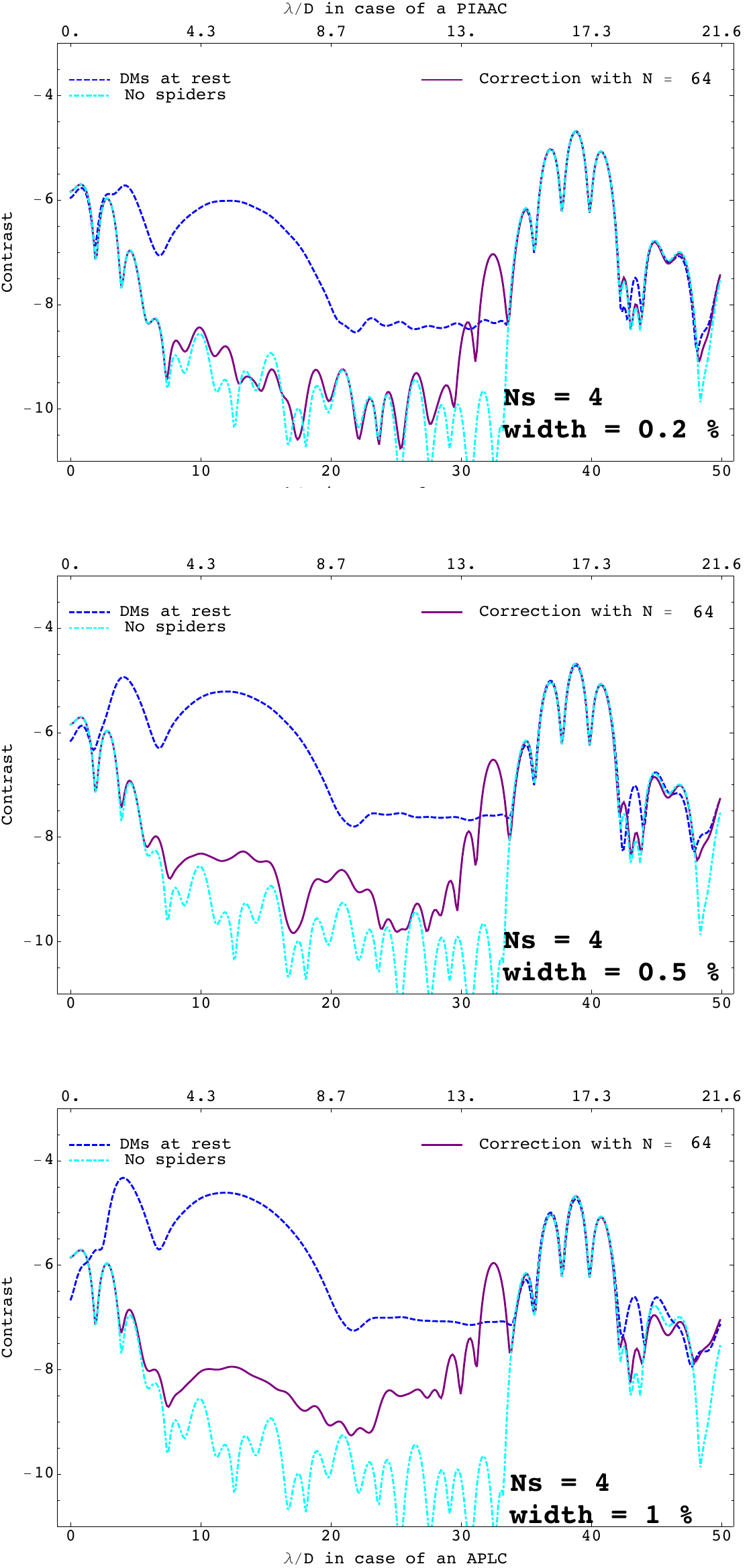}
\end{center}
\caption[]{PFSs resulting from ACAD when varying the number and thickness of secondary support structures. As the spiders get thinner their impact on raw contrast becomes lesser and the starlight suppression after DM correction becomes greater. In this case $\omega$ was optimized on a very fine grid and the aperture we clocked in a favorable direction with respect to the Fourier basis. Even for spiders as thick as $0.5 \%$ of the telescope aperture the designed contrast of the coronagraph is retrieved.}
\label{fig:SlicesThickness}
\end{figure}

So far, none of the examples in this manuscript demonstrate that ACAD can yield corrections all the way down to the theoretical contrast floor that is set by the coronagraph design. When seeking to image exo-earths from space, future missions will need to reach this limit. In order to explore this regime, we conducted a detailed study of an hypothetical on-axis monolithic telescope with four secondary support struts.  To establish the true contrast limits we varied the thickness of the spiders and for each geometries. The surfaces were calculated by solving the reverse problem over an even grid of 70 cutoff low-pass spatial frequencies ranging between $30$ and $70$ cycles per apertures for the tapering kernel. The value yielding the best contrast was chosen. In all cases we used a coronagraph with a central obscuration of $0.15 \;D$, with a  focal plane mask of $6 \lambda /D$ diameter, and IWA of $4 \lambda/D$, and an OWA of $30 \lambda / D$. Note that when using coronagraphs relying on pupil apodization these results can be readily generalized to larger circular secondary obscurations, at a loss in IWA (as shown on Fig.~\ref{fig:corono}). Moreover we clocked the telescope aperture by $45^{\circ}$ with respect to the grid of Fourier modes. We found that, indeed, the theoretical contrast floor set by the coronagraph design is met for thin spiders ($0.02 \;D$) and it is very close to be met for spiders only half the thickness of the the ones currently equipping the Hubble Space Telecope ($0.05 \;D$),  see Fig.~\ref{fig:ResultsThickness} and Fig.~\ref{fig:SlicesThickness}. Even in the case of thick struts ($0.1 \;D$) we find contrasts an order of magnitude higher than in the similar configuration on the top panel of Fig.~\ref{fig:SlicesArea1}, due to our thorough optimization of  the cutoff frequency of the tapering kernel and careful clocking of the aperture with respect to the actuators. On-axis telescopes are thus a viable option to image earth-analogs from space: their secondary support structures can be corrected down to contrast levels comparable to the target contrast of recent missions concept studies \citep{2008SPIE.7010E..59G,2010SPIE.7731E..67T}. Since the baseline wavefront control architecture for future space coronagraphs relies on two sequential DMs, ACAD does not add any extra complexity to such missions and merely consists of controlling the DMs in order to optimally compensate for the effects of asymmetric aperture discontinuities. 

\section{Discussion and future work}
\label{sec:Discuss}
\subsection{Field dependent distortion}
\begin{figure}[b!]
\begin{center}
\includegraphics[width=4in]{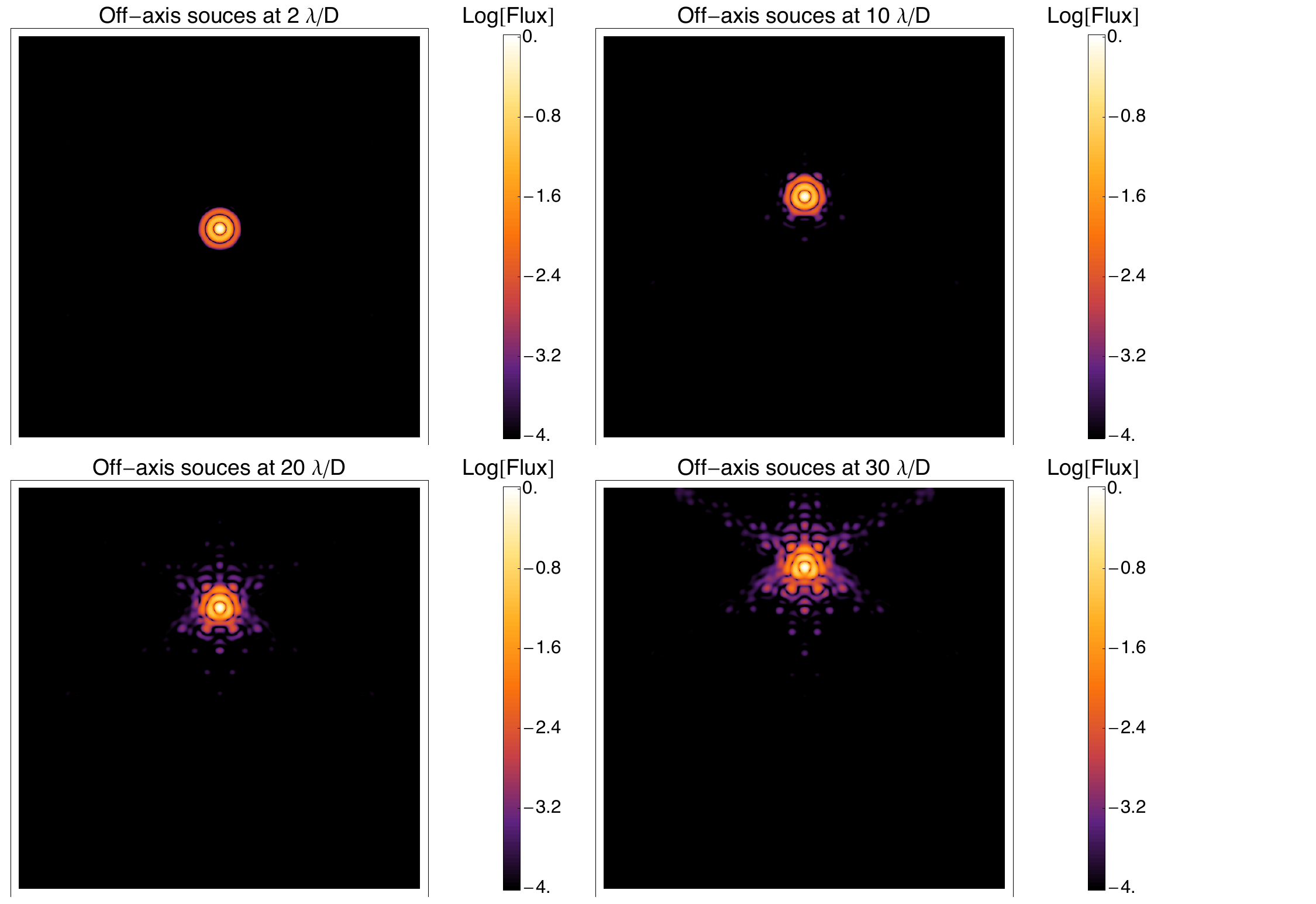}
\end{center}
\caption[]{Off-axis PSF after ACAD in the case of a geometry similar to JWST. The aspheric surface of the DMs introduce a slight field-dependent distortion. However the core of the PSF is still concentrated within the central airy disk and the DMs only have an effect on the PSF tail. Field distortion does not thus hamper the detectability of faint off-axis sources. }
\label{fig:OffAxisPSFs}
\end{figure}

Because ACAD relies on deforming the DMs surfaces in an aspherical fashion, off-axis wavefronts seen through the two DMs apparatus will be distorted, just as in a PIAA coronagraph \citep{2006ApJ...639.1129M}. However the asphericity of the surfaces in the case of ACAD operating on reasonably thin discontinuities, is much smaller than in a PIAA remapping unit. Fig.~\ref{fig:OffAxisPSFs} shows the impact on off-axis PSFs of such a distortion in the worse-case scenario of a geometry similar to JWST. We demonstrate that most of the flux remains in the central disk of radius $\lambda/D$ for all sources in the field of view of the coronagraphs considered here (all the way to $30$  cycles per capture). We conclude that, because of the small deformations of the DMs, PSF distortion will not be a major hindrance in exo-planet imaging instruments whose DMs are controlled in order to mitigate for discontinuities in the aperture. 

\subsection{Impact of wavefront discontinuities in segmented telescopes.}
\label{sec:wavefrontcontrol}
\subsubsection{General equations in the presence of incident phase errors and discontinuities}
So far we have treated primary mirrors' segmentation as a pure amplitude effect. In reality the contrast floor in segmented telescopes will be driven by both phase and amplitude discontinuities: here we explore the impact of phase errors and discontinuities occurring before two DMs whose surfaces have been set using ACAD. There are two main phenomena to be considered. The first is the conversion of the incident wavefront phase before DM1: $\frac{2 \pi}{\lambda}\Delta h_1$ into amplitude at the second mirror. The second is the projection of this wavefront phase into a remapped phase errors at DM2: $\frac{2 \pi}{\lambda}\Delta h_1(f_1(x_1,y_2),g_1(x_1,y_2))$. Since the remapping unit is designed using deformable mirrors, both DM1 and DM2  a complete correction could be attained in principle. However, the deformable mirrors are continuous while $\Delta h_1$ presented discontinuities. Thus, complete corrections for segmented mirrors might not be achieved  in practice. Below we discuss the following two main points. (1) Even if the phase wavefront error $\Delta h_1$ has discontinuities, the phase errors within in segment still drive the phase to amplitude conversion and thus the propagated amplitude at DM2. In that case treatments of these phenomenons that have already been discussed in the literature for monolithic apertures are still valid for small enough phase errors Eqs.~\ref{eq:DM1EquaDiffTilde1}-\ref{eq:DM1EquaDiffTilde} and smooth enough remapping functions. For ACAD remapping this smoothness constraint is naturally enforced by the limited number of actuators across the DM surface. In this case phase to amplitude conversion between can in principle be corrected using DM1. (2) Remapped phase discontinuities can be corrected for a finite number of spatial frequencies using a continuous phase sheet deformable mirror. We illustrate this partial correction over a $20 \%$ bandwidth using numerical simulations of a post-ACAD half dark hole created by superposing a small perturbation, computed using a linear wavefront control algorithm, to the ACAD DM2 surface.\\

\vspace{7cm}

If the incoming wavefront is  written as $\Delta h_1$ and the solution of the Monge Ampere Equation for DM1 as $h_1^{0}$ then one can conduct the analysis in Eq.~\ref{huygens} to \ref{eq:DM2EquaDiff} using $\tilde{h_1} = h_1^{0} + \Delta h_1$. Under the assumption that surface of DM2 is set as  $\tilde{h_2}$ in order to conserve Optical Path Length then one can re-write the remapping as $(\tilde{f_1},\tilde{g_1})$ defined by: 
\begin{eqnarray}
\frac{\partial \tilde{h_1}}{\partial x}\Big|_{(\tilde{f_1}(x_2,y_2),\tilde{g_1}(x_2,y_2))} &=&  \frac{\tilde{f_1}(x_2,y_2) -x_2}{Z} \nonumber \label{eq:DM1EquaDiffTilde1}\\
\frac{\partial \tilde{h_1}}{\partial y}\Big|_{(\tilde{f_1}(x_2,y_2),\tilde{g_1}(x_2,y_2))} &=&  \frac{\tilde{g_1}(x_2,y_2) - y_2}{Z}. \label{eq:DM1EquaDiffTilde}
\end{eqnarray}

\begin{widetext}
Moreover if edge ringing has been properly mitigated then the ray optics solution is valid and the field at DM2 can be written as:
\begin{equation}
E_{DM2}(x_2,y_2)  =  \left \{ \left(  \frac{ E_{DM1}}{(1+\frac{\partial^2 \tilde{h_1}}{\partial x^2})(1+\frac{\partial^2 \tilde{h_1}}{\partial y^2}) - (\frac{\partial^2 \tilde{h_1}}{\partial x \partial x})^2} \right) \bigg|_{(\tilde{f_1},\tilde{g_1})} e^{i \frac{2 \pi }{\lambda} \left( S(\tilde{f_1},\tilde{g_1}) + \tilde{h_{1}}(\tilde{f_1},\tilde{g_1}) -\tilde{ h_{2}} \right)} \right \}\bigg|_{(x_2,y_2)}
\label{eq:remappedfieldTilde}
\end{equation}
\end{widetext}
\subsubsection{Impact on the amplitude after ACAD}
\begin{figure*}[h!]
\begin{center}
\includegraphics[width=8in]{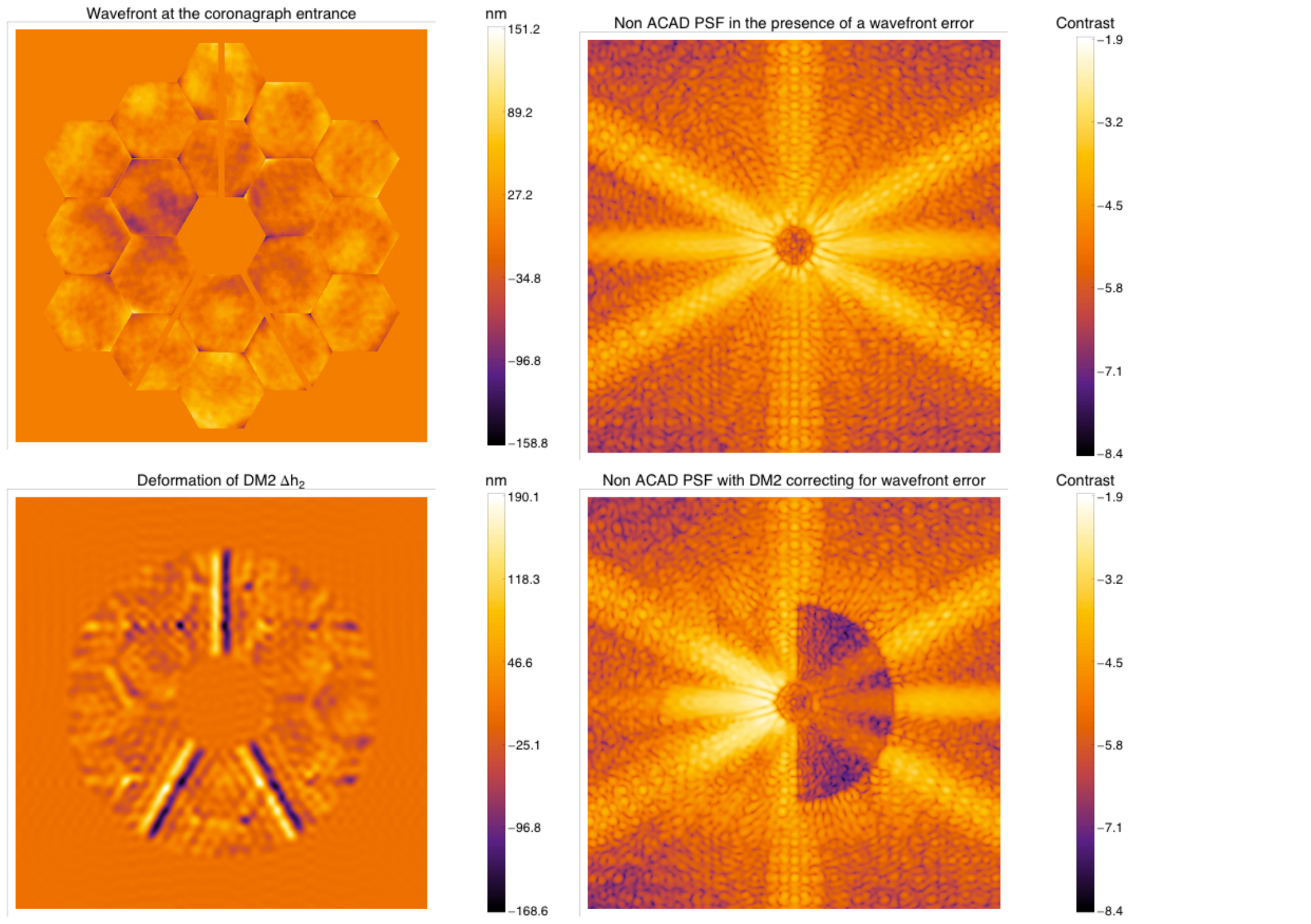}
\end{center}
\caption[]{Broadband wavefront correction  ($20 \%$ bandwidth around $700$ nm) with a single DM in segmented telescope with discontinuous surface errors. Top Left: wavefront before the coronagraph. Top Right: broadband aberrated PSF with DM at rest. Bottom Left: DM surface resulting from the wavefront control algorithm. Bottom Right: broadband corrected PSF. Note that the wavefront control algorithm seeks to compensate for the diffractive artifacts associated with the secondary support structures: it attenuates them on the right side of the PSF while it strengthens them on  the left side of the PSF. As a result the DM surface becomes too large at the pupil spider's location and the quasi-linear wavefront control algorithm eventually diverges.}
\label{fig:wavecontrolNoACAD}
\end{figure*}

We first consider the amplitude profile in Eq.~\ref{eq:remappedfieldTilde}: it is composed of two factors the remapped telescope aperture, $E_{DM1}(\tilde{f_1},\tilde{g_1})$, and the determinant of $Id + \mathcal{H}[\tilde{h_1}]$.

The first condition necessary for the incoming wavefront not to perturb the ACAD solution is:  $\Delta h_1$ is such that the remapping is not modified {\em at the pupil locations where the telescope aperture is not zero $E_{DM1} \neq 0$}. This results into the conditions
\begin{eqnarray}
\frac{\partial \Delta h_1}{\partial x} &\ll& \frac{\partial \Delta h_1^{0}}{\partial x} \mbox{ for } (x,y) \mbox{such that} E_{DM1}(x,y)\neq 0 \nonumber  \\
\frac{\partial \Delta h_1}{\partial y} &\ll& \frac{\partial \Delta h_1^{0}}{\partial y} \mbox{ for } (x,y) \mbox{such that} E_{DM1}(x,y)\neq 0 \nonumber
\end{eqnarray}
At the locations where $E_{DM1} = 0$ there is no light illuminating the discontinuous wavefront and thus the large local slopes at these location  have no impact on the remapping functions $(f_1,g_1)$. These conditions are not true in segmented telescopes that are not properly phased, for which the tip-tilt error over each segment can reach several waves. However under the assumption that the primary has been properly phased (for instance the residual rms wavefront after phasing is expect to be $\sim 1/10$ th of a wave, similar to values expected for JWST NIRCAM) then these conditions are true within the boundaries of each segment. Moreover, while the local wavefront slopes at  the segment's discontinuities do not respect this condition the incident amplitude at these points is $E_{DM1}(x,y) = 0$ and they thus do not perturb the ACAD remapping solution. 

The second necessary condition resides in the fact that  the determinant of $Id + \mathcal{H}(\tilde{h_1})$ is not equal to $det[Id + \mathcal{H}(h_1^{0})]$ {\em at the pupil locations where the telescope aperture is not zero} ought not have a severe impact on contrast. One can use the linearization in \cite{Loeper2005319} to show that:
\begin{widetext}
\begin{eqnarray}
\frac{1}{det[Id + \mathcal{H}(\tilde{h_1})]} &=& \frac{1}{det[Id + \mathcal{H}(h_1^{0})](1+\frac{(1+h_{1xx}^{0}) \Delta h_{1yy}+ (1+h_{1yy}^{0}) \Delta h_{1xx}- 2*h_{1xy}^{0}\Delta h_{1xy}}{det[Id + \mathcal{H}(h_1^{0})]})} \\
\frac{1}{det[Id + \mathcal{H}(\tilde{h_1})]} &=& \frac{1}{det[Id + \mathcal{H}(h_1^{0})](1+\Delta A(\Delta h_{1}))} 
\end{eqnarray}
\end{widetext}

The perturbation term $\Delta A(\Delta h_{1})$ corresponds to the full non-linear expression of the phase to amplitude conversion of wavefront errors that occurs in pupil remapping units. In \citet{Pueyo:11} we derived a similar expression in the linear case, when $\Delta h_{1} \ll \lambda$ and showed that in the pupil regions where the the beam is converging this phase to amplitude conversion was enhanced when compared to the case of a Fresnel propagation.  In a recent study \citet{2011SPIE.8151E..12K} presented simulations predicting that this effect was quite severe in PIAA coronagraphs and can limit the broadband contrast after wavefront control unless DMs where placed before the remapping unit. In principle ACAD will not suffer from this limitation as the first aspherical surface of the remapping unit is actually a Deformable Mirror that can actually compensate for $\Delta h_1$, before any phase to amplitude wavefront modulation occurs. Devising a wavefront controller that relies on DM1 requires moreover a computationally efficient model to propagate arbitrary wavefronts thought ACAD. Such a tool was developed in \citet{krist:77314N} assuming azimuthally symmetric geometries.  Since devising such a tool in ACAD's case, in the asymmetric case, represents a substantial effort well beyond the problem of prescribing ACAD DM shapes, we chose not to include such simulations in the present manuscript. Since we are using Deformable Mirrors with a limited number of actuators, ACAD remapping is in general less severe than in the case of PIAA. We thus expect the results regarding the  wavefront correction before the remapping unit reported in \citet{2011SPIE.8151E..12K} to hold. This is provided that the DM actuators can adequately capture the high spatial frequency content of $\Delta h_1$  to create a dark hole in the coronagraphic PSF. We tackle this particular aspect next when discussing the case of phase errors, in the absence of wavefront phase to amplitude conversion. Once again note that while the local wavefront curvatures are very large at the segment's discontinuities, the incident amplitude at these points is $E_{DM1}(x,y) = 0$ and they thus they do not have an impact on the ACAD phase to amplitude modulation. In practice if the DM is not exactly located at a location conjugate to the telescope pupil the actual wavefront discontinuities will be slightly illuminated and might perturb the remapping functions and the phase to amplitude conversion. While this might tighten requirements regarding the positioning of DM1 in the direction of the optical axis we do not expect this effect to be a major obstacle to successful ACAD implementations. 
\subsubsection{Impact on the phase after ACAD}
In practice, when $\Delta h_1$ presents discontinuities, the surface of DM2 cannot be set to the deformation $\tilde{h_2}$ that  conserves Optical Path Length, since we work under the assumption that  the DMs has a continuous phase-sheet. While this has no impact on the discussion above regarding the amplitude of $E_{DM2}$, since DM1 is solely responsible for this part,  it ought to be taken into account when discussing the phase at DM2. Under the assumption that  $\Delta h_1$ does not perturb the nominal ACAD remapping function then one can show that the phase at DM2 is: 
\begin{eqnarray}
& & arg[E_{DM2} (x_2,y_2)] = \nonumber \\
 &=& \frac{2 \pi}{\lambda} (\Delta h_1(f_1^{0}(x_2,y_2),g_1^{0}(x_2,y_2)) + \Delta h_2 (x_2,y_2))
\end{eqnarray}

where $\Delta h_2 (x_2,y_2)$ is a small continuous surface deformation superposed to the ACAD shape of DM2 and $\Delta h_1(f_1^{0}(x_2,y_2),g_1^{0}(x_2,y_2))$ is the telescope OPD seen through the DM based remapping unit. This second term presents phase discontinuities whose spatial scale has been contracted by ACAD. When these discontinuities are very small then their high spatial frequency content does not disrupt the ability of DM2 to correct for low to mid-spatial frequency wavefront errors wavefront errors. However as the discontinuities become larger their high spatial frequency content can fold into the region of the PSF that the DMs seek to cancel. These ``frequency folding'' speckles are highly chromatic \citep{Giveon:06} and can have a severe impact on the spectral bandwidth of a coronagraph whose wavefront is corrected using a continuous DM.

In order to assess the impact of this phenomenon, we conducted a series of simulations based on single DM wavefront control algorithm that seeks to create a dark hole in one half of the image plane at in as in \citet{2006ApJ...638..488B}. We use the example of a geometry similar to JWST and work under the assumption that the discontinuous wavefront incident to the coronagraph has the same spatial frequency content as a JWST NIRCAM Optical Path Difference that has been adjusted to $70 $ nm rms in order to mimic a visible Strehl similar to the near-infrared Strehl of JWST.  The non-linear wavefront and sensing and control problem associated with phasing a primary mirror to such level of precision is undoubtedly a colossal endeavor and is well beyond the scope of this paper. In this section we work under the assumption that the primary mirror either has been phased to such a level, that the wavefront discontinuities are no larger than $200$ nm peak to valley or that the wavefront has been otherwise corrected down to this specification using a segmented Deformable Mirror that is conjugate with the primary mirror. Moreover we assume (1) that the residual post-phasing wavefront map has been characterized and can be used in order to build the linear model underlying the wavefront controller (2) the focal plane wavefront estimator (carried using DM diversity as in \citet{2006ApJ...638..488B} for instance)  is capable to yield an exact estimate of the complex electrical field at the science camera. Underlying this last assumption is the overly optimistic premise that wavefront will remain unchanged over the course of each high-contrast exposure. While this is not a realistic assumption one could envision the introduction of specific wavefront sensing schemes, with architectures similar to the one currently considered for low order wavefront sensors on monolithic apertures \citep{2009ApJ...693...75G,2011SPIE.8126E..11W}, or using a separate metrology system. The results presented here are thus limited to configurations for which segment phasing will be dynamically compensated using specific sensing and control beyond the scope of this paper. As this section merely seeks to address the controllability of wavefront errors in segmented telescopes we chose to conduct our simulations with a perfect estimator. Finally we use the stroke minimization wavefront control algorithm presented in \citet{Pueyo:09} to ensure convergence for as many iteration as possible.  We first tested the case of a segmented telescope in the absence of  ACAD, using a azimuthally symmetric coronagraph and a single DM. We sough to create a Dark Hole between $5$ and $28 \; \lambda_0/D$ under a $20 \%$ bandwidth with $\lambda_0 = 700$ nm. Fig~\ref{fig:wavecontrolNoACAD} shows the results of such a simulation. The DM can indeed  correct for the discontinuities over a broadband in one half of the image plane. However the wavefront control algorithm seeks to compensate for the diffractive artifacts associated with the secondary support structures: it attenuates them on the right side of the PSF while it strengthens them on  the left side of the PSF. As a result the DM surface becomes too large at the pupil spider's location and the quasi-linear wavefront control algorithm eventually diverges for contrasts $\sim 10^{6}$. 
\begin{figure*}[t!]
\begin{center}
\includegraphics[width=8in]{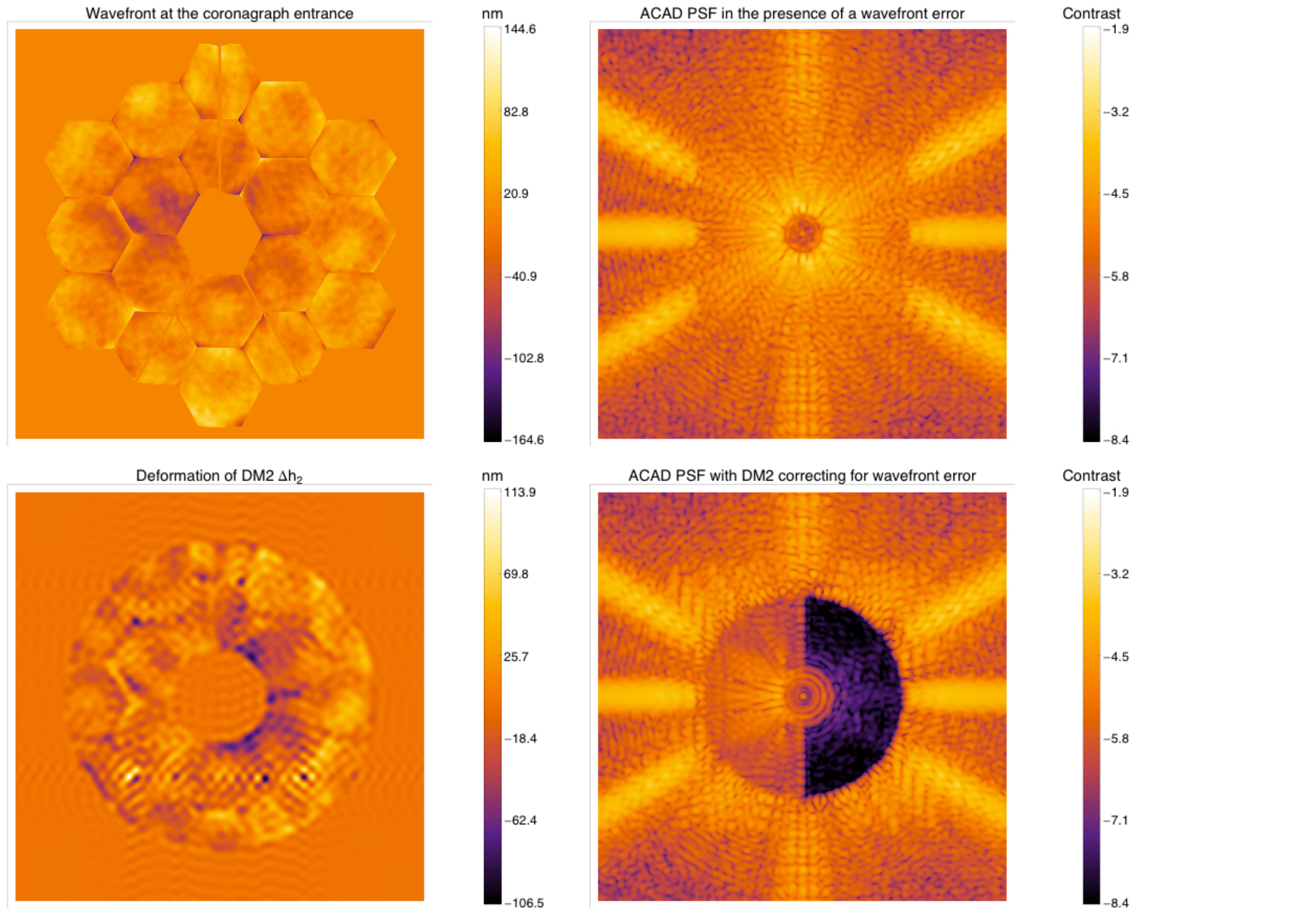}
\end{center}
\caption[]{Broadband wavefront correction  ($20 \%$ bandwidth around $700$ nm) in a segmented telescope whose pupil has been re-arranged using ACAD. The surface of the first DM is set according to the ACAD equations. The surface of the second DM  is the sum of the ACAD solution and a small perturbation calculated using a quasi-linear wavefront control algorithm . Top Left: wavefront before the coronagraph. Note that the ACAD remapping has compressed the wavefront errors near the struts and the segment gaps. Top Right: broadband aberrated PSF with DMs set to the ACAD solution. Bottom Left: perturbation of DM2's surface resulting from the wavefront control algorithm. Bottom Right: broadband corrected PSF. The wavefront control algorithm now yields a DM surface that does not feature prominent deformations at the location of the spiders.  Most of the DM stroke is located at the edge of the segments, at location of the wavefront discontinuities. There, the DM surface eventually becomes too large and the quasi-linear wavefront control algorithm diverges. However this occurs higher contrasts than in the absence of ACAD.}
\label{fig:wavecontrolACAD}
\end{figure*}
We then proceeded to simulate the same configuration in the presence of two DMs whose surface at rest was calculated using ACAD. Since there does not exist a model yet to propagate arbitrary wavefronts through ACAD (the models in \citet{2011SPIE.8151E..12K} only operate under the assumption of an azimuthally symmetric remapping) we can only use the second DM for wavefront control. We work under the assumption that the incident wavefront does not perturb the nominal ACAD remapping (which is true in the case of the surface map we chose for our example) and that the arguments in \citet{krist:77314N} hold so that phase to amplitude conversion in ACAD can be compensated by actuating DM1. Frequency folding will then be the phenomenon responsible for the true contrast limit. In this section we are  interested in exploring how this impacts the controllability of wavefront discontinuities using continuous phase-sheet DMs. We used a azimuthally symmetric coronagraph and  superposed our wavefront control solution to DM2. We sough to create a Dark Hole between $5$ and $28 \; \lambda_0/D$ under a $20 \%$ bandwidth with $\lambda_0 = 700$ nm.  Fig~\ref{fig:wavecontrolNoACAD} shows the results of such a simulation. When the incident wavefront is small enough it is indeed possible to superpose a ``classical linear wavefront control'' solution to the non-linear ACAD DM shapes in order to carve PSF dark holes. The wavefront control algorithm now yields a DM surface that does not feature prominent deformations at the location of the spiders.  Most of the DM stroke is located at the edge of the segments, at location of the wavefront discontinuities and seek to correct the frequency folding terms associated with such discontinuities. At these locations the DM surface eventually becomes too large and the linear wavefront control algorithm diverges. However this divergence occurs at contrast levels much higher than when the ACAD solution is not applied to the DMs.  These simulations show that indeed discontinuous phases can be corrected using the second DM of a ACAD whose surfaces have preliminary been set to mitigate the effects of spiders and segment gaps. 

\begin{figure}[t!]
\begin{center}
\includegraphics[width=3.5in]{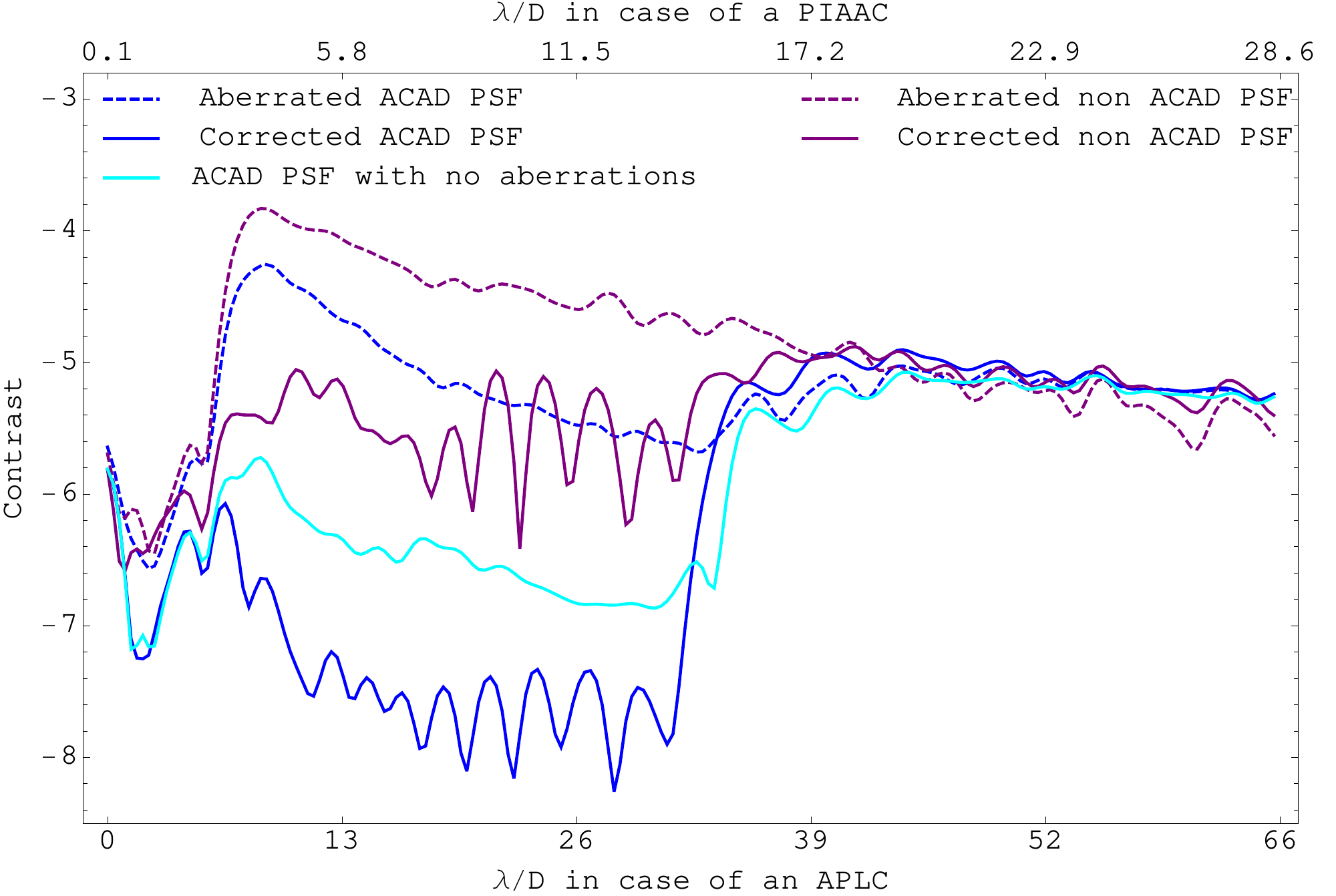}
\end{center}
\caption[]{Radial average in the half dark plane of the PSFs on Fig~\ref{fig:wavecontrolNoACAD} and Fig.~\ref{fig:wavecontrolACAD}. In the presence of wavefront discontinuities corrected using a continuous membrane DM, ACAD still yields, over a $20 \%$ bandwidth around $700$ nm, PSF with a contrast 100 larger than in a classical segmented telescope. Moreover this figure illustrates that since it is based on a true image plane metric, the wavefront control algorithm can be used ( within the limits of its linear regime) to improve upon the ACAD DM shapes derived solving the Monge Ampere Equation.}
\label{fig:wavecontroSlices}
\end{figure}

\subsection{Ultimate contrast limits}
Assuming that edge ringing has been properly mitigated, so that the ray optics approximation underlying the calculation of the DMs shapes is valid, one can wonder about the ultimate contrast limitations of the results presented in this manuscript. Increasing the number of actuators would have dramatic effects on contrast if the actuator count would be such that $N>D/d$ where $d$ is the scale of the aperture discontinuities. Unfortunately current DM technologies are currently far from such a requirement and the solutions presented here are in the regime where $ N \ll D/d$. In this regime $N$ only has a marginal influence on contrast when compared to the impact of the cutoff frequency of the tapering kernel. In the regime described here varying the actuator count only changes the size of the corrected region. 

The residual PSF artifacts in Figs.~\ref{fig:ResultsJWST} to \ref{fig:SlicesThickness} follow the direction of the initial diffraction pattern associated with secondary support structures and segments. When addressing the problem of aperture discontinuities by solving the Monge-Ampere Equation, ACAD calculates the DM shapes based on a pupil plane metric and thus mostly focuses on attenuating these structures with little regard to the final contrast. It is actually quite remarkable that such a pupil-only approach yields levels of starlight extinction of two to three orders of magnitude. A more appropriate metric would be the final intensity distribution in the  post-coronagraphic image plane. However, as discussed in \S.~\ref{sec:physicsofAM} classical wavefront control algorithms based on a linearization of the DMs deformations around local equilibrium shapes (such as the ones presented in \citet{2006ApJ...638..488B,Giveon:07} in the one DM case or \citet{Pueyo:09} for one or two DMs) cannot be used to compensate the full aperture discontinuities. This is illustrated in Fig~\ref{fig:wavecontrolNoACAD}, where the DM surface in the vicinity of spiders becomes too large  after a certain number of iterations, which leads the iterative algorithm to diverge. When attempting to circumvent this problem by recomputing the linearization at each iteration, we managed to somewhat stabilize the problem for a few iterations and reached marginal contrast improvements, but the overall algorithm remained unstable unless a prohibitively small step size was used. This is the problem which motivated our effort to calculate the DM shapes as the full non-linear solution of the Monge-Ampere Equation. While doing so yields significant contrast improvements in both the case of JWST like geometries, TMT and on axis-monolithic apertures similar, this approach does not give a proper weight to the spatial frequencies of interest for high contrast imaging. We mitigated this effect by giving a strong weight to the spatial frequencies of interest (in the Dark Hole) when solving the Monge Ampere Equation. \\

\begin{figure}[t!]
\begin{center}
\includegraphics[width=3.5in]{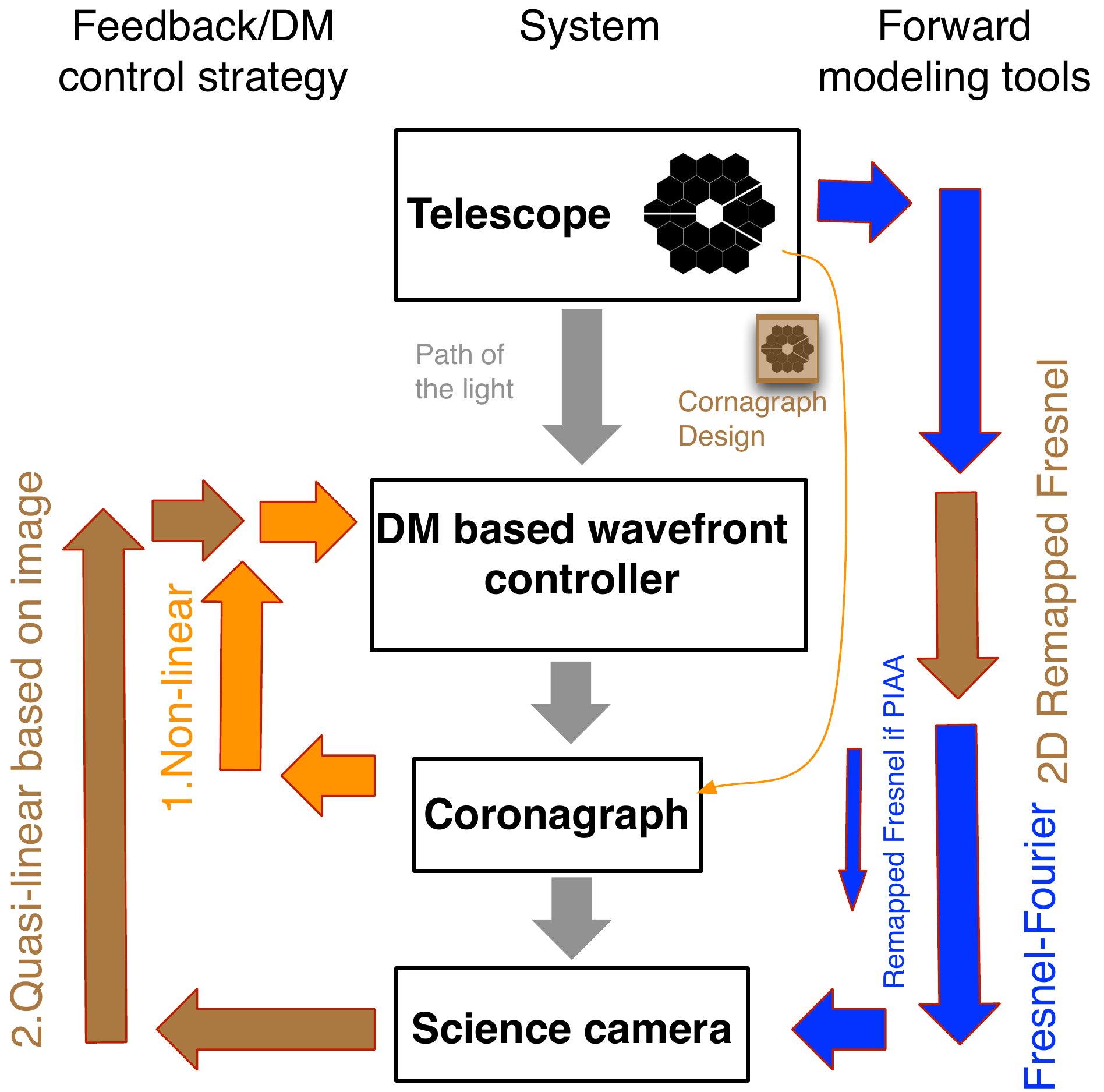}
\end{center}
\caption[]{Future work towards higher contrasts with ACAD. The blue and orange colors respectively represent the current state of the art in wavefront control and the work described in the present manuscript, as in Fig.~\ref{fig:State}. In brown are listed the potential avenues to further the contrasts presented herein: 1) combining ACAD with coronagraphs designed on segmented and/or on-axis apertures, 2) using diffractive models to close a quasi-linear focal plane based loop using a metric whose starting point corresponds to the DM shapes calculated in the non-linear regime. }
\label{fig:Future}
\end{figure}

The next natural step is thus to use non-linear solutions presented herein to correct for the {\em bulk} of the aperture discontinuities and to serve as a starting point for classical linearized waveform control algorithms, as illustrated on Fig.~\ref{fig:Future}. Fig.~\ref{fig:wavecontroSlices} indeed illustrates that when superposing an image plane based wavefront controller to the Monge Ampere ACAD solution, the contrast can be improved beyond the floor shown on Fig.~\ref{fig:SlicesJWST}. However one DM solutions, are of limited interest as they only operate efficiently over a finite bandwidth and over half of the image plane. ACAD yields a true broadband solution, and consequently it would be preferable to use the two DMs in the quasi-linear regime to quantify the true contrast limits of ACAD . In such a scheme the DM surfaces are first evaluated as the solution of the Monge-Ampere Equation and then adjusted using the image plane based wavefront control algorithm presented in \citet{Pueyo:09}. However such an exercise requires efficient and robust numerical algorithms to evaluate Eq.~\ref{eq::propint}. Such tool only exist so far in the case of azimuthally symmetric remapping units \citep{krist:77314N}.  Developing such numerical tools is thus of primary interest to both quantifying the chromaticity and the true contrast limits achievable with on-axis and/or segmented telescopes. 
 
\section{Conclusion}

We have introduced a technique that takes advantage of the presence of Deformable Mirrors in modern high-contrast coronagraph to compensate for amplitude discontinuities in  on-axis and/or segmented telescopes. Our calculations predict that  this {\em high throughput} class of solutions operates {\em under broadband illumination} even in the presence of reasonably small wavefront errors and discontinuities. Our approach relies on controlling two sequential Deformable Mirrors in a non-linear regime yet unexplored in the field of high-contrast imaging. Indeed the mirror's shapes are calculated as the solution of the two-dimensional pupil remapping problem, which can be expressed as a non-linear partial differential equation called Monge Ampere Equation. We called this technique Active Compensation of Aperture Discontinuities. While we illustrated the efficiency of ACAD using Apodized Pupil Lyot  and Phase Induced Amplitude Coronagraph, it is is applicable to all types of coronagraphs and thus enables one to translate the past decade of investigation in coronagraphy with unobscured monolithic apertures to a much wider class of telescope architectures. Because ACAD consists of a simple remapping of the telescope pupil, it is a true broadband solution. Provided that the coronagraph chosen operates under a broadband illumination, ACAD allows high contrast observations over a large spectral bandwidth as pupil remapping is an achromatic phenomenon. We showed that wavelength edge diffraction artifacts, which are the source of spectral bandwidth limits in PIAA coronagraphs (also based on pupil remapping), are no larger than classical Fresnel ringing. We thus argued that they will only marginally impact the spectral bandwidth of a coronagraph whose input beam has been corrected with ACAD. The mirror deformations we find can be achieved, both in curvature and in stroke, with technologies currently used in Ex-AO ground based instruments and in various testbeds aimed at demonstrating high-contrast for space based applications. Implementing ACAD on a given on-axis and/or segmented thus does not require substantial technology development of critical components. 

For geometries analogous to JWST we have demonstrated that ACAD can achieve at least contrast $\sim 10^{-7}$, provided that dynamic high precision segment phasing can be achieved. For TMT and ELT,  ACAD can achieve at least contrasts $\sim 10^{-8}$. For on-axis monolithic observatories the design contrast of the coronagraph can be reached with ACAD when the secondary support structures are $5$ times thinner than on HST. When they are just as thick as HST contrasts as high as $10^{8}$ can be reached. These numbers are, however, conservative: an optimal solution can be obtained by fine tuning the control term in the  Monge Ampere Equation to the characteristic scale of each discontinuity. As our goal was to introduce this technique to the astronomical community and emphasize its broad appeal to a wide class of  architectures  (JWST,ATLAST,HST,TMT,E-ELT) we left this observatory specific exercise for future work. 

The true contrast limitation of ACAD resides in the fact that the Deformable Mirrors are controlled using a pre-coronagraph pupil based metric. However, as illustrated in Fig.~\ref{fig:Future}, the solution provided by ACAD can be used as the starting point for classical linearized waveform control algorithms based in image plane diagnostics.  In such a control strategy, the surfaces are first evaluated as the solution of the Monge- Ampere Equation and then adjusted using the quasi-linear method presented in \citet{Pueyo:09}. This control strategy requires efficient and robust numerical algorithms to evaluate the full diffractive propagation in the remapped Fresnel regime. All the contrasts reported here are achieved without aberrations and we showed that in practice, quasi-linear DM controls based on images at the science camera will have to be superposed to the ACAD solutions. Finally, as ACAD is broadly applicable to all types of coronagraphs, the remapped pupil can be used as the entry point to relax the design of coronagraphs that do operate on segmented apertures such as discussed in \citet{2011OExpr..1926796C,2010ApJS..190..220G}, also illustrated in Fig.~\ref{fig:Future}. ACAD is thus a promising tool for future high contrast imaging instruments on a wide range of observatories as it will allow astronomers to devise high throughput broadband solutions for a variety of coronagraphs. It only relies on hardware (Deformable Mirrors) that have been extensively tested over the past ten years. Finally since ACAD can operate with all type of coronagraphs and it renders the last decade of research on high-contrast imaging technologies with off-axis unobscured apertures applicable to much broader range of telescope architectures.

\acknowledgements 

The authors thank Dr Bruce Macintosh who steered our attention towards this problem and provided invaluable guidance in the early stages of this manuscript. The authors also thank the anonymous referee for insightful comments which greatly improved this manuscript, Dr Remi Soummer, Dr Marshall Perrin, Prof. N. Jeremy Kasdin and Prof. Robert Vanderbei for very helpful discussions and Matt Sheckells for his help coding the linear wavefront control algorithms. This material is partially based upon work supported by the National Aeronautics and Space Administration under Grant NNX12AG05G issued through the Astrophysics Research and Analysis (APRA) program . This work was performed in part under contract with the California Institute of Technology funded by NASA through the Sagan Fellowship Program executed by the NASA Exoplanet Science Institute.



\end{document}